\documentclass[12pt,a4paper,twoside,shownumbers,manualsort]{article}

\usepackage[english]{babel}
\usepackage{graphicx}
\usepackage{rotating}
\usepackage{amsmath}
\usepackage{amssymb}
\usepackage{flafter}
\def\today{\ifcase\month\or
  January\or February\or March\or April\or May\or June\or
  July\or August\or September\or October\or November\or December\fi
        \space \number\year}

\def\versionno{S.~Profumo}
\catcode`\@=11
{\count255=\time\divide\count255 by 60
\xdef\hourmin{\number\count255}
\multiply\count255 by-60\advance\count255 by\time
\xdef\hourmin{\hourmin:\ifnum\count255<10 0\fi\the\count255}}
\def\ps@draft{\let\@mkboth\@gobbletwo
    \def\@oddhead{}
    \def\@oddfoot{\hbox to 7 cm{\tiny \versionno
       \hfil}\hskip -7cm\hfil\rm\thepage \hfil {\tiny\draftdate}}
    \def\@evenhead{}\let\@evenfoot\@oddfoot}
\def\draftdate{\number\day.\ \today \ \ \hourmin }

\catcode`\@=12
\newcommand{\thisfile}[1]{\renewcommand{\versionno}[0]%
{S.~Profumo --- File: bt.tex}}

\renewcommand{\thefootnote}{\fnsymbol{footnote}}

\def\sss{\scriptscriptstyle}

\def\beq{\begin{equation}}
\def\eeq{\end{equation}}

\newcommand{\neut}{\tilde\chi}
\newcommand{\stau}{\tilde\tau_{\sss 1}}
\newcommand{\sneu}{\tilde\nu_\tau}
\newcommand{\sbot}{\tilde b_{\sss 1}}

\def\mneut{m_{\tilde \chi^0_{\sss 1}}}

\begin{document}

\begin{titlepage}
\hfill
\vbox{
    \halign{#\hfil        \cr
           SISSA 29/2003/EP \cr
           } 
      }  
\vspace*{15mm}
\begin{center}
\LARGE{{\bf Neutralino dark matter,\\ $b$-$\tau$ Yukawa unification and\\  non-universal sfermion masses}}

\vspace*{7mm}
{\large S.~Profumo}\footnote{E-mail: profumo@sissa.it}\\

\vspace*{.5cm}

\small

{\it Scuola Internazionale Superiore di Studi
Avanzati (SISSA-ISAS), I-34014 Trieste, Italy\\}
\vskip .15cm
{\it and Istituto Nazionale di Fisica Nucleare, Sezione di Trieste,
I-34014 Trieste, Italy\\}
\end{center}
\vspace{1cm}
\begin{abstract}

We study the
implications of minimal non-universal boundary conditions in the
sfermion soft SUSY breaking (SSB) masses of mSUGRA. We impose asymptotic $b$-$\tau$ Yukawa coupling
unification and we resort to a
parameterization of the deviation from universality in the SSB motivated by the multiplet structure of $SU(5)$ GUT.
A set of cosmo-phenomenological constraints,
including the recent results from WMAP, determines the allowed
parameter space of the models under consideration. We highlight a ``new coannihilation
corridor'' where $\neut-\sbot$
and $\neut-\stau-\sneu$ coannihilations significantly contribute
 to the reduction of the neutralino relic density.

\vskip 15.mm \noindent\
\vskip -0mm \noindent\
\end{abstract}
\end{titlepage}
\newpage
\renewcommand{\thefootnote}{\arabic{footnote}}
\setcounter{footnote}{0}
\setcounter{page}{1}


\section{Introduction}

The minimal supersymmetric extension of the standard model (MSSM) and grand unification are often regarded as the main ingredients of the physics beyond the standard model \cite{langa,deboer,GunionHaber}. The phenomenological implications of the MSSM and of grand unification have been investigated since decades, and, though direct evidence of such theories is still missing, a large set of increasingly stringent constraints has put important bounds on the parameter space of these so-called SUSY-GUTs.\\
Nonetheless, the theoretical frameworks of SUSY-GUTs are countless, and typically characterized by a huge number of parameters. The common phenomenological practice is to make a certain number of (hopefully) theoretically motivated assumptions in order to deal with a reduced set of parameters. The first assumption one has to make in the context of SUSY is how to parameterize the mechanism of SUSY breaking.\\
Here, we resort to one of the most widely studied such contexts, the one of the so called supergravity models. In this framework, SUSY is broken in a {\em hidden sector}, whose fields couple only gravitationally to the MSSM fields \cite{mSUGRA}. Under some further assumptions (gaugino and scalar universality), the soft SUSY breaking part of the Lagrangian of supergravity models is determined by four continuous parameters plus one sign (mSUGRA).\\
In this paper, based on the results presented in \cite{profumo}, we study a ({\em minimal}) deviation from universality in the scalar fermions sector of the theory, inspired by the multiplet structure of the simplest GUT, $SU(5)$ (see sec.\ref{msnu}). We impose to the model the constraint of $b$-$\tau$ Yukawa coupling unification, which is a common prediction of a wide set of theories of grand unification, among which $SU(5)$ and $SO(10)$ GUTs.\\
One of the virtues of $R$-parity conserving SUSY models is to produce ideal candidates for cold dark matter \cite{WIMPreview}, in the present case the lightest neutralino (a linear superposition of the neutral gauge and Higgs boson superpartners). The neutralino is, in fact, a color and electromagnetically neutral, weakly interacting massive particle (WIMP), as required to be a good cold dark matter candidate. The recent results from the WMAP satellite \cite{WMAParticolo}, combined with other astrophysical data, determined with unprecedented accuracy the dark matter content range of the Universe. It has been shown \cite{WMAPrelicdensityEllis,WMAPrelicdensity} that the cosmologically allowed parameter space of mSUGRA is dramatically restricted by the considerably lowered upper bound on the neutralino relic density. Moreover, as recently pointed out in ref. \cite{Salati}, in cosmological scenarios involving quintessential fields, the relic density could undergo a further significant enhancement, even of several orders of magnitude.  The necessity of efficient relic density suppression mechanisms is therefore by now an uncontroversial point.\\
One of the known mechanisms of neutralino relic density suppression occurs when the next-to-lightest supersymmetric particle (NLSP), is close in mass to the neutralino. In this case, the neutralino relic density is not only suppressed by neutralino-neutralino annihilations, but also by {\em coannihilations} with the NLSP, and indirectly also by the annihilations of the NLSP itself \cite{GriestSeckel,GriestKamTurner,GondoloGelmini}.\\
This mechanism can involve in mSUGRA various sparticles \cite{djouadi}, such as the lightest stau \cite{EllisStau1,baerstau,pallisstau}, stop \cite{EllisStopCoann,cboehm}, or chargino \cite{Ullio,baerstau}.
In the present paper we address the possibility that minimal non-universality in the sfermion sector can produce ``new'', unusual coannihilation partners. We find that this possibility is effectively realized, in what we call a new coannihilation corridor, where the sbottom and the tau sneutrino play the role of NLSP. We show that this pattern is compatible with $b$-$\tau$ exact (``top-down'') Yukawa unification, and with all known phenomenological requirements.\\
In the next section we introduce and motivate the model we analyse. We discuss the issue of $b$-$\tau$ YU and of a GUT-motivated minimal sfermion SSB masses non-universality. We then show in sec.\ref{particlespectrum} the features of the resulting particle spectrum, which gives rise to ``new coannihilation corridors'' involving, besides the stau, the lightest sbottom and the tau sneutrino. sec.\ref{constraints} is devoted to the description of the cosmo-phenomenological requirements we apply. We demonstrate in particular that the $\mu>0$ case is not compatible with $b$-$\tau$ ``top-down'' YU. In sec.\ref{munegativecorridors} we describe the new coannihilation regions. After our final remarks, we give in the Appendix the complete list of the coannihilation processes  involving neutralino, sbottom, tau sneutrino and stau. Finally, an approximate treatment of the neutralino relic density contributions coming from sbottom-sbottom, sbottom-neutralino and stau-tau sneutrino coannihilations is provided.

\section{The Model}

\subsection{$b$-$\tau$ Yukawa unification}

One of the successful predictions of grand unified theories is the asymptotic unification of the third family Yukawa couplings \cite{langa}. The issue of Yukawa unification (YU) has been extensively studied, see e.g. \cite{YukUni,btau4,ferrandisbaer}. In particular, in this paper we address the issue of YU of the bottom quark and of the tau lepton, which is a prediction of some of the minimal grand unification gauge group, such as $SU(5)$. $b$-$\tau$ YU is a consequence of the fact that the two particles belong to the same $SU(5)$ multiplet, and therefore, at the scale of grand unification $M_{\sss GUT}$ they are predicted to have the same Yukawa coupling. The experimental difference between $m_{\tau}$ and $m_{b}$ is then mainly explained by two effects. First, the renormalization group (RG) running from $M_{\sss GUT}$ to the electroweak scale drives the two masses to different values. Second, in the minimal supersymmetric standard model, the supersymmetric sparticles affect with different finite radiative corrections the values of the masses, in particular the one of the $b$ quark \cite{bquark}.\\
Previous investigations of $b$-$\tau$ YU include ref. \cite{btauNS} as regards non supersymmetric GUTs and ref. \cite{btau2,btau3,btaucarena,btau5}, and more recently ref. \cite{btauRec,deboeretal} for the SUSY-GUT case. In particular, in \cite{btauRec} the implications of the recent experimental, and theoretical, results on the muon anomalous magnetic moment and on the inclusive branching ratio $b\rightarrow s\gamma$ were also taken into account, while in \cite{btauNonUniversalGaugino} the neutralino relic density constrain was examined, in the context of gaugino non-universality. In ref. \cite{btauNeutrinos,senja,masiero} the puzzle of neutrino masses and mixing has been tackled within the framework of $b$-$\tau$ YU.\\
A possible approach to $b$-$\tau$ YU is of the ``bottom-up'' type \cite{btau5,btauRec}. It consists in defining some parameter which evaluates the \emph{accuracy} of YU, such as
\begin{equation*}
\displaystyle \delta_{b\tau}\ \equiv\ \frac{h_b(M_{\sss GUT})-h_\tau(M_{\sss GUT})}{h_\tau(M_{\sss GUT})}.
\end{equation*}
The procedure we take here is instead a ``top-down'' approach \cite{deboeretal,btau3,btaucarena}: for a given set of SUSY parameters we fix the value $h_\tau(M_{\sss GUT})=h_b(M_{\sss GUT})$ requiring the resulting $m_{\tau}$ to be equal to its central experimental value. We then compute $m_{b}(M_Z)$ through RG running and taking into account the SUSY corrections. A model giving a value of the $b$-quark mass lying outside the experimental range is ruled out. With this procedure, we perform \emph{exact} $b$-$\tau$ YU at the GUT scale, and directly check whether a given model can, or cannot, be compatible with it.

\subsection{Minimal sfermion non-universality}
\label{msnu}

The parameter space of the minimal supersymmetric extension of the standard model, in its most general form, includes more than a hundred parameters \cite{MSSMrev,MSSMparamspace}. Therefore, it is commonly assumed that some underlying  principle reduces the number of parameters appearing in the soft supersymmetry breaking (SSB) Lagrangian. In particular, in the case of gravity mediated supersymmetry breaking, one can theoretically motivate \cite{mSUGRA} the assumption that there exists, at some high energy scale $M_{\sss X}$ a {\em common} mass $m_0$ for all scalars as well as a {\em common} trilinear coupling term $A_0$ for all SSB trilinear interactions. Moreover, in SUSY GUT scenarios, the additional assumption that the vacuum expectation value of the gauge kinetic function does not break the unifying gauge symmetry yields a common mass $M_{1/2}$ for all gauginos. One is then left with four parameters ($m_0,\ A_0,\ M_{1/2},\ \tan\beta$) and one sign (${\rm sign} \mu$), which define the so-called constrained MSSM, or mSUGRA, parameter space.\\
Much work has been done in the investigation of non-universality in the gaugino sector, see e.g. ref. \cite{nonuniversalgaugino,NEZRI}. As regards the SSB scalar masses, it has been since long known that universality is not a consequence of the supergravity framework, but rather an additional assumption \cite{scalarUniversality}. This justified an uprising interest in the possible consequences of non-universality in the scalar sector \cite{nonuniversalscalar,EllisNUHM}. In particular, in \cite{NONUNIVERSALITYFINALE} an analysis of various possible deviations from universality in the SSB was carried out.\\
In this paper we focus on a simple model exhibiting minimal non-universal sfermion masses (mNUSM) at the GUT scale. Our model is inspired by an $SU(5)$ SUSY GUT where the scale of SSB universality $M_{\sss X}$ is higher than the GUT scale $M_{\sss GUT}$ \cite{NONUNIVERSALITYFINALE}. The RG evolution of the SSB from $M_{\sss X}$ down to $M_{\sss GUT}$ induces a pattern of non-universality in the sfermion sector dictated by the arrangement of the matter fields into the supermultiplets:
\begin{eqnarray}
\left(\hat{L},\ \hat{D}^c\right)& \rightarrow & \overline{\bf 5}\\
\left(\hat{Q},\ \hat{U}^c,\ \hat{E}^c\right) & \rightarrow & {\bf 10}
\end{eqnarray}
This structure entails the following pattern of sfermion mass non-universality at the GUT scale:
\begin{eqnarray}
& m^2_L\ =\ m^2_D\ \equiv\ m^2_{\bf 5}& \\
& m^2_Q\ =\ m^2_U\ =\ m^2_E\ \equiv\ m^2_{\bf 10}&
\end{eqnarray}
The running between $M_{\sss X}$ and $M_{\sss GUT}$ will also produce two other effects. First, a typically large deviation from universality and a splitting between the up and down Higgs masses $m_{H_1}$ and $m_{H_2}$ is generated at the GUT scale (a detailed study of non-universal Higgs masses is presented in \cite{EllisNUHM} and \cite{EllisNUHMBis}). Second, a small splitting is also present between the SSB masses of the two lightest sfermion families and the third one. In the present paper we will however restrict to a \emph{phenomenological} parameterization of sfermion non-universality, simply setting
\begin{eqnarray}
&m^2_{\bf 10}\ \equiv \ m^2_0&\\
&m^2_{\bf 5}\ =\ K^2m^2_{\bf 10}\ =\ K^2m_0^2&\\
&m^2_{H_1}\ =\ m^2_{H_2}\ =\ m^2_{0}&
\end{eqnarray}
We are therefore left with a single parameter $K$, which scans this ``minimal'', GUT-inspired deviation from universality in the sfermion sector.\\
For our purposes, we let $K$ vary between 0 and 1: in this way we \emph{recover full universality} (i.e. the CMSSM) for $K=1$, while, for $K<1$, we can lower the spectrum of the sparticles belonging to the $ \overline{\bf 5}$ multiplet. Hence, we generate a spectrum with significantly lower masses for the tau sneutrino and the lightest stau and sbottom. Whenever the masses of these sparticles are close to the neutralino mass, they can play an important role in coannihilation processes.\\
\noindent Being inspired by a SUSY GUT $SU(5)$ framework, it is
natural to mention, within the proposed mNUSM model, the critical
question of the proton decay \cite{protondecay}. First, the
non-universality pattern of mNUSM is not derived from a definite
$SU(5)$ GUT: it simply inherits from such theory a plausible
asymptotic soft sfermion mass structure ($m^2_{\bf 10}$ and
$m^2_{\bf 5}$) and the feature of $b$-$\tau$ YU. Nonetheless, it
has been shown that {\em consistent} $SU(5)$ models exist
\cite{consistent}, where suitable structures for the leptoquarks Yukawa
couplings $h_{QQ}$, $h_{UE}$, $h_{UD}$ and $h_{QL}$ drastically suppress the proton decay
rate, even at large $\tan\beta$. The resulting proton life time is then well below the current
experimental limits \cite{SKprotondecaylimits}. These consistent
models are compatible with the present mNUSM in the soft SUSY
breaking Lagrangian, though for computational ease we take here
into account only the third generation Yukawa couplings $h_t$ and
$h_b=h_\tau$. Hence, we conclude that mNUSM models, within an
$SU(5)$ framework, are viable and are not in contrast with the
present constraints on the proton life time.

\section{Numerical procedure}
\label{procedure}

The mNUSM model we propose is defined by the following parameters:
\begin{equation}
m_0,\ A_0,\ M_{1/2},\ \tan\beta,\ {\rm sign} \mu \ \ {\rm and}\ K.
\end{equation}
We impose gauge and $b$-$\tau$ Yukawa coupling unification at a GUT scale $M_{\sss GUT}$ self-consistently determined by gauge coupling unification through two-loops SUSY renormalization group equations \cite{bertolini}, both for the gauge and for the Yukawa couplings, between $M_{\sss GUT}$ and a common SUSY threshold \mbox{$M_{\sss SUSY}\simeq\sqrt{m_{\tilde t_1}m_{\tilde t_2}}$} ($\tilde t_{1,2}$ are the stop quark mass eigenvalues). At $M_{\sss SUSY}$ we require radiative electroweak symmetry breaking, we evaluate the SUSY spectrum and calculate the SUSY corrections to the $b$ and $\tau$ masses \cite{btaucarena}. For the latter we use the approximate formula of ref. \cite{MTAU}:
\begin{eqnarray}
&\displaystyle \frac{\Delta m_\tau}{m_\tau}\ =\ \frac{g^2_1}{16 \pi^2}\frac{\mu M_2 \tan\beta}{\mu^2-M_2^2}\left(F(M_2,m_{\tilde\nu_\tau})-F(\mu,m_{\tilde\nu_\tau})\right), \label{mtaueq}&\\[0.2cm]
&\displaystyle F(m_1,m_2)=-\ln\left(\frac{M^2}{M_{\sss SUSY}^2}\right)+1+\frac{m^2}{m^2-M^2}\ln\left(\frac{M^2}{m^2}\right),&\\[0.2cm]
&M={\rm max}(m_1,m_2),\ m={\rm min}(m_1,m_2).&
\end{eqnarray}
From $M_{\sss SUSY}$ to $M_Z$ the running is continued via the SM one-loop RGEs. We use fixed values \cite{PDG} for the running top quark mass $m_t(m_t)=166\ {\rm GeV}$, for the running tau lepton mass $m_{\tau}(M_Z)=1.746\ {\rm GeV}$ and for $\alpha_s(M_Z)=0.1185$, all fixed to their central experimental values. The asymptotic Yukawa couplings $h_\tau(M_{\sss GUT})=h_b(M_{\sss GUT})$ and $h_t(M_{\sss GUT})$ are then consistently determined to get the correct top and tau masses, while the tree level $m_b^{\rm tree}$ and the SUSY-corrected $m_b^{\rm corr}$ masses of the running bottom quark at $M_Z$ are outputs.\\
The neutralino relic density is computed interfacing the output of the RGE running with the publicly available code {\tt micrOMEGAs} \cite{micromegas}, which includes thermally averaged exact tree-level cross-sections of all possible (co-)annihilation processes, an appropriate treatment of poles and the one-loop QCD corrections to the Higgs coupling with the fermions. The output of {\tt micrOMEGAs} also produces the relative contributions of any given final state to the reduction of the neutralino relic density.\\
The direct and indirect detection rates are estimated through another publicly available numerical code, {\tt darkSUSY} \cite{darksusy}.\\
As regards the phenomenological constraints, the Higgs boson
masses are calculated using {\tt micrOMEGAs} \cite{micromegas},
which incorporates the {\tt FeynHiggsFast} \cite{FeynHiggsFast}
code, where the SUSY contributions are calculated at two-loops.
The inclusive $BR(b\rightarrow s\gamma)$ is again calculated with
the current updated version of the {\tt micrOMEGAs} code
\cite{micromegasnew}, where the SM contributions are evaluated
using the formalism of ref. \cite{bsg1} and the charged Higgs
boson SUSY contributions are computed including the
next-to-leading order SUSY QCD re-summed corrections and the
$\tan\beta$ enhanced contributions (see ref. \cite{bsg2}). The
SUSY contributions to the muon anomalous magnetic moment $\delta
a_\mu$ are  directly calculated from the formul\ae \ of ref.
\cite{dam1} and compared with the output of the {\tt micrOMEGAs}
code.

\section{The particle spectrum with mNUSM}
\label{particlespectrum}

In minimal supergravity, especially after the very precise results from the WMAP satellite \cite{WMAParticolo}, giving a considerably reduced upper limit on the CDM density of the Universe, the cosmologically allowed regions of the parameter space are strongly constrained \cite{WMAPrelicdensityEllis,WMAPrelicdensity}. As pointed out in \cite{ellisolive,lahanas}, mainly two mechanisms can suppress the neutralino relic density to sufficiently low values. The first one are coannihilations of the neutralino with the next-to-lightest sparticles (NLSP), which are effective whenever the mass of the latter lies within $10$-$20\%$ of the neutralino mass \cite{BGSalati,GriestSeckel,GriestKamTurner}. The second mechanism is given by direct, rapid $s$-channel annihilation of the neutralino with the CP-odd Higgs boson $A$ \cite{dreesnojiri}, which takes place if $m_A\simeq2m_{\neut}$, i.e. if the channel is enhanced by pole effects\footnote{Since here $m_A\simeq m_H$, where $H$ is the heaviest CP-even neutral Higgs boson, the condition $m_A\approx2m_{\neut}$ implies pole effects also for $H$. However, since the CP quantum number of the exchanged Higgs boson must match that of the initial state, only $A$ exchange contributes in the $S$ wave, while $H$ (and $h$) contribute to the $P$ wave. This implies a suppression, in the thermal averaged cross section, of a factor $3/x_F\approx0.1\div0.2$ (see the Appendix).}. In particular, we choose to focus here on the case of coannihilations, which in the present scenario of mNUSM are expected to exhibit a rich pattern.\\
In order to understand the possible pattern of coannihilations emerging from the parameterization of sfermion masses non-universality outlined in sec.\ref{msnu}, we study the spectrum of the candidate NLSPs, namely the lightest sbottom and the tau sneutrino. The dependence of the respective masses of these two sparticles on the high energy SSB inputs can be parameterized as follows:
\begin{eqnarray}
m^2_{\sbot}&\simeq&a_{\sbot}K^2m_0^2+\left(b_{\sbot}m_0^2+c_{\sbot}M_{1/2}^2+\Delta_{\sbot}\right)\label{massapprox}\\[0.2cm]
m^2_{\sneu}&\simeq&a_{\sneu}K^2m_0^2+\left(b_{\sneu}m_0^2+c_{\sneu}M_{1/2}^2+\Delta_{\sneu}\right)
\end{eqnarray}
The contributions $\Delta_{\sbot,\sneu}$ originate in part from the $SU(2)_L$ and $U(1)_Y$ $D$-term quartic interactions of the form $({\rm squark})^2({\rm Higgs})^2$ and  $({\rm slepton})^2({\rm Higgs})^2$, and are  $\propto\cos(2\beta)M_Z^2$ \cite{deboer}. The sbottom mass gets a further contribution in $\Delta_{\sbot}$ arising from the $LR$ off-diagonal elements of the mass matrix. The mixing terms are generated by the typically large values of $A_b$, in its turn induced, even for $A_0(M_{\sss GUT})=0$, by RG running, and by a $\tan\beta$-enhanced contribution $\propto\mu m_b$. In the case of the tau sneutrino, instead, a further, though small, contribution to $\Delta_{\sneu}$ analogously arises from $A_\tau$.\\
\begin{figure}[!t]
\begin{center}
\begin{tabular}{cc}
\includegraphics[height=6.55cm,width=6.4cm,angle=-90]{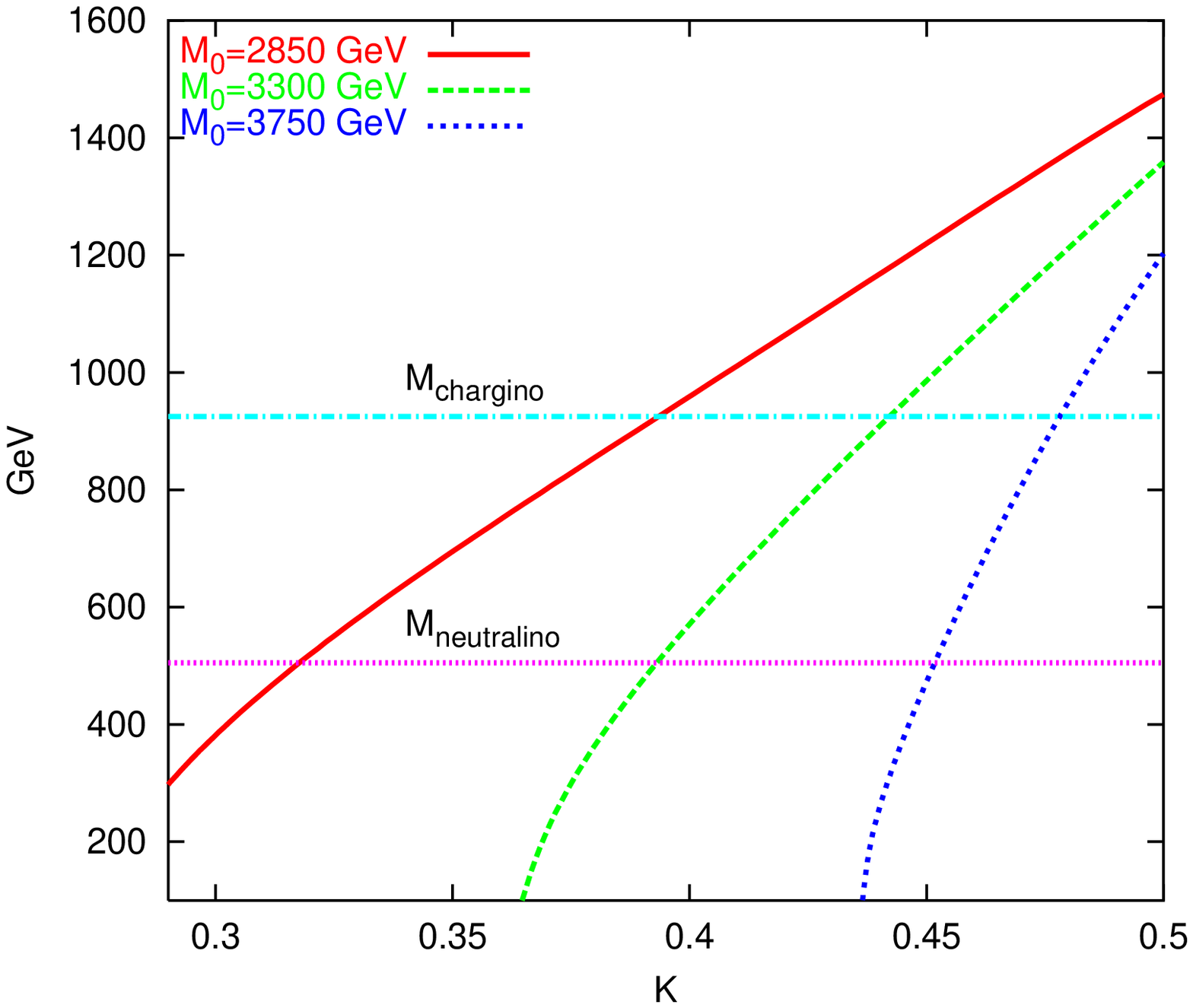} &
\includegraphics[scale=0.45,angle=-90]{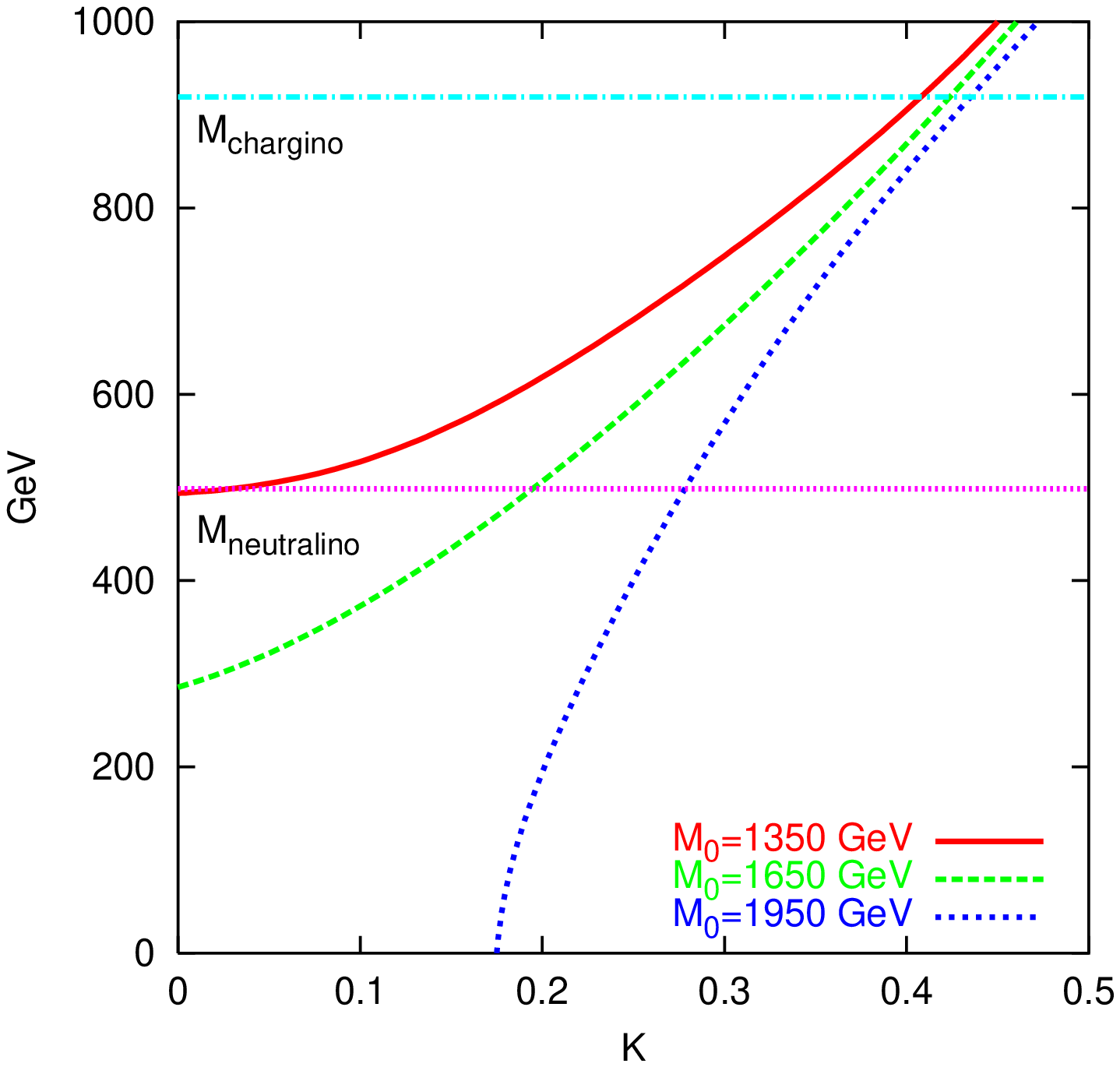}\\
&\\[-0.2cm]
\hspace{1.cm} (\emph{a}) & \hspace{1.cm} (\emph{b})\\
\end{tabular}
\caption{\em\small The sbottom ($a$) and sneutrino ($b$) spectrum at $M_{1/2}=1.1\ {\rm TeV}$, $\tan\beta=38.0$ and $A_0=0$ for different values of $m_0$.}
\label{spectraSneuSbot}
\end{center}
\end{figure}
\begin{figure}[!t]
\begin{center}
\includegraphics[scale=0.55,angle=-90]{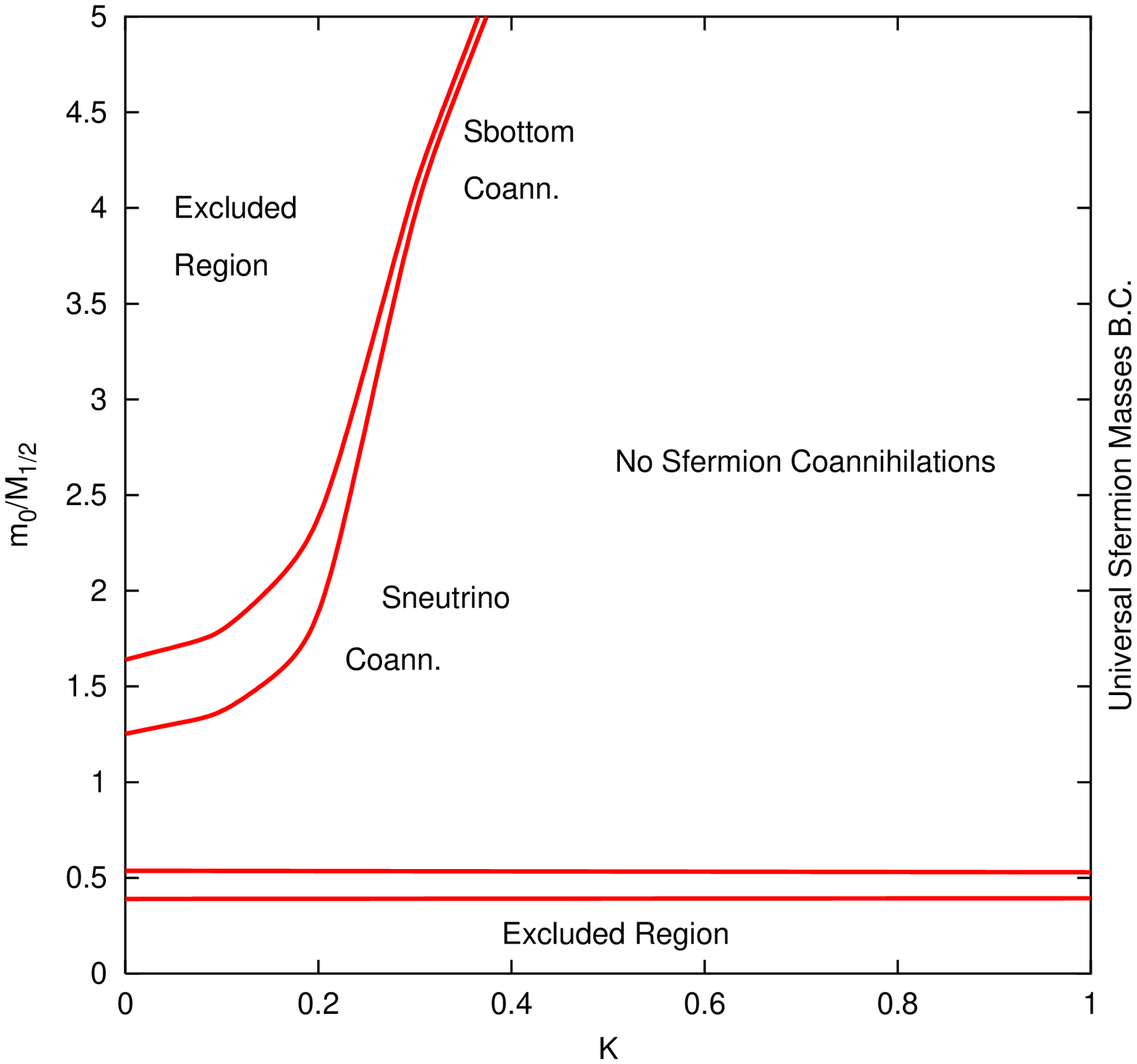}
\caption{\em\small Coannihilation regions in the $(\displaystyle m_0/M_{1/2},K)$ plane at $M_{1/2}=1.1\ {\rm TeV}$, $\tan\beta=38.0$ and $A_0=0$. Points within the red lines are characterized by $\displaystyle \frac{m_{\sss NLSP}-m_{\neut}}{m_{\neut}}\lesssim 20\%$. In the top left and lower part of the figure the LSP is not a neutralino, while in the top right region no coannihilations take place between the neutralino and the sleptons.}
\label{corridorMAIN}
\end{center}
\end{figure}
We plot in fig.\ref{spectraSneuSbot} ($a$) and ($b$) the typical behaviors of the sparticle masses as functions of $K$, at fixed $M_{1/2},\ m_0$ and $\tan\beta$. Frame ($a$) shows the case of the mass of the sbottom, (we also plot the corresponding masses of the lightest neutralino and chargino). The high energy parameters are fixed at $M_{1/2}=1100\ {\rm GeV}$, $\tan\beta=38.0$, $A_0=0$ and $m_0=2850,\ 3300$ and 3750 GeV. We clearly see from the figure that the behavior is, as expected, $m_{\sbot}\approx\sqrt{\alpha+\beta K^2}$, where $\alpha<0$ is the sum of the terms in parenthesis in eq.(\ref{massapprox}) and $\beta= a_{\sbot}m_0^2>0$. Increasing the value of $m_0$ induces higher negative values for $\alpha$, as well as obviously higher values for $\beta$. Notice in any case that for sufficiently low values of $K$ the mass of the sbottom is driven {\em below} the mass of the neutralino and also to negative values.\\
Frame ($b$) shows instead the mass of the tau sneutrino, at $M_{1/2}=1100\ {\rm GeV}$, $\tan\beta=38.0$, $A_0=0$ and $m_0=1350,\ 1650$ and 1950 GeV. We see that also here $m_{\sneu}\approx\sqrt{\alpha+\beta K^2}$, but in this case the interplay between $b_{\sneu}$ and $c_{\sneu}$ generates either $\alpha$ positive  ($m_0=1350,\ 1650\ {\rm GeV}$) or negative ($m_0=1950\ {\rm GeV}$). The outcome is therefore that the sneutrino mass can be lowered towards the mass of the neutralino for low values of $K$, depending on $m_0$; the typical range of $m_0$ for which this is possible is always lower than in the sbottom case.\\
We carried out a thorough investigation of the possible coannihilation regions, and we found that a good parameter is represented by the ratio $(m_0/M_{1/2})$ at fixed $\tan\beta$, $M_{1/2}$ and $A_0$. In fig.\ref{corridorMAIN} we plot the coannihilation corridors which we found scanning the plane $(\displaystyle m_0/M_{1/2},K)$ at $M_{1/2}=1.1\ {\rm TeV}$, $\tan\beta=38.0$ and $A_0=0$. Within the red solid lines \mbox{$\displaystyle \frac{m_{\sss NLSP}-m_{\neut}}{m_{\neut}}\lesssim 20\%$}. In the lower part of the figure, which we indicate as 'Excluded Region', the low value of $m_0$ implies that the stau becomes lighter than the neutralino, or even gets a negative (unphysical) mass. Very stringent bounds \cite{isotopes} indicate that the LSP has to be electrically and color neutral, and therefore this region is excluded. The strip above this excluded region represents a first coannihilation corridor, where the NLSP is the stau. We notice that this region survives up to $K=1$, i.e. fully universal boundary conditions. It actually represents a slice of the narrow band, in the $(m_0,M_{1/2})$ planes, which is cosmologically allowed thanks to neutralino-stau coannihilations (see e.g. fig.1 and 2 of ref. \cite{WMAPrelicdensityEllis}).\\
\begin{figure}[!t]
\begin{center}
\begin{tabular}{cc}
\includegraphics[scale=0.45,angle=-90]{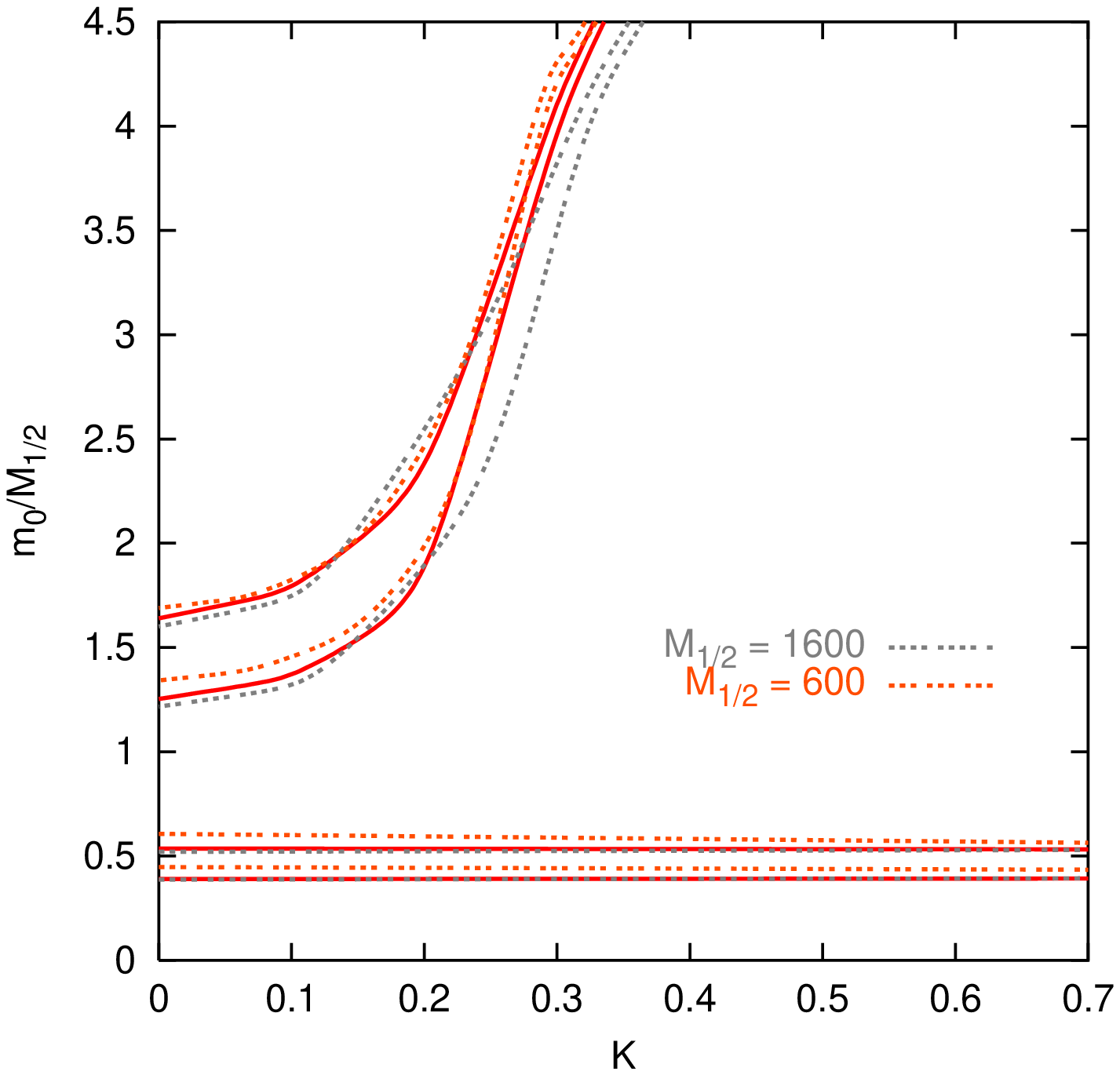} &
\includegraphics[scale=0.45,angle=-90]{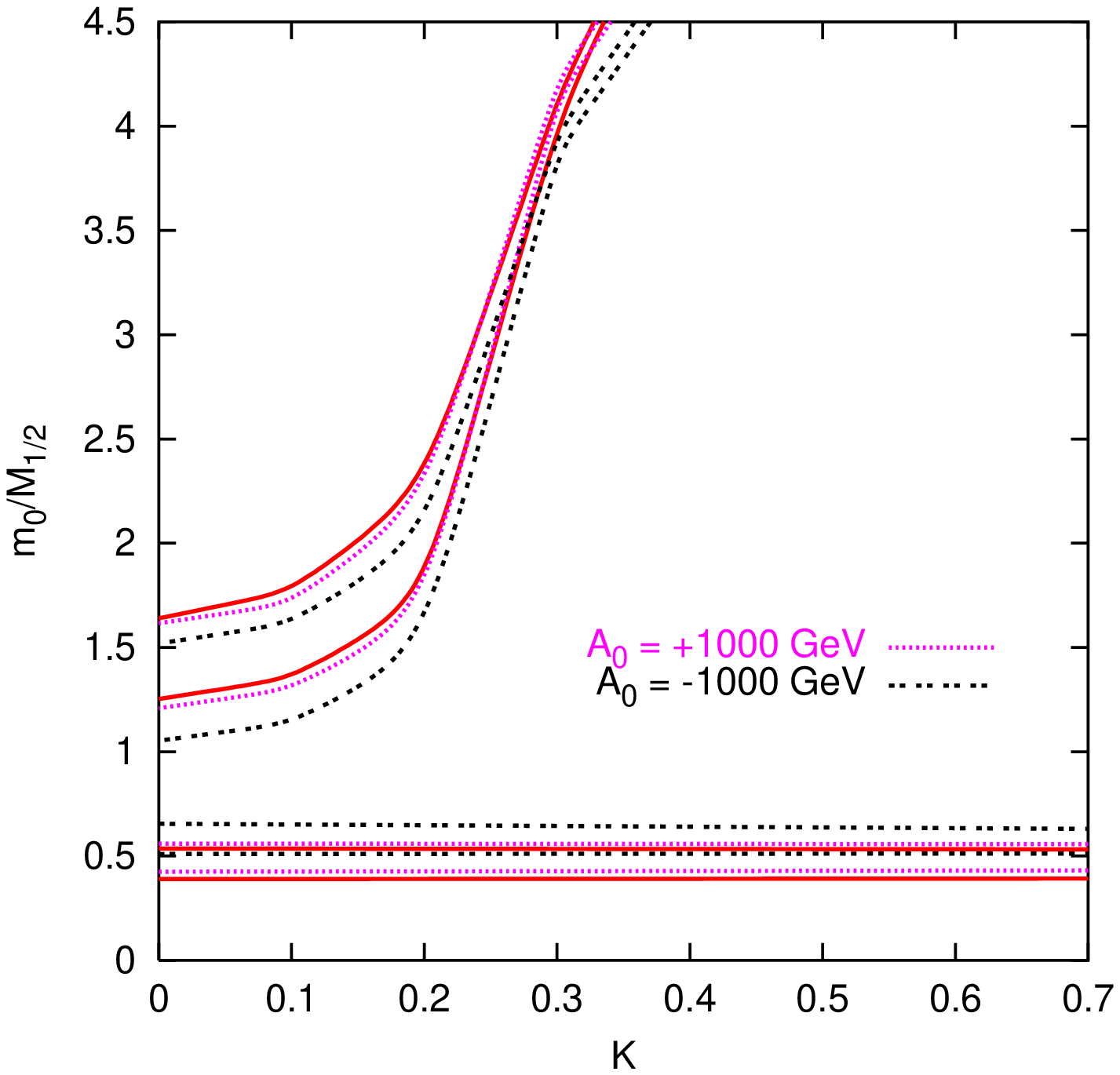}\\
&\\[-0.2cm]
\hspace{1.cm} (\emph{a}) & \hspace{1.cm} (\emph{b})\\
\end{tabular}
\caption{\em\small Coannihilation regions in the $(\displaystyle m_0/M_{1/2},K)$ plane at $\tan\beta=38.0$ and $A_0=0$. In frame (a) the dependence on $M_{1/2}$ is studied at $A_0=0$. In frame (b) two different values for the trilinear coupling $A_0=\pm\ 1 \ {\rm TeV}$ are plotted at $M_{1/2}=1100\ {\rm GeV}$.}
\label{corridorother}
\end{center}
\end{figure}
For values of $K\lesssim0.5$ we find a second, distinct, branch where the mass of the NLSP lies within 20\% of the LSP mass. In the lower part of the branch (in fig.\ref{corridorMAIN} up to $m_0/M_{1/2}\approx 3$) the NLSP turns out to be the tau sneutrino, with the lightest stau which is quasi degenerate with it (say within few percent, see the discussion in sec.\ref{neutstausneut}). Increasing $m_0$ and moving to the upper part of the branch, the lightest sbottom becomes the NLSP, while the tau sneutrino and the stau, still quasi degenerate, become by far heavier. In the upper left part of the figure, once again indicated as 'Excluded Region', either the sbottom or the tau sneutrino become lighter than the neutralino, or even get negative values for their masses (see fig.\ref{spectraSneuSbot}).
\begin{figure}[!t]
\begin{center}
\includegraphics[scale=0.55,angle=-90]{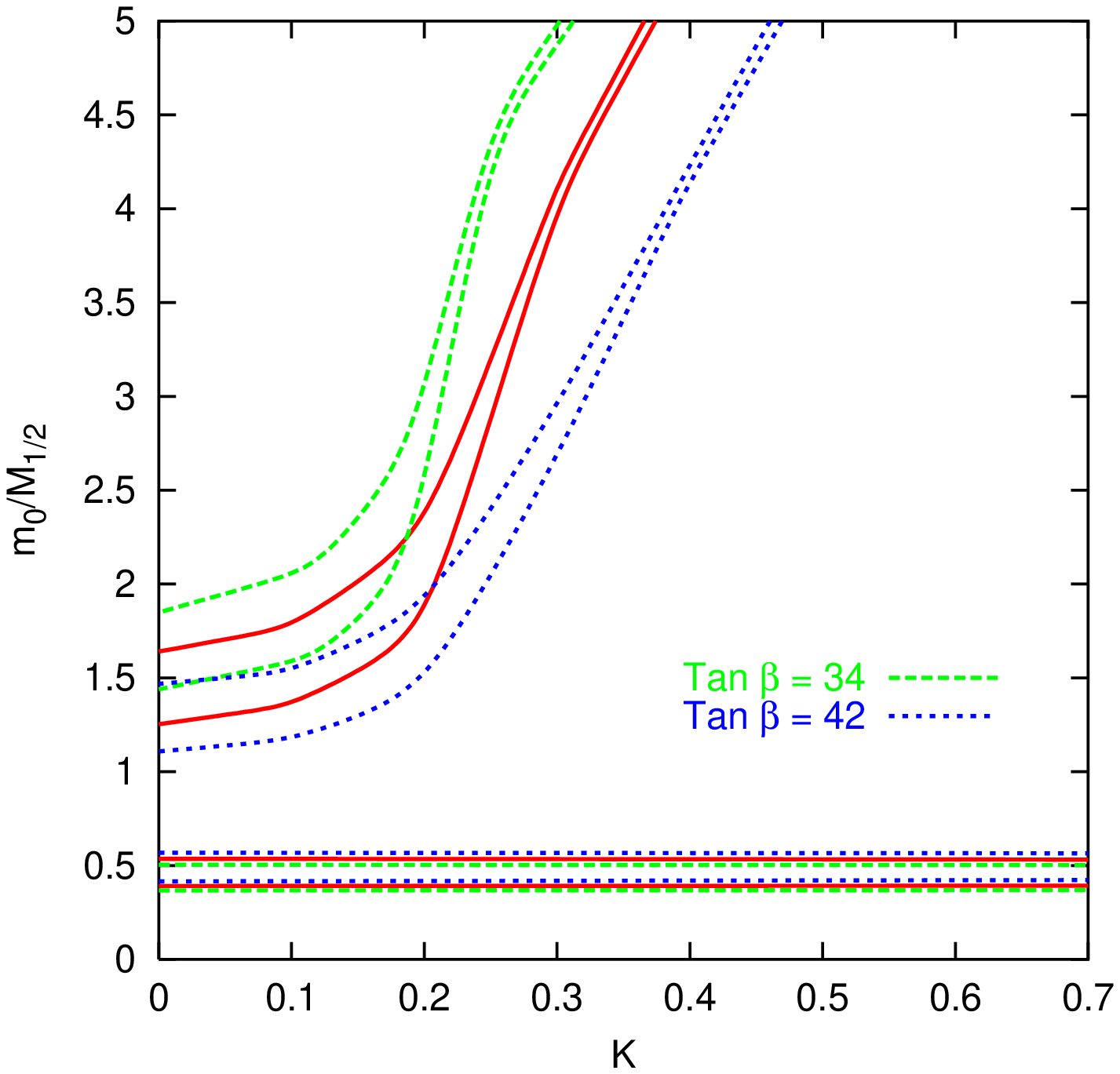}
\caption{\em\small Coannihilation regions in the $(\displaystyle m_0/M_{1/2},K)$ plane at $M_{1/2}=1.1\ {\rm TeV}$ and $A_0=0$. The red solid lines indicate the coannihilation corridor for $\tan\beta=38.0$, while the green dashed and the blue dotted lines respectively $\tan\beta=34,\ 42$.}
\label{corridortb}
\end{center}
\end{figure}
In fig.\ref{corridorother} we study the dependence of the shape of the coannihilation corridors on the various parameters which we fixed in fig.\ref{corridorMAIN}. In frame ({\em a}) we vary the values of $M_{1/2}$, comparing the $M_{1/2}=1100\ {\rm GeV}$ case of fig.\ref{corridorMAIN} with respectively $M_{1/2}=1600\ {\rm GeV}$ and $M_{1/2}=600\ {\rm GeV}$. We see that there is practically no significant dependence of the shape on the value of $M_{1/2}$, and therefore we conclude that the chosen parameter $m_0/M_{1/2}$ is good, being $M_{1/2}$-quasi independent. In frame ({\em b}) we vary instead the value of the universal trilinear coupling $A_0$. We plot again with a red solid line the $A_0=0$ case, as well as the $A_0=1\ {\rm TeV}$ and $A_0=-1\ {\rm TeV}$ cases, always at $\tan\beta=38.0$ and $M_{1/2}=1100\ {\rm GeV}$. Once again we do not see any significant effect, apart from a common shift of the lower coannihilation branch upwards and of the upper downwards. These shifts can be traced back to the effect of the off diagonal $A_0$ (sign-independent) entries in the sfermion mass matrices.\\
Fig.\ref{corridortb} illustrates the dependence on $\tan\beta$, highlighting this time a significant effect on the upper branch: increasing $\tan\beta$ yields higher values for $K$, i.e. the branch is moved to the right, and vice-versa. This effect can be qualitatively understood from the approximate expression for the mass eigenvalues of the relevant sfermions, eq.(\ref{massapprox}), where an increase in $\tan\beta$ is compensated by an increase in the effective boundary value of the scalar masses, tuned by the parameter $K$. For instance, $\Delta_{\sbot}$ contains a negatively contributing term $\propto\tan\beta m_b \mu$, which is compensated, when the value of $m_0$ and of $m_{\neut}$ is fixed, by an increase of $K$ .\\

\section{Cosmo-phenomenological bounds}
\label{constraints}
In this paper we apply two classes of constraints: on the one hand the cosmological bounds coming from the limits on the cold dark matter content of the Universe and from direct and indirect neutralino searches; on the other hand we impose the most stringent ``accelerator'' constraints, such as the inclusive $BR(b\rightarrow s \gamma)$ and the Higgs boson mass. In this second class we also include the bound on the $b$-quark mass, direct sparticle searches and a conservative approach \cite{superconservative} to the constraint  coming from the SUSY corrections to the muon anomalous magnetic moment $\delta a_\mu$.\\
For illustrative purposes, we show the behavior of three {\em benchmark} cases, pertaining the three regions of coannihilations highlighted in the previous section. Namely, we choose three representative values of $K$ and require the mass splitting between the neutralino and the NLSP to be 10\%. We then scan the parameter space, setting $A_0=0$ and $\mu<0$ for simplicity and varying $m_{\neut}$. The features of the three cases considered are summarized in Tab.\ref{Tabellamodelli}.
\begin{table}[!t]
\begin{center}
\begin{tabular}{|c|c|c|c|}
\hline
NLSP & $\Delta_{\sss NLSP}$ & $K$ & $\tan\beta$\\
\hline
stau & 0.1 & 0.8 & 36.0\\
\hline
sbottom & 0.1 & 0.4 & 36.0\\
\hline
tau sneutrino & 0.1 & 0.2 & 36.0\\
\hline
\end{tabular}
\caption{\small \em The three benchmark scenarios described in the text}
\label{Tabellamodelli}
\end{center}
\end{table}
\begin{figure}[!t]
\begin{center}
\includegraphics[scale=0.65,angle=-90]{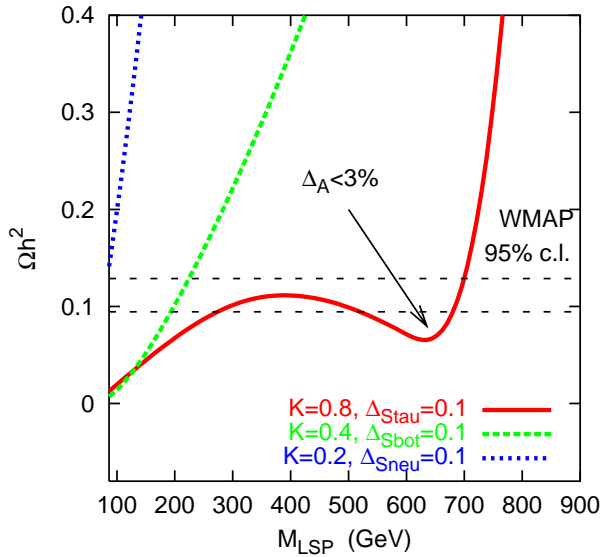}
\caption{\em\small Neutralino relic density for the three benchmark cases of Tab.\ref{Tabellamodelli}. The upper and lower bounds on $\Omega_{\sss CDM}h^2$ are the 95\% C.L. from WMAP global fit \cite{WMAParticolo}.}
\label{omegaUBC}
\end{center}
\end{figure}

\subsection{Neutralino relic density facing WMAP results}\label{secomega}
Supersymmetric models with conserved $R$ parity generate ideal candidates for cold dark matter, as, in the present case, the lightest neutralino. It is therefore natural to require, besides the fulfillment of the other phenomenological constraints, that the cosmological relic density of the neutralinos lies within the bounds indicated by cosmology. In particular, the recent results from the WMAP satellite \cite{WMAParticolo}, combined via a global fit procedure with other astrophysical data (including other CMB experiments, LSS surveys and the Ly$\alpha$ data), give a compelling bound on the cosmological relic density of Cold Dark Matter
\begin{equation}
\displaystyle \Omega_{\sss CDM}\ h^2\ =\ 0.1126^{+0.00805}_{-0.00905}
\label{omegarange}
\end{equation}
We take here the $2$-$\sigma$ range, and we require that $\Omega_{\neut}h^2\lesssim0.1287$. Strictly speaking, the lower bound cannot be directly imposed, if we suppose that the neutralinos are not the only contributors to the cold dark matter of the Universe. For illustrative purposes, in fig.\ref{k035tb38} we will plot also the lower bound on the relic density.\\
In fig.\ref{omegaUBC} we show instead the behavior of the neutralino relic density as a function of $m_{\neut}$ for the three benchmark cases. We see that in the case of tau sneutrino the coannihilations only weakly contribute to the reduction of the relic density (see sec.\ref{neutstausneut}), which, as expected, quadratically diverges with $m_{\neut}$. In the case of the sbottom coannihilation, SUSY QCD effectively enhance the relic density suppression, which nonetheless still exhibits a divergent behavior. In the case of the stau, instead, we notice how the interplay between coannihilation and direct rapid annihilation through the $A$-pole $s$-channel can drastically reduce the relic density. In the dip located between 600 and 700 GeV, in fact, the splitting between the mass of the $CP$-odd Higgs boson and twice the mass of the neutralino $\displaystyle \frac{m_A-2m_{\neut}}{2m_{\neut}}\lesssim3\%$, and therefore direct pole annihilations are extremely efficient, leading to viable values of $\Omega_{\neut} h^2$ for rather high $m_{\neut}\lesssim700\ {\rm GeV}$.

\subsection{Direct and indirect WIMP searches}
\begin{figure}[!t]
\begin{center}
\begin{tabular}{cc}
\includegraphics[scale=0.5,angle=-90]{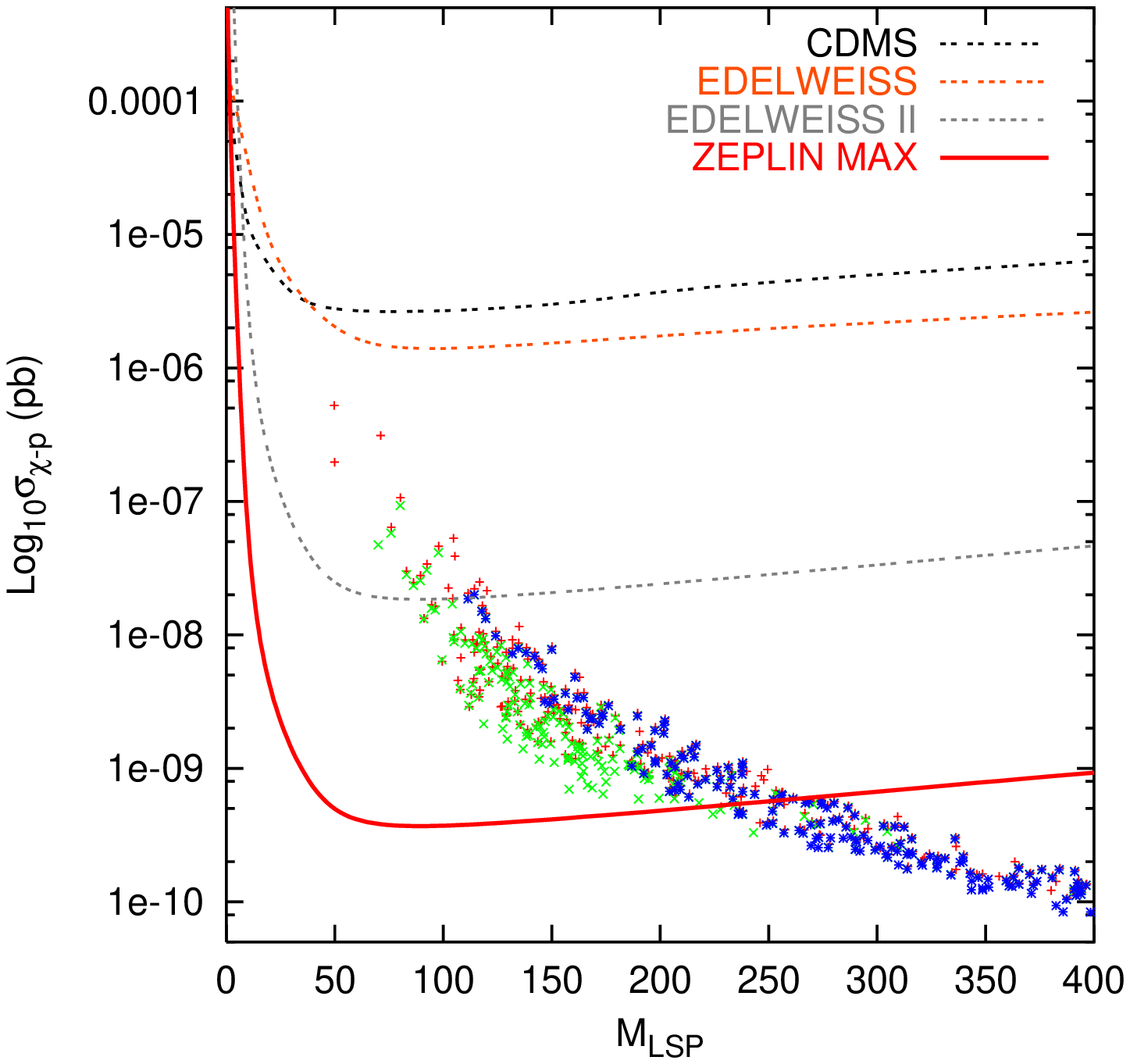} &
\includegraphics[scale=0.5,angle=-90]{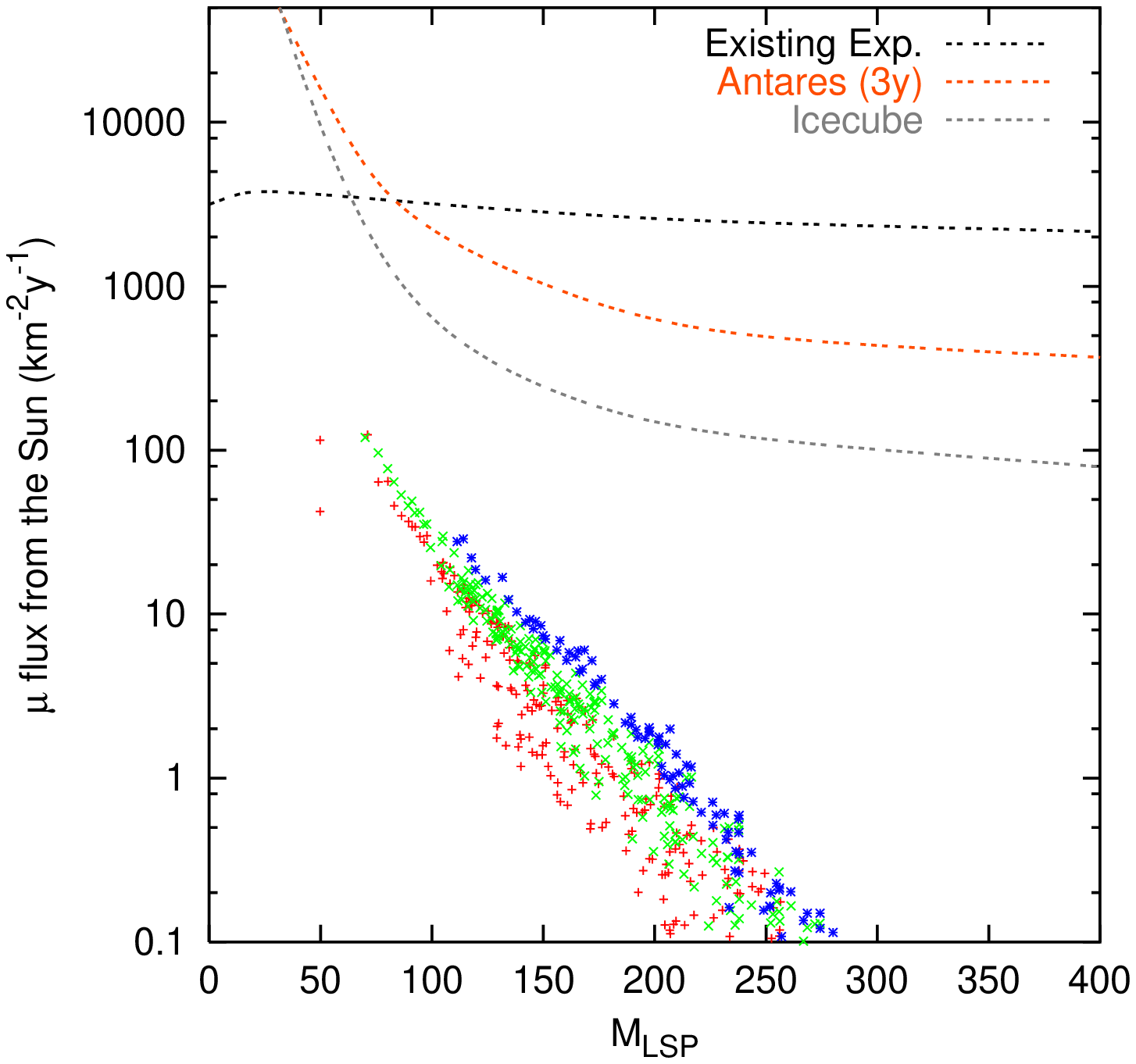}\\
&\\[-0.2cm]
\hspace{1.cm} (\emph{a}) & \hspace{1.cm} (\emph{b})\\
\end{tabular}
\caption{\em\small Direct ($a$) and indirect ($b$) detection rates for $K=1.0$ (red $+$, full universality), $K=0.35$ (green $\times$) and $K=0.1$ (blue $*$). The points refer to $50<M_{1/2}<1000\ {\rm GeV}$, $20<m_0<3000\ {\rm GeV}$ and $-500<A_0<500\ {\rm GeV}$, at $\tan\beta=38.0$. In frame ($a$) we plot $\sigma^{\rm scal}_{\neut-q}$ and the sensitivity bounds of present and future direct detection experiments \cite{directdet}. In frame ($b$) we report the neutralino annihilation induced muon flux from the Sun, in ${\rm Km}^{-2}y^{-1}$, and we compare again with current and future indirect detection experiments sensitivities \cite{indirectdet}.}
\label{Wimpd}
\end{center}
\end{figure}
As pointed out in \cite{NEZRI}, non-universality in the sfermion sector should not give rise to substantial modifications to $\sigma^{\rm scal,\ spin}_{\neut-q}$. In fact, owing to the small Yukawa couplings of the lightest generation of quarks, NUSM do not greatly affect the masses of the up and down squarks, leading to substantially unchanged values for the direct annihilation cross sections, sensitive to processes like $\neut\ q\xrightarrow{\tilde{q}}\neut\ q$. Further, we find that Higgs exchange processes do not give rise, in the mNUSM models, to sizeable contributions, as compared with the fully universal case. In fig.\ref{Wimpd} ($a$) we scatter plot the spin-independent cross section $\sigma^{\rm scal}_{\neut-q}$ as a function of $m_{\neut}$ for $K=1.0$ (red $+$, full universality case), $K=0.35$ (green $\times$) and $K=0.1$ (blue $*$). We randomly scan the parameter space for a given value of $K$ with $50<M_{1/2}<1000\ {\rm GeV}$, $20<m_0<3000\ {\rm GeV}$ and $-500<A_0<500\ {\rm GeV}$, fixing $\tan\beta=38.0$, and requiring the fulfillment of the phenomenological constraints. We also include the sensitivities of present and future direct detection experiments \cite{directdet}. Notice that some mNUSM models lie within the reach of future direct detection experiments. As far as indirect detection (e.g. the muon flux from the Sun) is concerned, non-universality do not significantly affect the detection rates, though $t$-channel sfermions exchange is enhanced by lighter third generation sfermions for $K<1$, as we can see in fig.\ref{Wimpd} ($b$): for smaller values of $K$ we obtain larger muon fluxes from the Sun. Present and future indirect detection experiments \cite{indirectdet} are however sensitive to muon fluxes well above what we obtain for mNUSM models.\\
To summarize, we find, both for direct and indirect detection, that the mNUSM scenario produces detection rates in the same range of the CMSSM \cite{WIMPdetection}, therefore (minimal) sfermion non-universality can hardly be inferred from WIMP searches. Further, present experimental bounds do not constrain the models under consideration.

\subsection{SUSY corrections to the $b$-quark mass}

It has been already pointed out, see e.g. \cite{btauRec,deboeretal}, that the requirement of $b$-$\tau$ Yukawa unification favors negative\footnote{We use here the standard sign conventions of ref. \cite{msugracollaboration}} values of $\mu$. We recall that ${\rm sign}\mu$ is one of the parameters included both in the CMSSM and in our mNUSM model. In this section we demonstrate that, even with mNUSM, $\mu>0$ is not compatible with YU. We then discuss the $\mu<0$ case which allows, in a suitable $\tan\beta$ range, the fulfillment of  $b$-$\tau$ YU.\\
The main problem of  $b$-$\tau$ YU with $\mu>0$ is that one
typically obtains a tree level mass for the $b$ quark which is
close to the experimental upper bound, and has to add on top of it
large {\em positive} SUSY corrections (eq.(\ref{mbcorr})), which
drive $m_b^{\rm corr}$ outside the experimental range (or, the
other way round, $h_b$ far away from $h_\tau$, in the bottom-up
approach). We impose  $b$-$\tau$ YU at the GUT scale, we fix
$h_\tau(M_{\sss GUT})=h_b(M_{\sss GUT})$ from the properly
corrected and RG evolved $m_\tau(M_Z)$, obtaining as outputs the
tree level $m_b^{\rm tree}$ and the SUSY-corrected $m_b^{\rm
corr}$ masses of the running bottom quark at $M_Z$. We compare
these numbers with the appropriately evolved $b$-quark pole mass
\cite{ExpBQUARK} up to the $M_Z$ scale, with
$\alpha_s(M_Z)\simeq0.1185$, following the procedure of ref.
\cite{baerBQUARK,YQU}:
\begin{equation}\label{mbexprange}
m_b(m_b)=4.25\pm0.3\ {\rm GeV} \ \ \ \Rightarrow \ \ \ m_b(M_Z)=2.88\pm0.2\ {\rm GeV}
\end{equation}
The largest SUSY corrections arise from sbottom-gluino and stop-chargino loops, frozen at the $M_{\sss SUSY}$ scale \cite{bquark,MTAU,morebquark}. They are \emph{non-decoupling effects} because one gets a finite contribution even in the infinite sparticle mass limit, and they can be cast in the following approximate form:
\begin{eqnarray}
\displaystyle \frac{\Delta m_b^{\tilde g}}{m_b}&\approx&\frac{2\alpha_s}{3\pi}M_3\mu\ I(m^2_{\tilde b_1},m^2_{\tilde b_2},M_3^2)\ \tan\beta, \label{mbcorr}\\[0.2cm]
\displaystyle \frac{\Delta m_b^{\tilde \chi^-}}{m_b}&\approx&\frac{h^2_t}{16\pi^2}\mu\ A_t\ I(m^2_{\tilde t_1},m^2_{\tilde t_2},\mu^2)\ \tan\beta,\\[0.2cm]
\displaystyle I(x,y,z)&\equiv&\frac{xy\ln(x/y)+xz\ln(z/x)+yz\ln(y/z)}{(x-y)(y-z)(x-z)}.
\end{eqnarray}
\begin{figure}[!t]
\begin{center}
\begin{tabular}{ccc}
\includegraphics[scale=0.6,angle=-90]{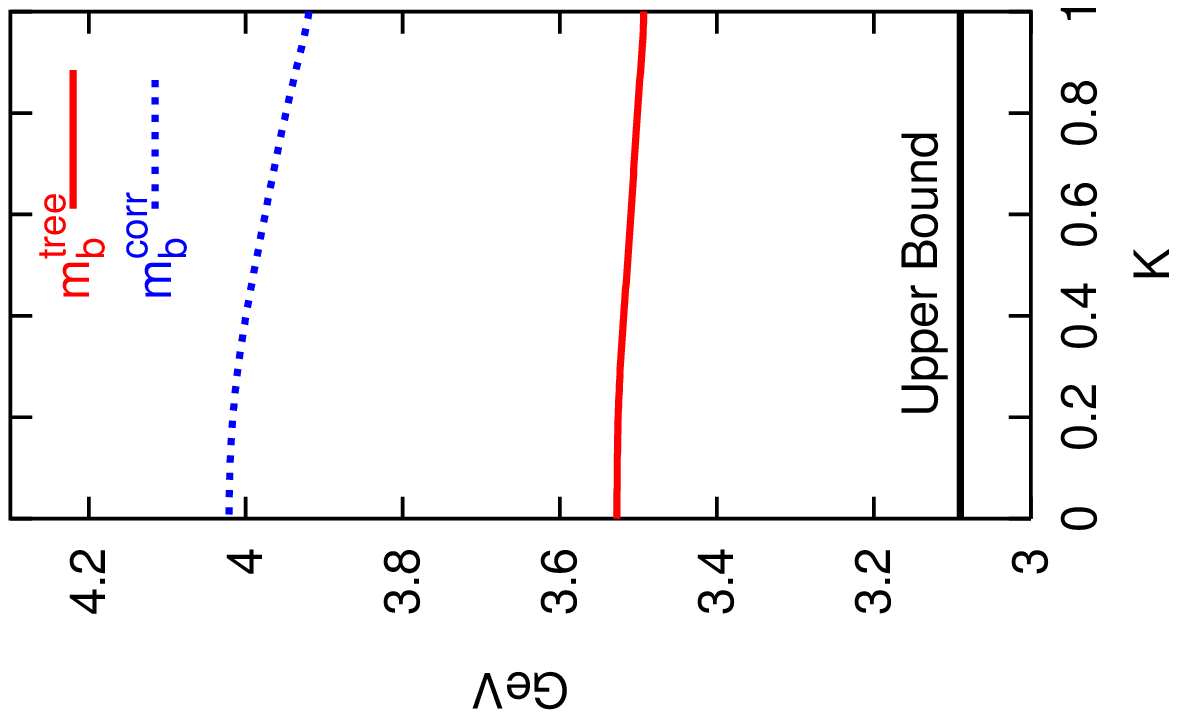} &
\includegraphics[scale=0.6,angle=-90]{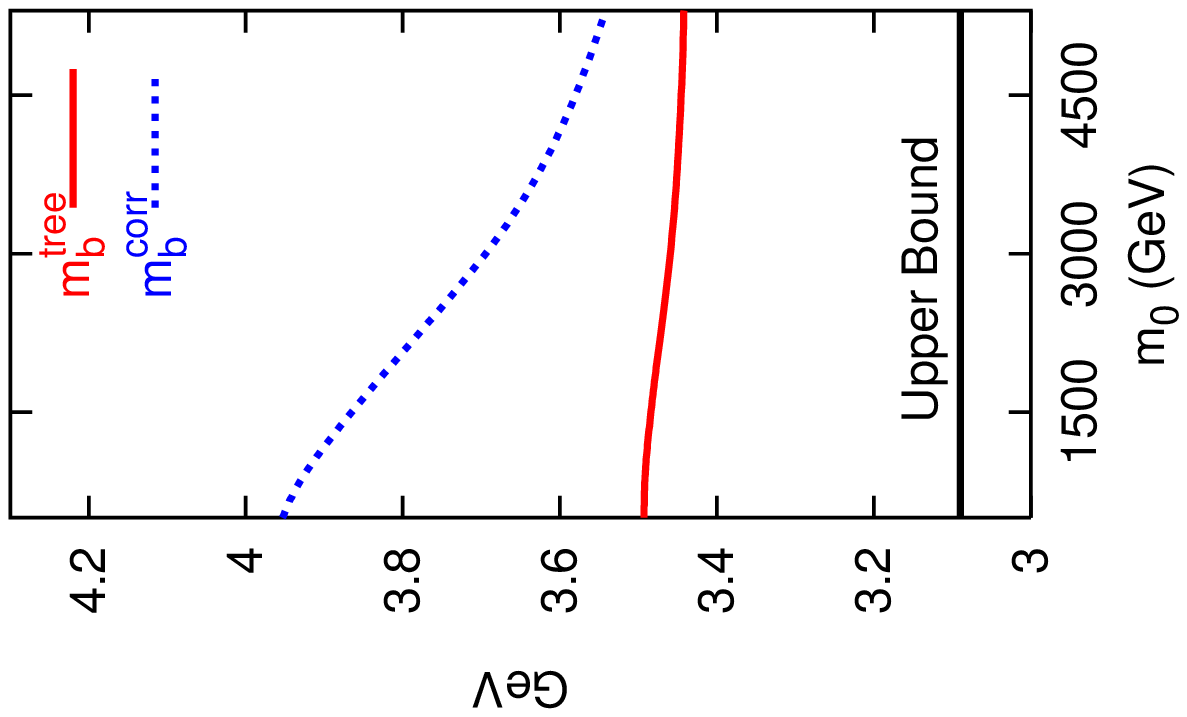} &
\includegraphics[scale=0.6,angle=-90]{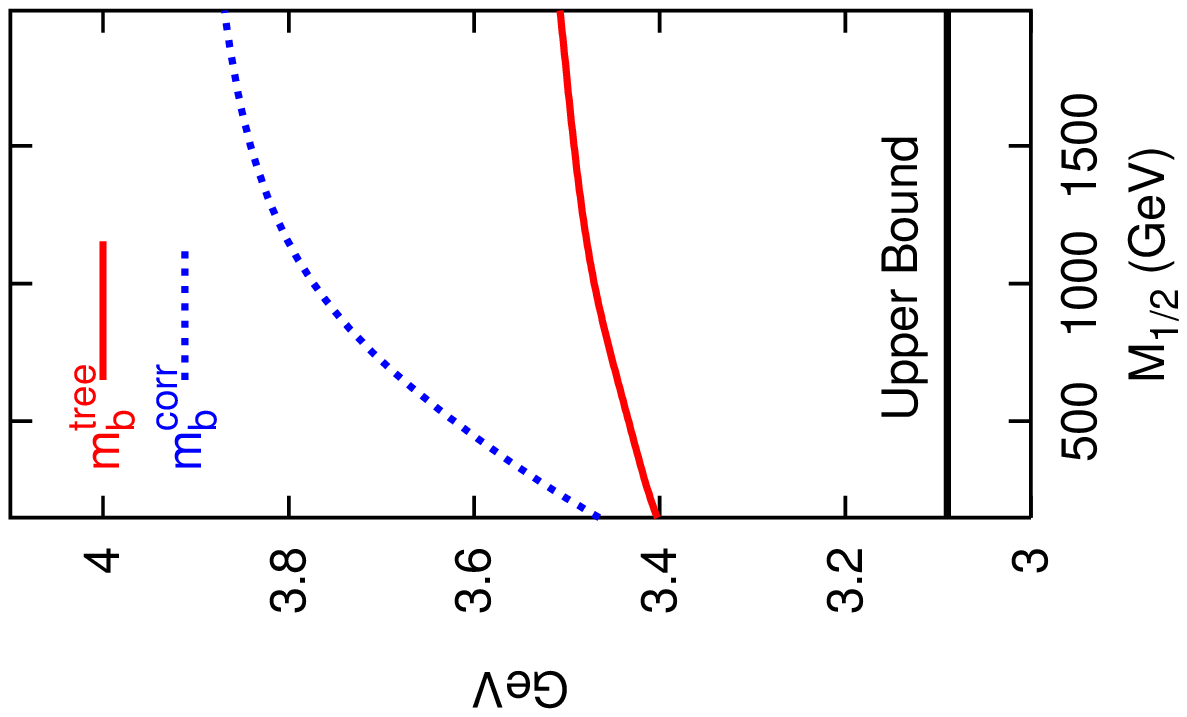} \\
&&\\[-0.2cm]
\hspace{1.cm} (\emph{a}) & \hspace{1.cm} (\emph{b}) & \hspace{1.cm} (\emph{c}) \\
\end{tabular}
\caption{\em\small Tree level and SUSY corrected values of the $b$-quark mass at $\tan\beta=38.0$ and $A_0=0$. In frame (a) the parameter $K$ is varied at fixed $M_{1/2}=1100\ {\rm GeV}$ and $m_0=1000\ {\rm GeV}$. In frame (b) $m_0$ is varied at fixed $M_{1/2}=1100\ {\rm GeV}$ and $K=1$. Frame (c) shows the dependence on $M_{1/2}$ at $m_0=2000\ {\rm GeV}$ and $K=1$.}
\label{Mb}
\end{center}
\end{figure}
Unless the trilinear coupling $A_t$ is very large, the gluino loop typically dominate (an exception is investigated in ref. \cite{controesempiobquark}) and the sign of the SUSY contribution is given by the sign of $M_3\mu$. Therefore, since we assume here gaugino universality, this implies that $b$-$\tau$ YU is favored in the $\mu<0$ case.\\
To numerically quantify this statement, we study the behavior of $m_b^{\rm tree}$ and $m_b^{\rm corr}$ varying the parameters of the mNUSM model. In fig \ref{Mb} ({\em a}) we fix $A_0=0$, $\tan\beta=38.0$, $m_0=1000\ {\rm GeV}$ and $M_{1/2}=1100\ {\rm GeV}$, and analyse the dependence on $K$. Wee see that the tree level value $m_b^{\rm tree}$ is roughly constant, while the SUSY corrections \emph{decrease} as $K$ increases. Both remain however well above the experimental upper bound. This can be understood from eq. (\ref{mbcorr}), since increasing $K$ means increasing $m_{\tilde b_{1}}$, with a fixed value for $M_3$ and, roughly, for $\mu$ and $m_{\tilde b_{2}}$, and the function $I(x,y,z)$, at fixed $y$ and $z$ is inversely proportional to $x=m_{\tilde b_{1}}$. Therefore, subsequently, we concentrate on the {\em universal} $K=1$ case, in order to check whether a parameter space allowing for ``top-down'' $b$-$\tau$ YU exists or not.\\
The second step is to study the dependence of the $b$-quark mass on the parameter $m_0$ (fig \ref{Mb} (b)). We take here $A_0=0$, $\tan\beta=38.0,\ M_{1/2}=1100\ {\rm GeV}$ and $K=1$, and we notice, as expected, that the size of the corrections decreases with increasing $m_0$. This is explained on the one hand by the fact that from the radiative EWSB condition the value of $\mu^2$ is slightly decreased by the increase of $m_0$ \cite{deboer}, and on the other hand because increasing the value of $m_{\tilde b_{1,2}}$ leads to a decrease of the function $I$. In fig.\ref{Mb} (c) we take instead  $A_0=0$, $\tan\beta=38.0,\ m_0=2000\ {\rm GeV}$ and $K=1$, and vary $M_{1/2}$. As can be easily inferred from eq. (\ref{mbcorr}), increasing $M_{1/2}$ leads to an increase both in $M_3$ and in $\mu$. In conclusion, the candidate parameter space for $b$-$\tau$ YU for $\mu>0$ is at low $M_{1/2}$ and high $m_{0}$ values. We choose therefore two trial values, $M_{1/2}=300\ {\rm GeV},\ m_0=2000\ {\rm GeV}$ and we show our results in fig.\ref{Mb2}. As readily seen from eq. (\ref{mbcorr}), we find that the SUSY contributions grow with $\tan\beta$. We notice however that the tree level mass strongly decreases with $\tan\beta$, owing to the fact that the positive SUSY contributions to $m_{\tau}$ (see eq. (\ref{mtaueq})) imply a smaller value for the asymptotic common $b$-$\tau$ Yukawa coupling. The overall conspiracy of these two effects is to maintain the corrected $b$ quark mass well above the experimental upper bound. This conclusion is further confirmed investigating extreme values of ($M_{1/2},\ m_0$) for high $\tan\beta$, always in the universal $K=1$ case, and resorting also to nonzero values of $A_0$.\\
To sum up, we demonstrated that, due to the SUSY corrections to the $b$-quark mass, top-down $b$-$\tau$ YU is excluded, both with universal and with minimal non-universal sfermion masses, in the case $\mu>0$.
\begin{figure}[!t]
\begin{center}
\includegraphics[scale=0.7,angle=-90]{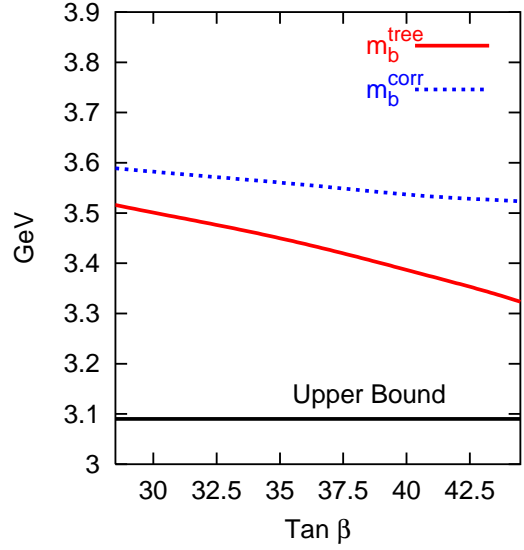}
\caption{\em\small Tree level and SUSY corrected values of the $b$-quark mass at $M_{1/2}=300\ {\rm GeV},\ m_0=2000\ {\rm GeV},\ A_0=0$ and $K=1$.}
\label{Mb2}
\end{center}
\end{figure}
\subsubsection{The $\mu<0$ case}
\begin{figure}[!t]
\begin{center}
\begin{tabular}{cc}
\includegraphics[scale=0.5,angle=-90]{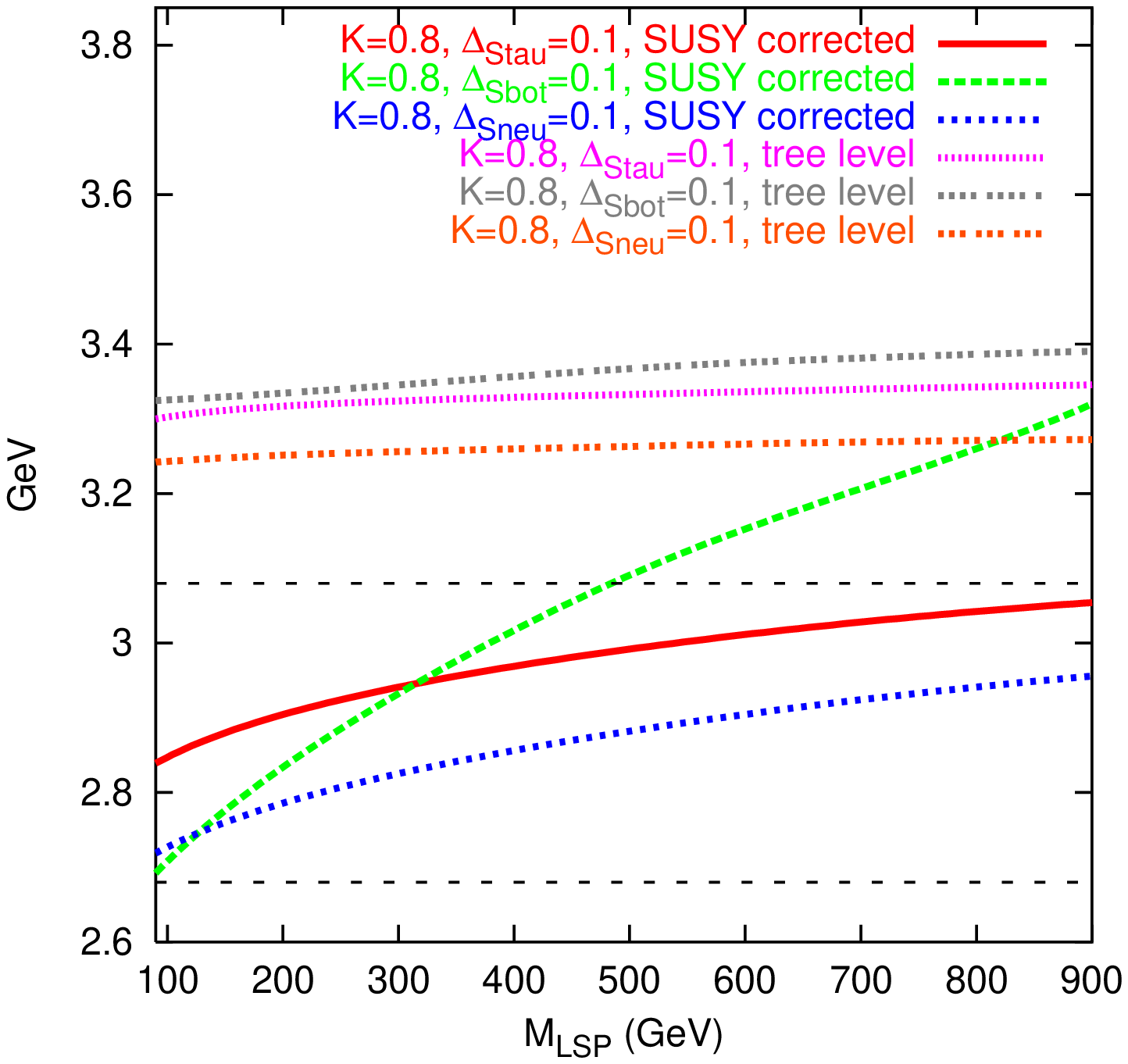} &
\includegraphics[scale=0.5,angle=-90]{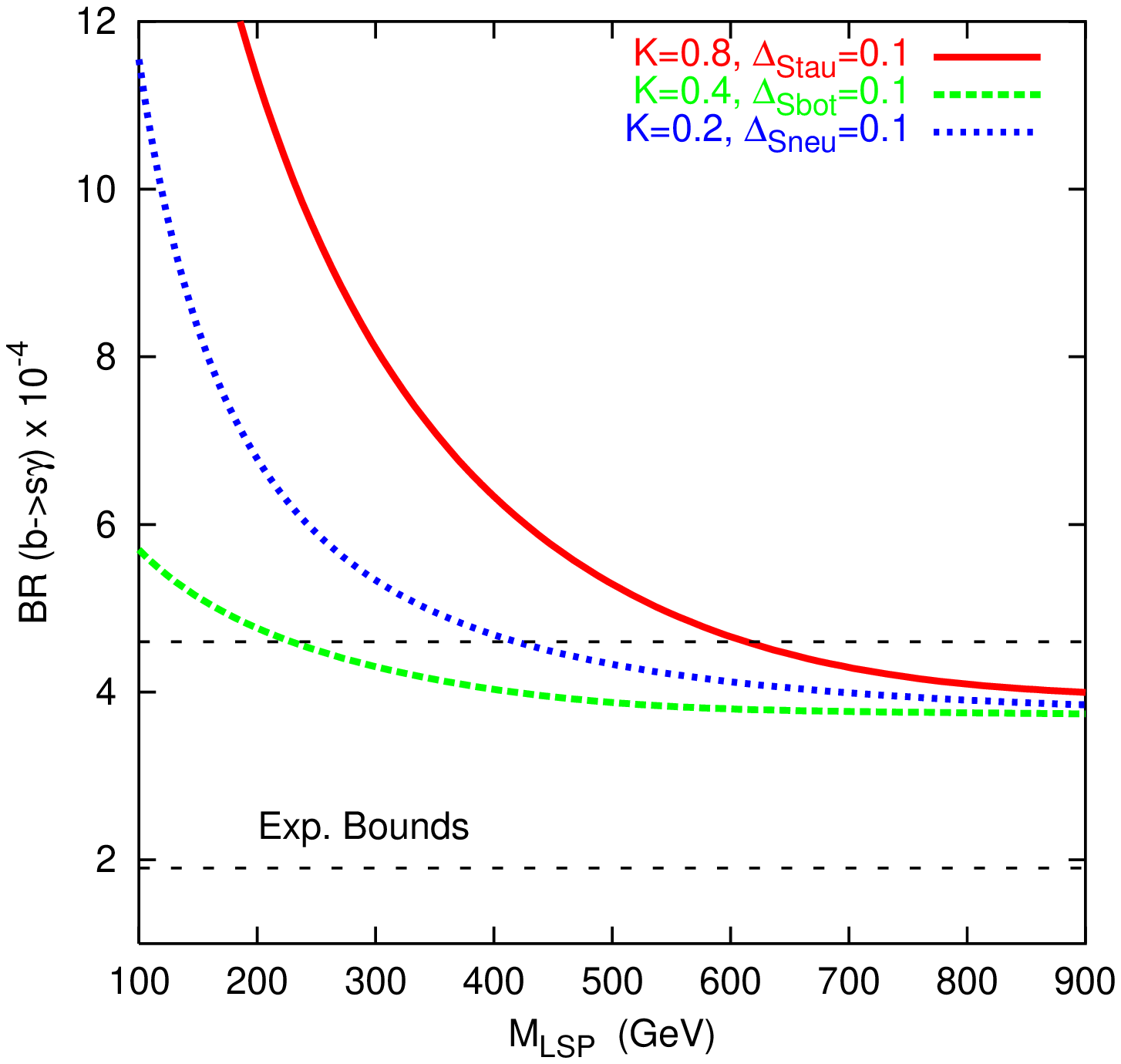}\\
&\\[-0.2cm]
\hspace{1.cm} (\emph{a}) & \hspace{1.cm} (\emph{b})\\
\end{tabular}
\caption{\em\small Tree level and SUSY corrected $b$-quark masses ($a$) and the $BR(b\rightarrow s\gamma)$ for the three benchmark scenarios of Tab.\ref{Tabellamodelli}. The upper and lower bounds are those of eq. (\ref{mbexprange}) for frame ($a$) and those of eq. (\ref{bsglimits}) for frame ($b$).}
\label{BsgUBC}\label{mbUBC}
\end{center}
\end{figure}

In the $\mu<0$ case the SUSY contributions to the $b$-quark mass are negative, and therefore conspire to bring the tree level mass dictated by $b$-$\tau$ YU {\em within} the experimental range (\ref{mbexprange}). We plot in fig.\ref{mbUBC} ($a$) the tree level and SUSY-corrected values of the $b$-quark mass for the three benchmark cases of Tab.\ref{Tabellamodelli}. We notice that the size of the SUSY corrections decrease when $m_{\neut}$ is increased, because of the mentioned behavior of the function $I(x,y,z)$ of eq.(\ref{mbcorr}), which is inverse-proportional to its arguments. In the case of the sbottom coannihilation (green dashed line) the effect is enhanced by the large size of $m_0$ which directly reduces, through the suppression of $\mu$, due to the EWSB condition, the size of the SUSY contributions.
\subsection{The inclusive branching ratio $b\rightarrow s \gamma$}
\label{secbsg}
We construct the 95\% C.L. range on the $BR(b\rightarrow s \gamma)$ starting from the recent experimental data of ref. \cite{bsgEXP} and properly combining the experimental and theoretical uncertainties. The resulting bound is
\begin{equation}
1.9\times10^{-4}\ \lesssim\ BR(b\rightarrow s \gamma) \ \lesssim \ 4.6\times 10^{-4}.
\label{bsglimits}
\end{equation}
We calculate the $BR(b\rightarrow s \gamma)$ using the current updated version of {\tt micrOMEGAs} \cite{micromegasnew}. The SM contribution is calculated through the formul\ae\ of ref. \cite{bsg1}, while the SUSY corrections through the ones of ref. \cite{bsg2}. We can estimate the dependence of the SUSY corrections from the approximate formul\ae\ of ref. \cite{Borzumati,bertolini,otherBSG}. They are given, respectively for the charged Higgs and for the chargino exchanges, by
\begin{eqnarray}
C^{H^+}&\approx&
\frac{1}{2}\frac{m^2_t}{m^2_{H^+}}
f^{H^+}\left(\frac{m^2_t}{m^2_{H^+}}\right),
\label{Hpluscontrib}
\\
C^{\tilde\chi^-}&\approx&
\left({\rm sign}\ A_t\right)  \frac{\tan\beta}{4} \frac{m_t}{\mu}
\left[
f^{\tilde\chi^-}\left(\frac{m^2_{\tilde{t_1}}}{\mu^2}\right)
-
f^{\tilde\chi^-}\left(\frac{m^2_{\tilde{t_2}}}{\mu^2}\right)
\right],
\label{chicontrib}
\end{eqnarray}
where
\begin{eqnarray}
f^{H^+}(x)&=&
\frac{3-5x}{6(x-1)^2} + \frac{3x-2}{3(x-1)^3} \ln x,
\\
f^{\tilde\chi^-}(x)&=&
\frac{7x-5}{6(x-1)^2} - \frac{x(3x-2)}{3(x-1)^3} \ln x.
\end{eqnarray}
We notice that the relative sign of the two contributions of eq. (\ref{Hpluscontrib}) and (\ref{chicontrib}) is given by ${\rm sign} (A_t\mu)=-{\rm sign} (M_3\mu)$, (the latter equality being valid for large top and bottom Yukawa couplings, as in the present case). Moreover, the SM contribution has the same sign as the charged Higgs one. All the contributions {\em decrease} as $m_{\neut}$ increases, therefore we expect to draw a lower bound on $m_{\neut}$ from the lower bound on $BR(b\rightarrow s\gamma)$ (\ref{bsglimits}) in the case $M_3\mu>0$ and from the upper bound of eq.  (\ref{bsglimits}) for $M_3\mu<0$, which is the present case. We notice that the lower bound on $m_{\neut}$ becomes more restrictive as $\tan\beta$ is increased, as can be read out from eq. (\ref{chicontrib}).\\
In fig.\ref{BsgUBC} ($b$) we plot the results we get for the three benchmark cases of Tab.\ref{Tabellamodelli}. We can see that as the SUSY spectrum becomes heavier (we recall that if the sbottom is the NLSP $m_0/M_{1/2}\gtrsim 3$, see fig.\ref{corridorMAIN}) the lower bound on $m_{\neut}$ weakens, as can be understood from eq. (\ref{Hpluscontrib}) and (\ref{chicontrib}) and from the explicit form of $f^{H^+}$ and $f^{\tilde\chi^-}$.

\subsection{The Higgs bosons masses}

\begin{figure}[!t]
\begin{center}
\begin{tabular}{cc}
\includegraphics[scale=0.5,angle=-90]{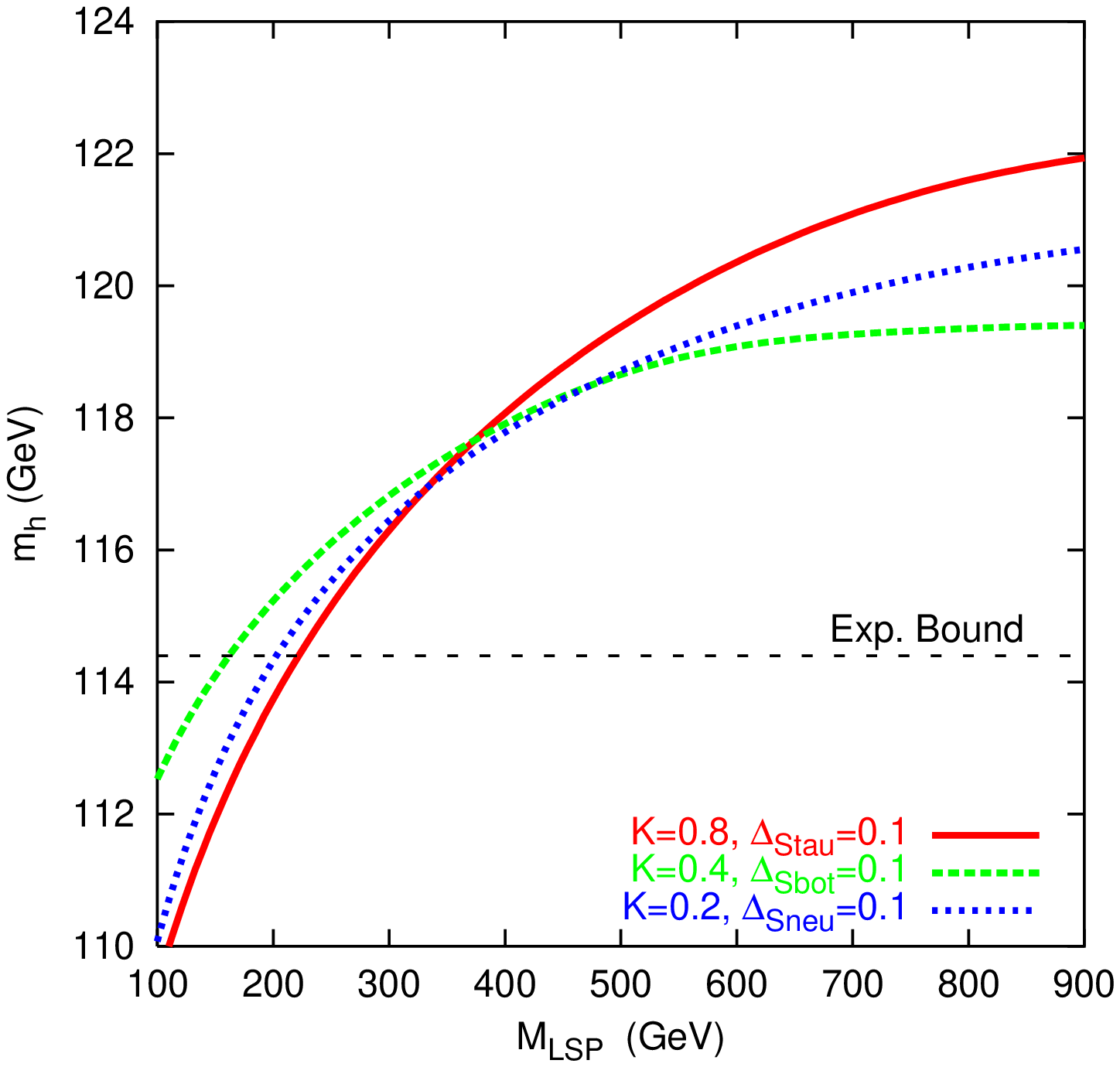} &
\includegraphics[scale=0.5,angle=-90]{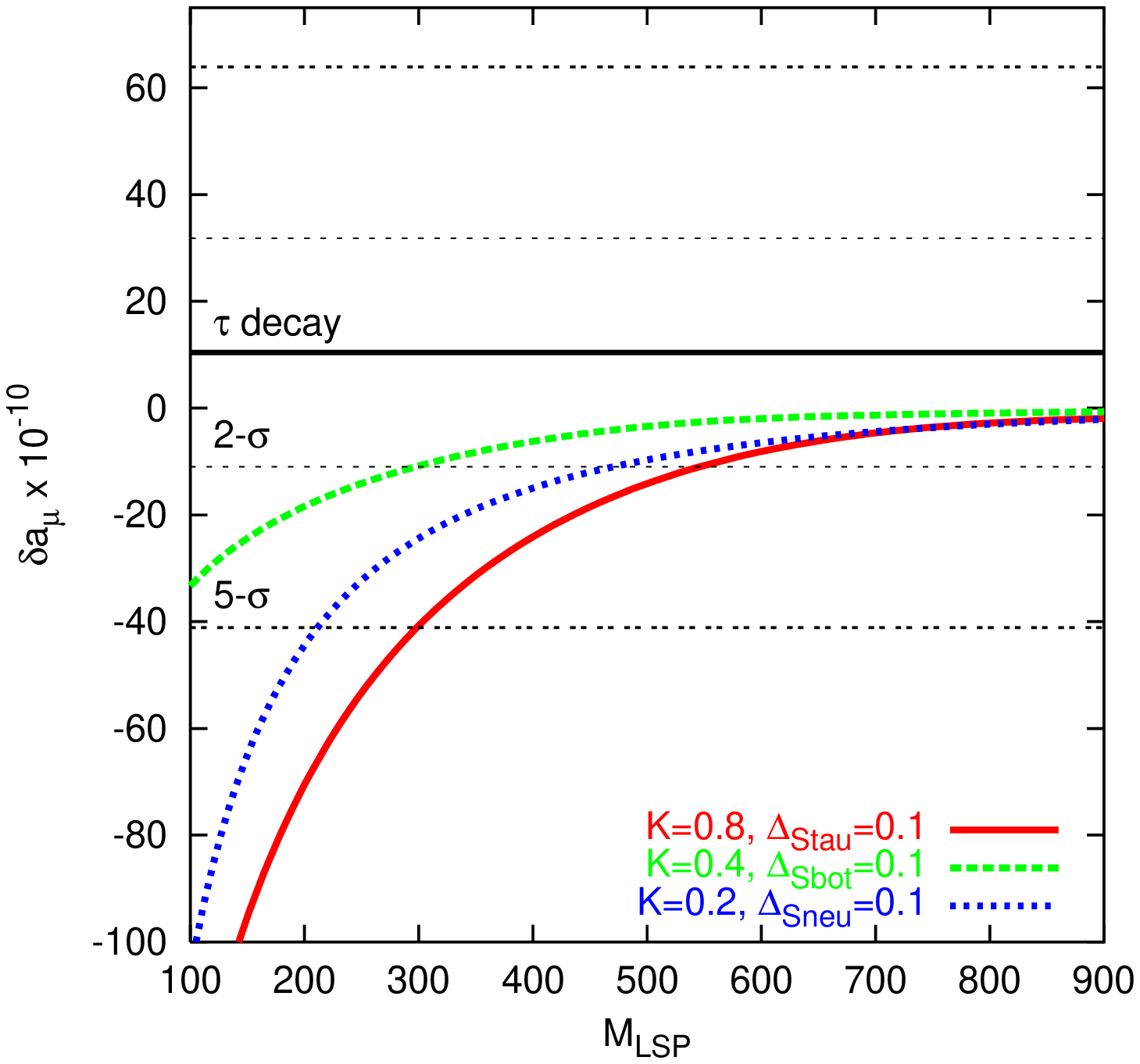}\\
&\\[-0.2cm]
\hspace{1.cm} (\emph{a}) & \hspace{1.cm} (\emph{b})\\
\end{tabular}
\caption{\em\small  $m_h$ ($a$) and $\delta a_\mu^{\sss SUSY}$
($b$) for the three benchmark scenarios of
Tab.\ref{Tabellamodelli}. The lower bound of frame ($a$) is given
in eq. (\ref{higgsbound}), while in frame ($b$) the $2$-$\sigma$
and $5$-$\sigma$ bounds are taken from eq. (\ref{damEXP2}).}
\label{damUBC}\label{mhUBC}
\end{center}
\end{figure}

We take in our analysis the 95\% C.L. LEP bound \cite{higgsbound} on the lightest CP-even neutral Higgs boson mass
\begin{equation}
m_h\ \gtrsim\ 114.3\ {\rm GeV},
\label{higgsbound}
\end{equation}
which gives a lower bound on the $m_{\neut}$. Two loop corrections are taken into account, as described in sec.\ref{procedure}. In fig.\ref{mhUBC} ($a$) we plot the results for the Higgs mass we obtain for the three benchmark scenarios.
\subsection{Direct accelerators sparticle searches}
We impose the current direct accelerator search limits on the sparticle masses \cite{PDG}, respectively
  $m_{\chi_1^+}\gtrsim 104$ GeV, $m_{\tilde{f}}\gtrsim 100$ GeV for
  $\tilde{f}=\tilde{t}_1,\tilde{b}_1,\tilde{l}^{\pm},\tilde{\nu}$,
  $m_{\tilde{g}}\gtrsim 300$ GeV, $m_{\tilde{q}_{1,2}}\gtrsim260$ GeV for
  $\tilde{q}=\tilde{u},\tilde{d},\tilde{s},\tilde{c}$.
We always find the constraints from  $BR(b\rightarrow s\gamma)$ and from $m_h$ more restrictive than the direct sparticle searches. Therefore we do not plot these bounds in the panels of sec.\ref{neutsbot} and \ref{neutstausneut}.
\subsection{The muon anomalous magnetic moment}
\label{damSEC}
The deviation $\delta a_\mu$ of the muon anomalous magnetic moment $a_\mu$ from its predicted value in the Standard Model can be interpreted as arising from SUSY contributions, $\delta a_\mu^{\sss SUSY}$, mainly given by neutralino-smuon ($a$) and chargino-sneutrino ($b$) loops.
The quantity $\delta a_\mu^{\sss SUSY}$, which we numerically compute with the {\tt micrOMEGAs} package, can be estimated through the formul\ae \  of ref. \cite{dam1}, assuming a common value for the SUSY sparticles masses $m_{\sss SUSY}$
\begin{equation}
\delta a^{\sss SUSY}_\mu\ \approx\ {\rm sign}(M_2\mu)\frac{\tan\beta}{192\pi^2}\frac{m_\mu^2}{m^2_{\sss SUSY}}(5g_2^2+g_1^2).
\label{damapprox}
\end{equation}
On the experimental side, the BNL E821 experiment recently delivered a high precision measurement (0.7 ppm) of $a_\mu^{\rm exp}=11659203(8)\times10^{-10}$. The theoretical computation of the SM prediction is plagued by the problem of estimating the hadronic vacuum polarization contribution. In particular, there is a persisting discrepancy between the calculations based on the $\tau$ decay data and those based on low-energy $e^+ e^-$ data. Recent evaluations \cite{damTH,damNyff,damNarison} give the following range for the deviation of the SM value of $a_\mu$ from the experimental one:
\begin{eqnarray}
a_\mu^{exp}-a_\mu^{\sss SM}&=&\left(35.6\ \pm\ 11.7\right)\ \times 10^{-10} \quad (e^+\ e^-)\label{damEXP1}\\
a_\mu^{exp}-a_\mu^{\sss SM}&=&\left(10.4\ \pm\ 10.7\right)\ \times 10^{-10} \quad (\tau\ {\rm decay})\label{damEXP2}
\end{eqnarray}
Following the lines of ref \cite{superconservative} we decided to take a conservative approach to the problem of choosing which bound should be culled from the tantalizing results of eq. (\ref{damEXP1}, \ref{damEXP2}). In what follows we will therefore just indicate the region determined by the  $2$-$\sigma$ and $5$-$\sigma$ range for the $e^+ e^-$ based approach and for the $\tau$ decay based approach, but we will not use this constraint to derive (lower) bounds on $m_{\neut}$.\\
In fig.\ref{damUBC} ($b$), where we present  the results we get for the three benchmark scenarios, we indicate only the $\tau$ decay based limit, since $\delta a_\mu^{\sss SUSY}$, being $\mu<0$, is negative (see eq.(\ref{damapprox})). In passing, we remark that it has been recently claimed \cite{damNarison} that the result from $e^+e^-$ data (\ref{damEXP1}) is less clean than the one from $\tau$-decay (\ref{damEXP2}). This would be due to intereferences of isoscalar $I=0$ mesons, not produced in $\tau$-decay, with vector mesons \cite{damNarison}. Therefore, theories requiring a negative sign of $\mu$, as those with exact $b-\tau$ YU, seem to be no longer disfavored by the $\delta a_\mu$ constraint.\\
Formula (\ref{damapprox}) allows us to interpret the results we show in fig.\ref{damUBC} ($b$) and in the figures of the next section. First, we see that as the SUSY spectrum becomes heavier (sbottom NLSP case) the corrections become smaller, as emerging from (\ref{damapprox}). Second, we can see that $\delta a^{\sss SUSY}_\mu$ decreases with increasing $m_{\neut}$, for the same region as above, and that it increases with $\tan\beta$.

\section{The new coannihilation corridor}
\label{munegativecorridors}

In the $\mu<0$ case the SUSY corrections to the $b$-quark mass are
negative, and therefore can naturally drive, for suitable values of $\tan\beta$, the corrected
$m_b^{\rm corr}(M_Z)$ within the experimental range.\\ 
\noindent In this section we focus on the extended sfermion coannihilatoin modes allowed by the particle spectrum of mNUSM. As pointed out in sec.\ref{particlespectrum}, the boundary conditions given by mNUSM produce a new ``coannihilation branch'', extending from $K=0$ up to $K\approx0.5$ and for values of $m_0$ greater than \mbox{$\approx c M_{1/2}$}, with $c$ of $o(1)$, increasing with decreasing values of $\tan\beta$. In the upper part of the branch, typically for $m_0/M_{1/2}\gtrsim3$, the NLSP turns out to be the lightest sbottom, while for $c\lesssim m_0/M_{1/2}\lesssim3$ we find $m_{\neut}\simeq m_{\sneu}\simeq m_{\stau}$. The following two subsections are devoted to the analysis of the coannihilations processes $\neut$-$\sbot$ and $\neut$-$\sneu$-$\stau$, taking into account the constraints described in the previous section. The final two subsections deal respectively with the determination of the $\tan\beta$ range allowed by $b$-$\tau$ YU, and with the question of fine-tuning in mNUSM models.

\subsection{$\neut-\sbot$ coannihilations}
\label{neutsbot}

\noindent Being a strong-interacting particle, we find that sbottom annihilation and coannihilations channels are very efficient, and substantially contribute to the neutralino relic density suppression.\\
\begin{figure}[!b]
\begin{center}
\includegraphics[scale=0.55,angle=-90]{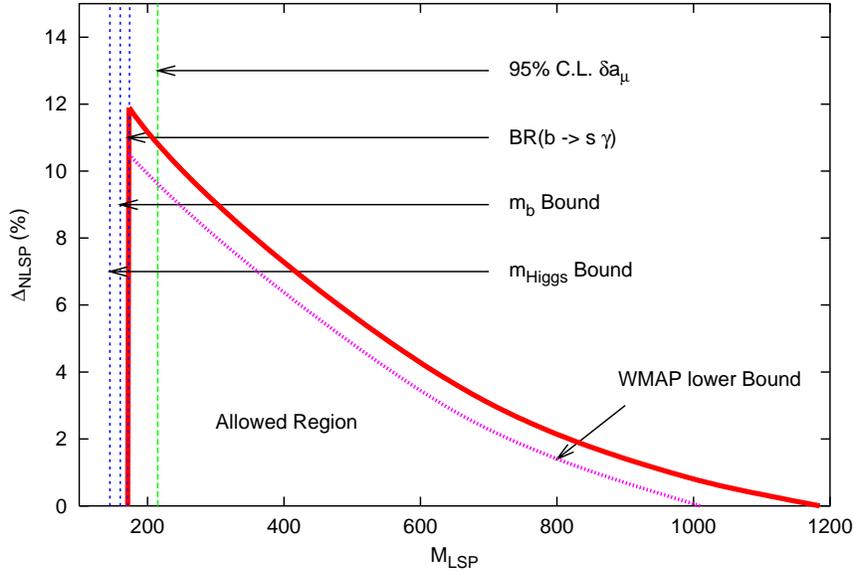}
\caption{\em\small Cosmologically allowed parameter space in the sbottom coannihilation region, at $K=0.35$ and $\tan\beta=38.0$. The region below the red solid line has $\Omega_{\neut}h^2<0.1287$. The magenta line shows the putative lower bound on the neutralino relic density (\ref{omegarange}). The blue dotted nearly vertical lines indicate the phenomenological constraints described in the text, while the green dashed line stands for the $2$-$\sigma$ bound on the muon anomalous magnetic moment.}
\label{k035tb38}
\end{center}
\end{figure}
\begin{figure}[!t]
\begin{center}
\begin{tabular}{cc}
\includegraphics[scale=0.45,angle=-90]{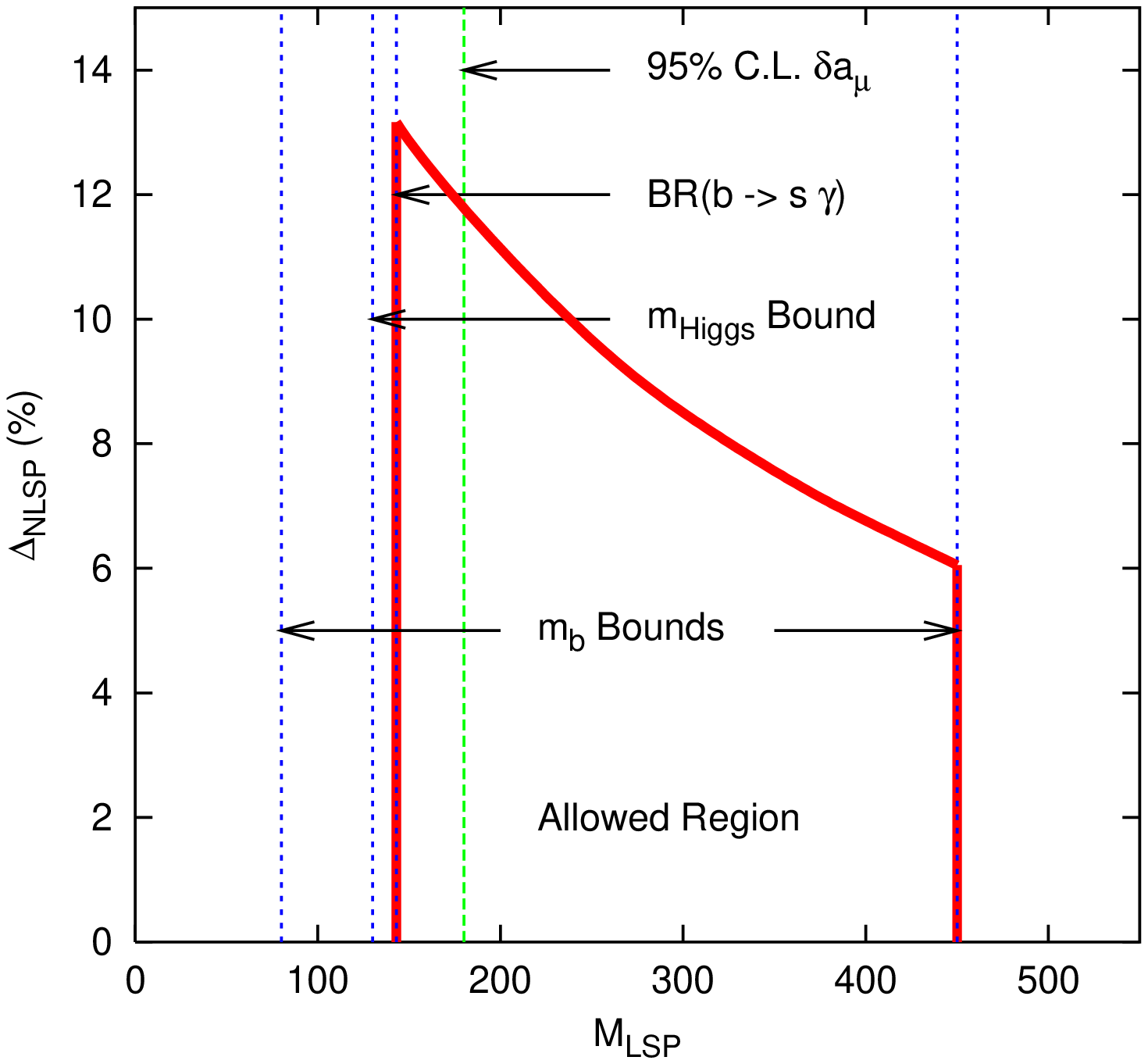} &
\includegraphics[scale=0.45,angle=-90]{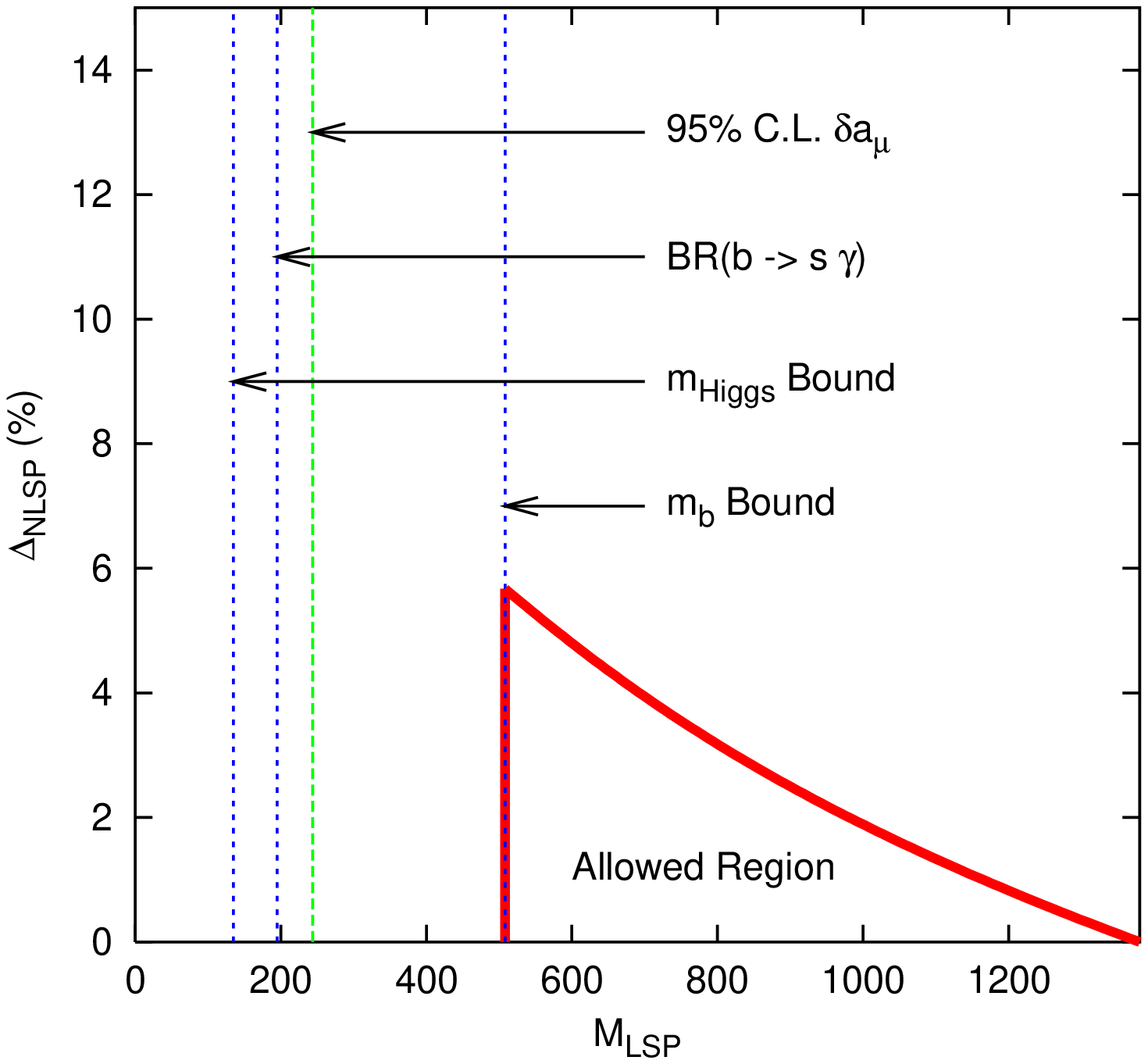}\\
&\\[-0.2cm]
\hspace{1.cm} (\emph{a}) & \hspace{1.cm} (\emph{b})\\
\end{tabular}
\caption{\em\small Cosmo-phenomenologically allowed parameter space in the sbottom Coannihilation region at $K=0.35$, $\tan\beta=34.0$ (a) and $\tan\beta=42.0$ (b).}
\label{k035tb34tb42}
\end{center}
\end{figure}
\noindent We plot in fig.\ref{k035tb38} and \ref{k035tb34tb42} ($a$) and ($b$) the cosmologically allowed regions for three representative values of $\tan\beta=34,\ 38$ and $42$ for a fixed value of $K=0.35$. The $y$ axis represents the relative percent splitting between the neutralino mass and the next-to-lightest SUSY particle, in this case the lightest sbottom:
\begin{equation}
\displaystyle \Delta_{\sss NLSP}\ =\ \frac{m_{\sss NLSP}-m_{\neut}}{m_{\neut}}.
\end{equation}
The relevant phenomenological constraints described in the previous sections determine lower, and sometimes upper, limits on $m_{\neut}$. They result in almost vertical lines, since they typically weakly depend on $m_0$, which moreover very little varies in the plotted regions. They are instead fixed by $M_{1/2}$ (and therefore, in the plots, by $m_{\neut}$). The regions allowed by a given constraint lie at the {\em right} of the respective lines. Points below the red lines are characterized by relic densities which fall within the 2-$\sigma$ range of eq. (\ref{omegarange}). We emphasize that the limits on  $m_{\neut}$ and $\Delta_{\sss NLSP}$ we find here cannot be regarded as being representative of the full parameter space: they show instead, for a few benchmark cases, how the interplay of the cosmo-phenomenological bounds described in sec.\ref{constraints} constrain the mass spectrum (here parameterized by $m_{\neut}$ and $\Delta_{\sss NLSP}$) of the models under consideration.\\ 
\noindent In fig.\ref{k035tb38} we plot with a magenta dotted line also the lower bound on $\Omega_{\sss CDM}h^2$. We see that the most stringent bound, for $\tan\beta\lesssim38$, is provided by the $BR(b\rightarrow s\gamma)$, while for higher values of $\tan\beta$ the $b$-quark mass constraint becomes more restrictive. As far as the upper bound on $m_{\neut}$ is concerned, we find that high values of $\tan\beta$ allow $m_{\neut}$ to extend up to 1.5 TeV (fig.\ref{k035tb34tb42} ($b$)). For smaller $\tan\beta$ the upper bound on $m_{\neut}$ is instead fixed by the upper bound on the $b$-quark mass: the SUSY corrections to $m_{\tau}$ (see eq.(\ref{mtaueq})) drive the common $b$-$\tau$ Yukawa coupling to higher values, therefore generating a higher tree level $b$-quark mass which is not compensated by the (smaller) negative SUSY corrections to $m_b$ (eq. (\ref{mbcorr})). The inverse mechanism pushes the lower bound on $m_{\neut}$ to higher values ($\approx0.5\ {\rm TeV}$) in the case of $\tan\beta=42$. We notice that the region determined by the 2-$\sigma$ $\delta a_\mu$ range from the $\tau$ decay approach always covers the great part (or the totality) of the allowed parameter space.\\
\begin{figure}[!t]
\begin{center}
\includegraphics[scale=0.55,angle=-90]{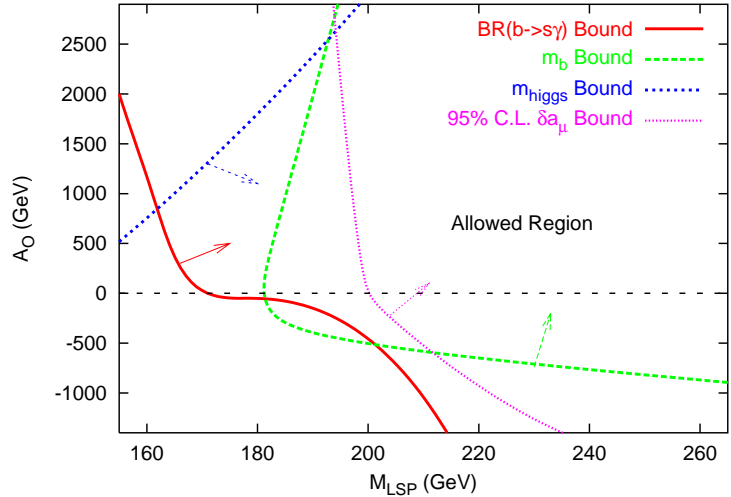}
\caption{\em\small The determination of the lower bound on $m_{\neut}$ in the case of non zero $A_0$, for $\tan\beta=38.0$, $\Delta_{\sbot}\simeq0$ and $K=0.35$. The allowed region is in the upper right part of the figure, as indicated.}
\label{Azeroplot}
\end{center}
\end{figure}
In fig.\ref{Azeroplot} we studied the sensitivity of our results to a non-zero value of the common trilinear coupling $A_0$ at the GUT scale, for $\tan\beta=38.0$, $\Delta_{\sbot}\simeq0$ and $K=0.35$. We notice that the lower limit on $m_{\neut}$ is fixed by the $A_0=0$ case. As for the cosmological bound on $\Omega_{\neut}h^2$, the variation of $A_0$ leads to very little modifications of the upper bounds on $m_{\neut}$ displayed in fig.\ref{k035tb38} and \ref{k035tb34tb42}: we therefore conclude that the only relevant change is the type of constraint that, depending on the sign and on the absolute value of $A_0$ determines the lower bound on $m_{\neut}$.\\
As far as the coannihilation channels are concerned, the case of the sbottom is characterized by a rather simple pattern, clearly dominated by strong interaction processes. We find that for sbottom masses quasi-degenerate with the neutralino mass, the neutralino pair annihilation rate is very low (less than few percent). The dominant channels concern instead neutralino-sbottom coannihilations into gluon-$b$ quark (up to 10\%) and sbottom-sbottom annihilations into a couple of $b$ quarks (up to 15\%) or into a couple of gluons (up to 80\%). We give in the Appendix (sec.\ref{acoefficients}) an approximate analytical treatment of these three most relevant processes.
\subsection{$\neut-\stau-\sneu$ coannihilations}
\label{neutstausneut}
\begin{figure}[!t]
\begin{center}
\includegraphics[scale=0.55,angle=-90]{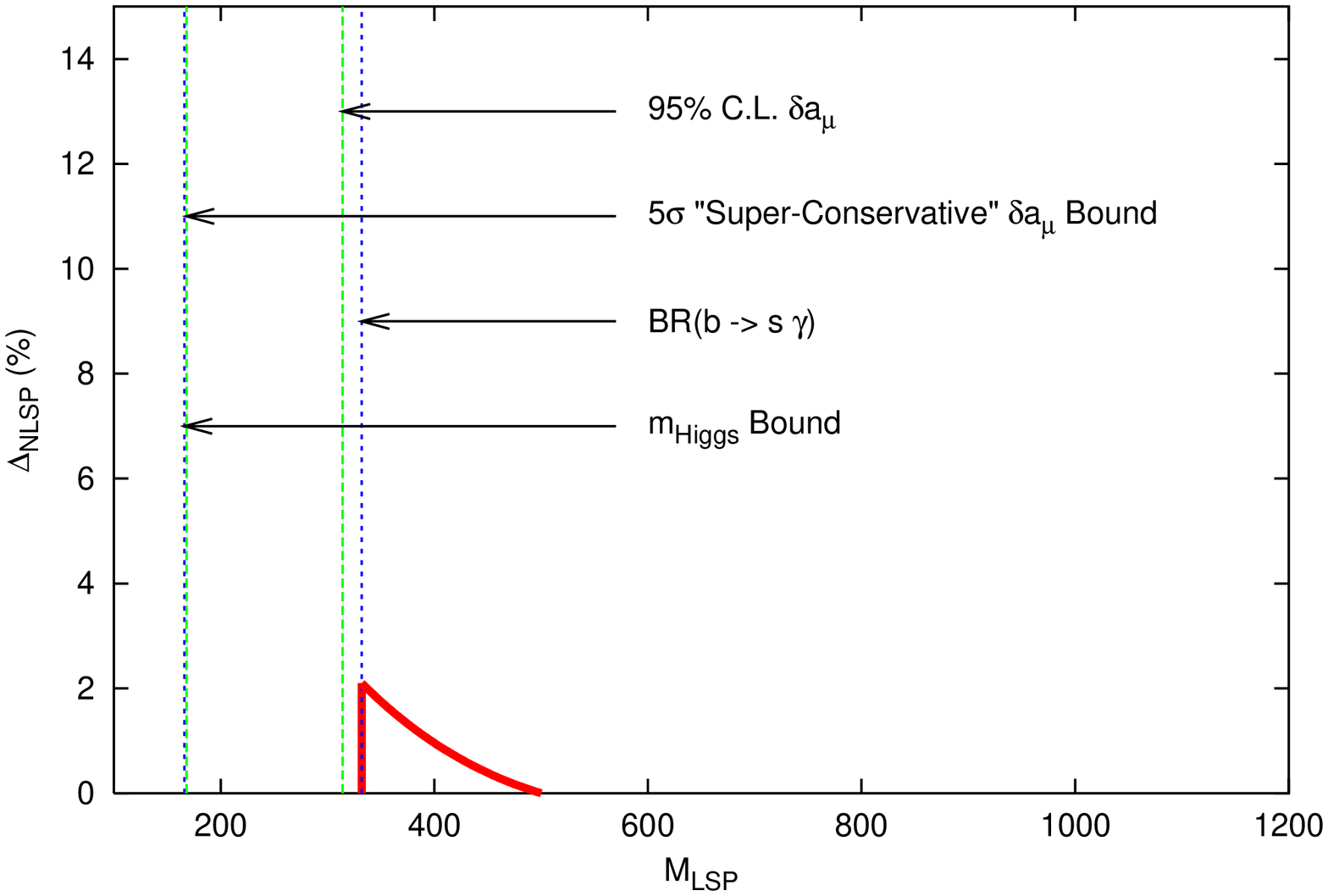}
\caption{\em\small Cosmologically allowed parameter space in the tau sneutrino-stau coannihilation region, at $K=0.1$ and $\tan\beta=38.0$. The scale of the axis, as well as the notation, are the same as in fig.\ref{k035tb38}.}
\label{k01tb38}
\end{center}
\end{figure}

\begin{figure}[!t]
\begin{center}
\begin{tabular}{cc}
\includegraphics[scale=0.45,angle=-90]{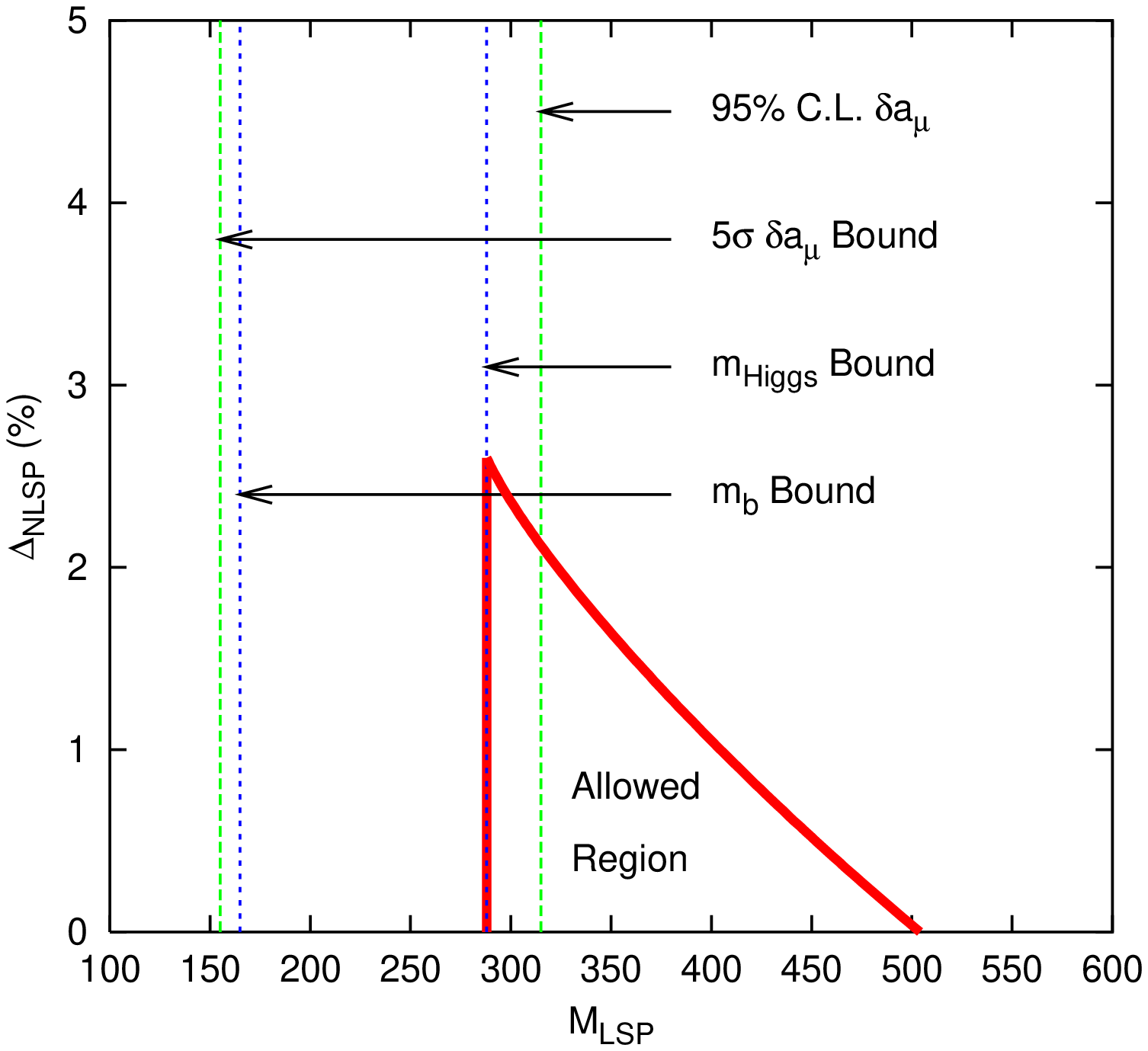} &
\includegraphics[scale=0.45,angle=-90]{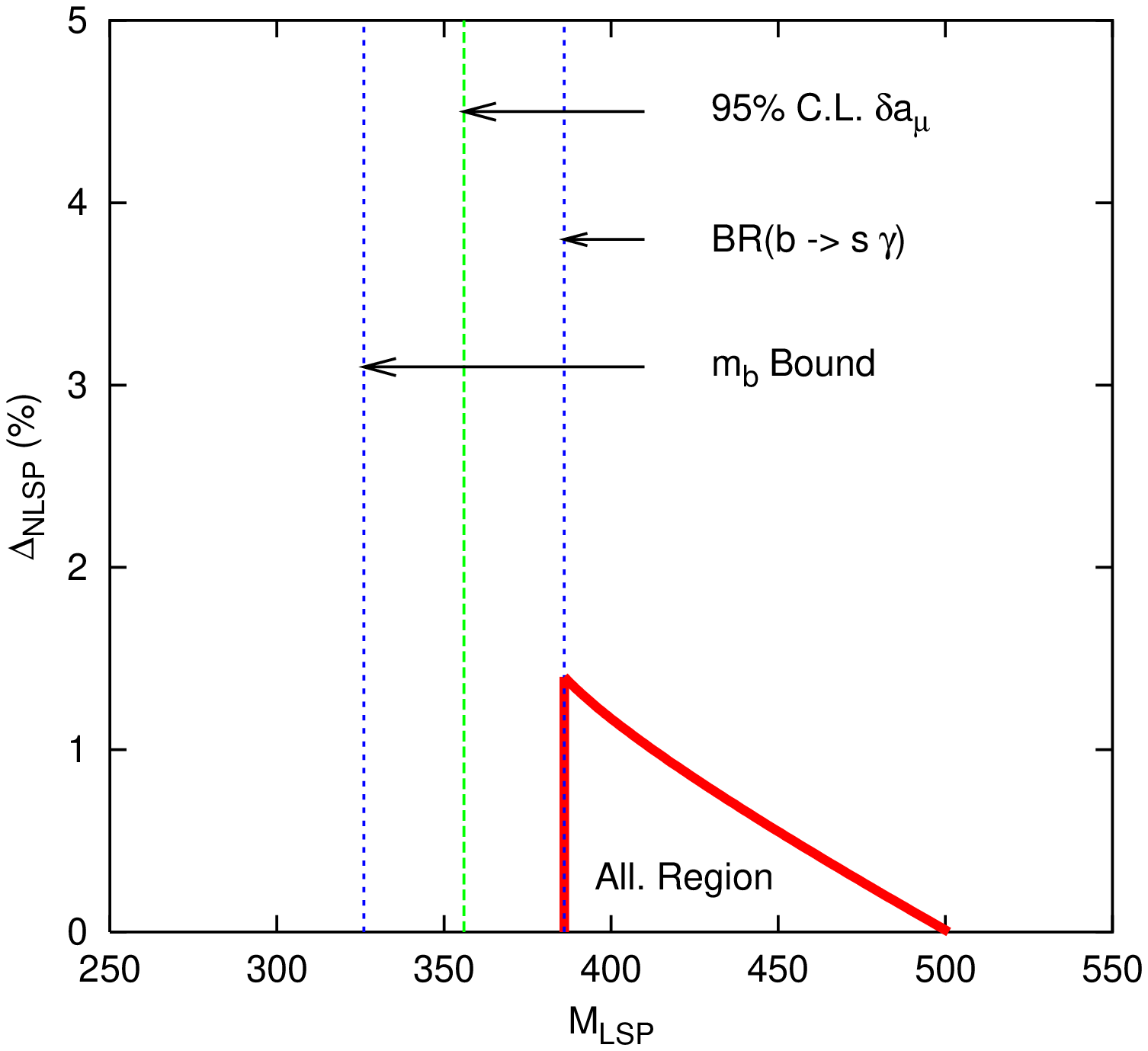}\\
&\\[-0.2cm]
\hspace{1.cm} (\emph{a}) & \hspace{1.cm} (\emph{b})\\
\end{tabular}
\caption{\em\small Cosmo-phenomenologically allowed parameter space in the tau sneutrino-stau Coannihilation region at $K=0.1$, $\tan\beta=34.0$ (a) and $\tan\beta=42.0$ (b).}
\label{k01tb34tb42}
\end{center}
\end{figure}

\begin{figure}[!t]
\begin{center}
\includegraphics[scale=0.6,angle=0]{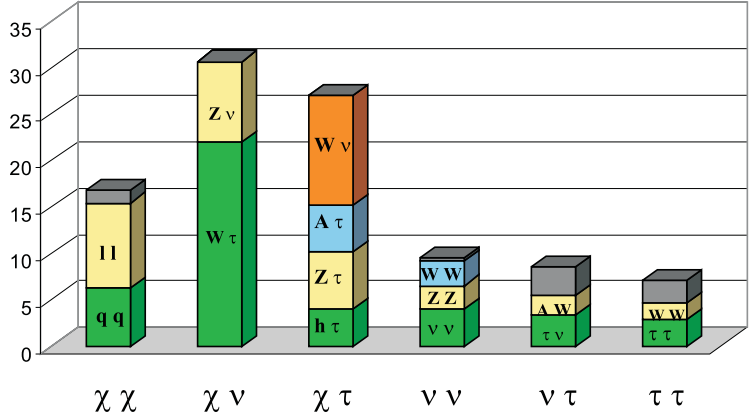}
\caption{\em\small A typical pattern of the relative contribution of coannihilation processes in the tau sneutrino coannihilation region. The plot refers to the case $\tan\beta=38$, $A_0=0$ and $m_{\neut}\simeq m_{\sneu}$. The upper gray part of the columns represents other contributing channels.}
\label{histo}
\end{center}
\end{figure}

In the lower of part of the new coannihilation branch of fig \ref{corridorMAIN}, for $(m_0/M_{1/2})\lesssim3$ and $0<K\lesssim0.25$, the NLSP switches from the sbottom to the tau sneutrino. In this region the lightest stau is always heavier than the sneutrino, but the relative mass splitting is within few percent. This small splitting results from the combination of two RG effects:
\begin{equation}
m^2_{\sneu}-m^2_{\stau}\approx\Delta^{\tilde\tau}_{LR}(A_\tau)-c_{\tilde l}\cos(2\beta)M_Z^2.\label{StauSneusplitting}
\end{equation}
The first contribution in eq.(\ref{StauSneusplitting}) comes from the mixing between $\tilde\tau_L$ and $\tilde\tau_R$, which depends on $A_\tau$, while the second term stems from the mentioned $D$-term quartic interaction, and the coefficient $c_{\tilde l}\simeq0.8$ \cite{deboer}. Even though $A_0=0$ at the GUT scale, RG effects can drive $A_\tau(M_{\sss SUSY})$ to large values (for the case of fig.\ref{corridorMAIN} we obtain typical values for $A_\tau\approx 0.5 \ {\rm TeV}$) and thus entail a non trivial $LR$ mixing $\Delta^{\tilde\tau}_{LR}$. In the case of fig.\ref{corridorMAIN} we get $m^2_{\sneu}-m^2_{\stau}\approx10\ {\rm GeV}$.\\
In fig.\ref{k01tb38} we plot the cosmo-phenomenologically allowed region at $K=0.1$ and $\tan\beta=38.0$, in analogy with fig.\ref{k035tb38}. We also show the ``super-conservative'' 5-$\sigma$ bound on $\delta a_\mu$, while the $m_b$ bounds lie outside the depicted range.\\
fig.\ref{k035tb38} and \ref{k01tb38} allow to compare the efficiency of coannihilation processes involving a strongly interacting sparticle (the sbottom) and a weakly, or electromagnetically interacting one (the tau sneutrino and the stau). We see that in the second case the maximum mass splitting between the NLSP and the neutralino is 2\%, while in the first one, at the same $m_{\neut}$, the cosmological bound allows a mass splitting up to $\approx8\%$. In fig.\ref{k01tb34tb42} we plot the allowed regions at $\tan\beta=34$ ($a$) and $42$ ($b$). We notice in all cases that slepton coannihilations constrain  $m_{\neut}$ to less than approx. $0.5\ {\rm TeV}$, while in the case of the sbottom the bound is three times higher.\\
Remarkably, we see in fig.\ref{k01tb38} and \ref{k01tb34tb42} that the 2-$\sigma$ bound on $\delta a_\mu$ coming from $\tau$ decay data is {\em always fulfilled} in the parameter space regions allowed by the other cosmo-phenomenological constraints.\\
As regards non-zero values for $A_0$, we draw in the present case the same conclusions as in the preceding section (see fig.\ref{Azeroplot}): the type of phenomenological constraint giving the lower bound on $m_{\neut}$ in general changes, but the lowest value of $m_{\neut}$ is determined by the $A_0=0$ case.\\
The coannihilation pattern is in this case by far more complicated than in the sbottom case. We show in fig.\ref{histo} a typical situation for the (percent) contributions of the possible coannihilating initial sparticles, and a detail of the most relevant final states, taken at $m_{\neut}\simeq m_{\sneu}$ and $\tan\beta=38$. This pattern is however rather dependent on the $\tan\beta$ value and on the relative mass splitting. The case of stau-tau sneutrino coannihilations is further discussed in the Appendix.\\
%

\subsection{The $\tan\beta$ range}\label{allowedtanbetarange}

\begin{figure}[!t]
\begin{center}
\includegraphics[scale=0.5,angle=-90]{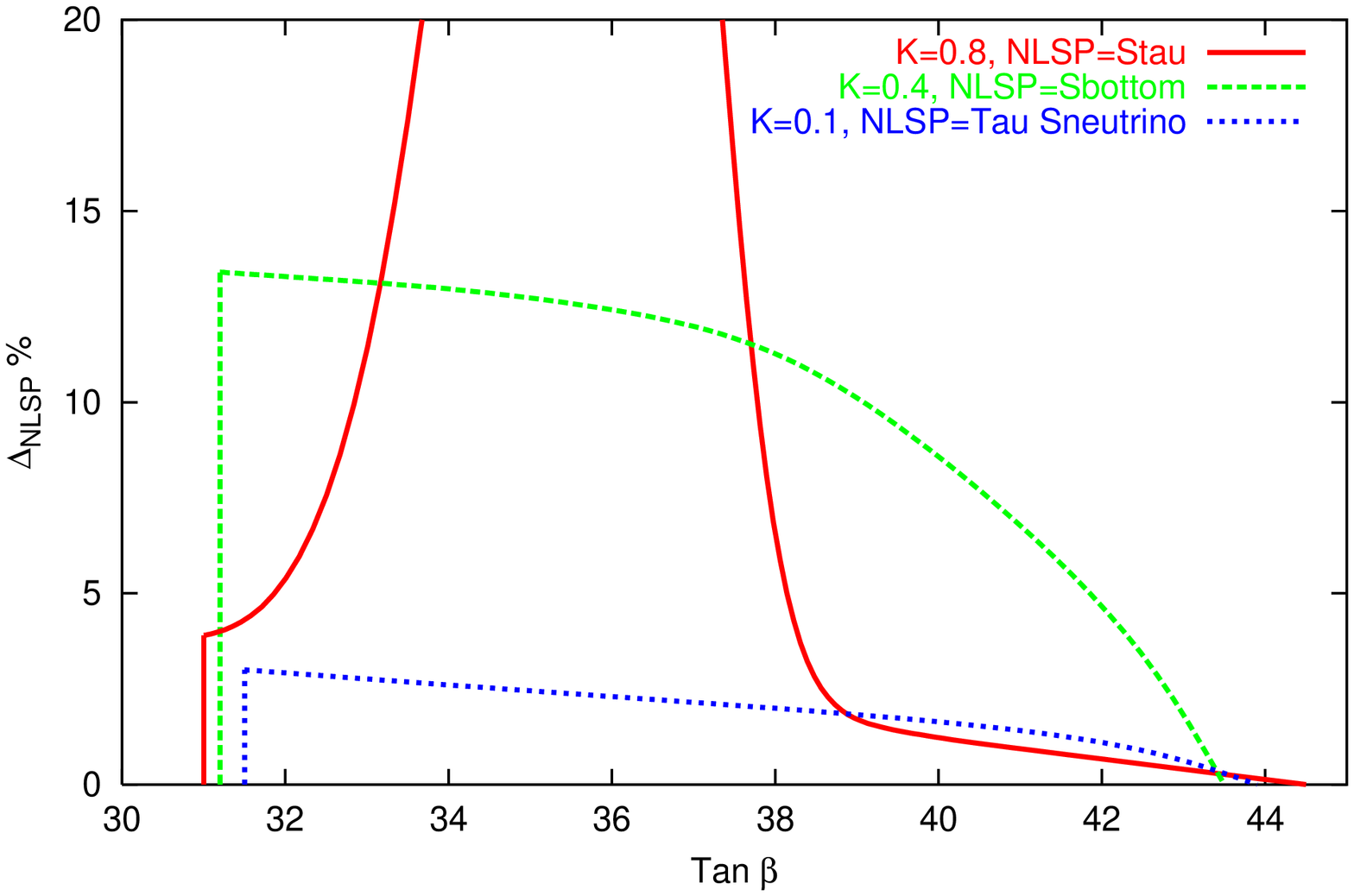}
\caption{\em\small The allowed ($\tan\beta,\Delta_{\sss NLSP}$) regions for three benchmark $K$ values, $K=0.8,\ 0.4,\ 0.1$, corresponding respectively to NLSP=$\stau,\ \sbot,\ \sneu$. Points below the lines are cosmo-phenomenologically allowed. For $K=0.8$ we find that, for $32\lesssim\tan\beta\lesssim38$, the coannihilation region is contiguous to the $A$ pole direct annihilation region, producing high values for $\Delta_{\sss NLSP}$ which are however unrelated to the efficiency of the coannihilation processes.}
\label{tanbetarangefigure}
\end{center}
\end{figure}

\noindent From eq.
(\ref{mbcorr}) we notice that the size of the corrections
(linearly) grows with $\tan\beta$, therefore we expect that the
upper bound on $m_b(M_Z)$ determine the lower bound on $\tan\beta$
and vice-versa. As emerging from the figures of
sec.\ref{neutsbot} and \ref{neutstausneut}, the $m_b$ constraint
concurs with the other cosmo-phenomenological bounds, 
entails the determination of the allowed range of $\tan\beta$
(sec.\ref{allowedtanbetarange}).
\noindent We determine the lowest limit on $\tan\beta$ combining the most restrictive phenomenological constraint, which turns out to be the $BR(b\rightarrow s\gamma)$ and which gives the lower bound on $m^{\rm min}_{\neut}\equiv m_{\neut}$, with the upper bound on $m_b^{\rm corr}$, which in its turn gives the upper bound $m^{\rm max}_{\neut}$. Requiring $m^{\rm min}_{\neut}=m^{\rm max}_{\neut}$ unambiguously gives the lowest allowed value of $\tan\beta$. In order to find the upper limit on $\tan\beta$, we notice that the bound on $m_b$ is weaker than the constraint coming from $BR(b\rightarrow s\gamma)$ in determining the lowest value $m^{\rm min}_{\neut}$ (see e.g. fig.\ref{k035tb34tb42} ($b$) and fig.\ref{k01tb34tb42}  ($b$)). In fact, the $BR(b\rightarrow s\gamma)$ constraint becomes more and more stringent as one increases $\tan\beta$. On the other hand, $m^{\rm max}_{\neut}$ is this time fixed by the cosmological constraint on the maximum allowed neutralino relic density (see sec.\ref{secomega} and fig.\ref{k035tb34tb42} ($b$) and \ref{k01tb34tb42} ($b$)). In fig.\ref{tanbetarangefigure} we depict the allowed $\tan\beta$ range in the ($\tan\beta,\Delta_{\sss NLSP}$) plane for three benchmark $K$ values, respectively $K=0.8,\ 0.4$ and 0.1. Points below the lines are cosmologically and phenomenologically viable. As emerging from fig.\ref{omegaUBC}, in the case of stau NLSP the coannihilation region can be connected to the $A$ pole region, hence the allowed $\Delta_{\sss NLSP}$ becomes large and no more meaningful to test the efficiency of coannihilation processes. Fig.\ref{tanbetarangefigure} also highlights that sbottom coannihilations allow a larger mass range for the coannihilating particle, since they involve strong interaction processes: the maximum $\Delta_{\sss NLSP}$ extends in this case up to $\approx13\%$, while in the case of $\neut-\stau-\sneu$ coannihilations its maximum value is around 3\%. In the high $\tan\beta$ tail of fig.\ref{tanbetarangefigure}, as easily understood, $\neut-\stau$ coannihilations are slightly less efficient than $\neut-\stau-\sneu$. Though our analysis was limited to benchmark $K$ values, we can conclude from fig.\ref{tanbetarangefigure} that in the $\mu<0$ case the allowed $\tan\beta$ range is
\begin{equation}
31\ \lesssim\ \tan\beta \ \lesssim\ 45, \qquad \mu<0.
\label{tanbetarange}
\end{equation}
We checked that the range given in (\ref{tanbetarange}) holds also for $A_0\neq0$. We find in fact that in this case an analogous procedure of determination of $(\tan\beta)^{\rm max,\ min}$ yields weaker limits. What happens at $A_0\neq0$ is mainly that the phenomenological bounds tend to push $m^{\rm min,\ max}_{\neut}$ to higher values, therefore  $(\tan\beta)^{\rm max}$ is lowered, while $(\tan\beta)^{\rm min}$ is left substantially unchanged (see e.g. fig.\ref{Azeroplot}).\\

\subsection{Fine tuning}
\begin{figure}[!t]
\begin{center}
\includegraphics[scale=0.55,angle=-90]{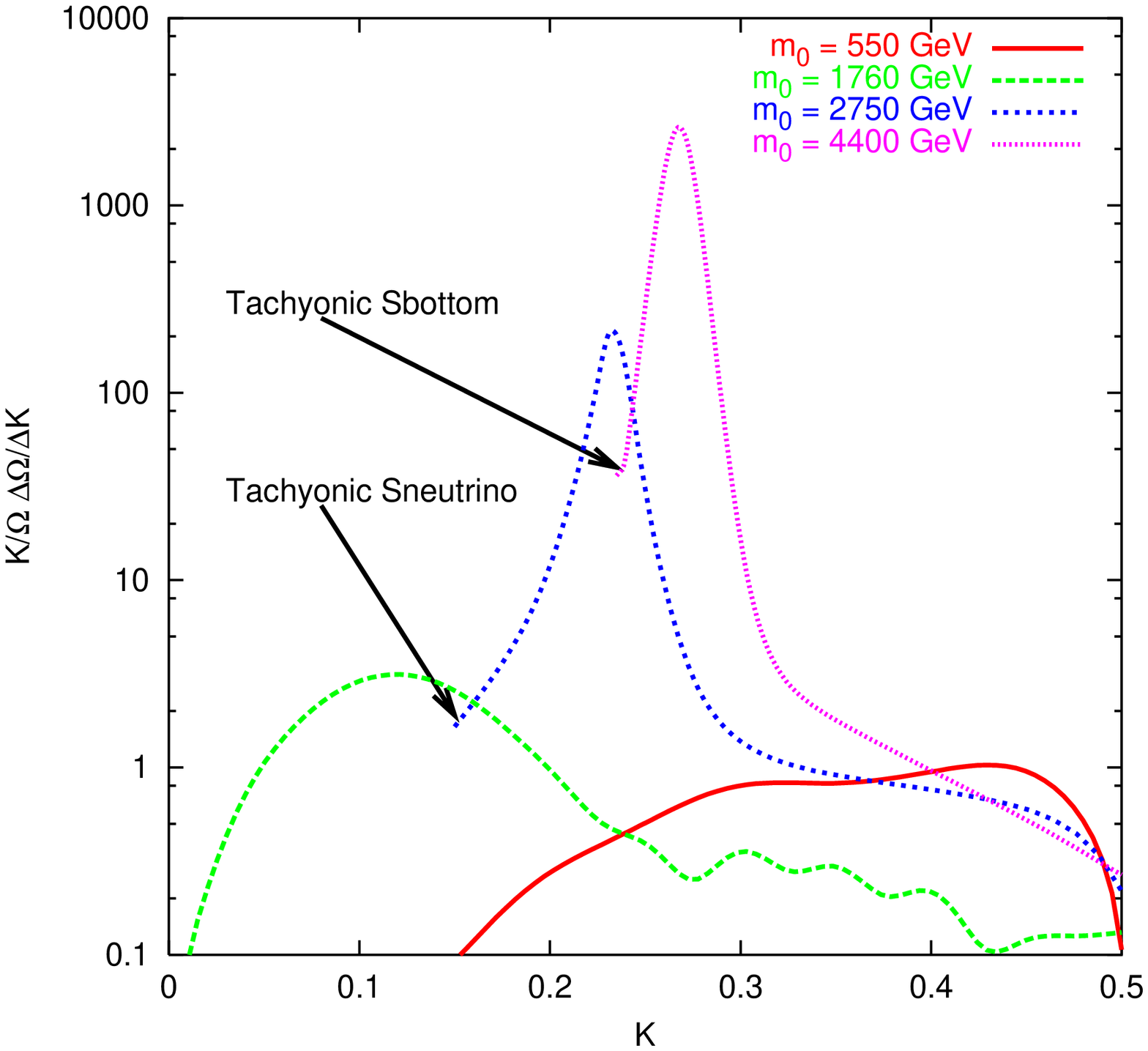}
\caption{\em\small The fine-tuning parameter $\Delta^\Omega(K)\equiv \frac{K}{\Omega_{\neut}}\frac{\partial \Omega_{\neut}}{\partial K}$ for the four benchmark values of $m_0=550,\ 1760,\ 2750,\ 4400 / {\rm GeV}$. The outher SUSY input parameters are fixed at $\tan\beta=38.0,\ M_{1/2}=1100/ {\rm GeV},\ A_0=0$, hence the benchmark $m_0$ correspond respectively to $(m_0/M_{1/2})=0.5,\ 1.6,\ 2.5,\ 4$ (see fig.\ref{corridorMAIN}). At the left end of the lines corresponding to $m_0=2750,\ 4400$ the respective NLSP becomes tachyonic, as indicated.}
\label{FT}
\end{center}
\end{figure}
\noindent The question of the quantification of fine-tuning in SUSY models
for dark matter, besides involving the naturalness of the models,
is also related to the possibility of reliably predicting the
relic density $\Omega_\chi h^2$ from accelerator measurements of
the SUSY input parameters \cite{FToliveellis}. We emphasize that
the present mNUSM models can hardly be regarded as realistic SUSY
models, being conceived in order to address the issue of extended
sfermion coannihilation processes in the presence of non-universal
sfermion masses boundary conditions. Nevertheless it is natural to
face here the problem of the fine-tuning related to
the extra parameter $K$ appearing in mNUSM models. In fact, as
emerging from fig.\ref{spectraSneuSbot} and
fig.\ref{corridorMAIN}, the new coannihilation regions appearing
at large $m_0$ for $K\lesssim0.5$ are expected to generate abrupt
changes in the neutralino relic density for small variations of
the mNUSM parameter $K$. We will follow here the formalism of
ref. \cite{FToliveellis} in order to study the logarithmic
dependence of the neutralino relic density . There, the amount of
fine-tuning was described through the overall logarithmic
sensitivity
\begin{equation}
\displaystyle
\Delta^\Omega\equiv\sqrt{\sum_i\left(\frac{a_i}{\Omega_\chi}\frac{\partial\Omega_\chi}{\partial
a_i}\right)^2},
\end{equation}
where $a_i$ are generic SUSY input parameters. As outlined in
\cite{baerstau}, the coannihilation regions tend to have higher
fine-tuning due to the steep rise of the cross sections at the onset of coannihilation processes. In the present case we concentrate only in the contribution $\Delta^\Omega(K)$
coming from the parameter $K$, and study four benchmark cases at
fixed $\tan\beta=38.0$, $M_{1/2}=1100 \ {\rm GeV}$ and $A_0=0$.
Motivated by fig.\ref{corridorMAIN}, we choose to plot the
following values for $(m_0/M_{1/2})$: 0.5, corresponding to the
ordinary stau NLSP coannihilation region; 1.6, intersecting the
low $K$ sneutrino coannihilation region, where fine-tuning of $K$
is expected to be reasonably low; 2.5 and 4.0, values which cut the steep
part of the new coannihilation branch, corresponding respectively to tau sneutrino and sbottom NLSP, and where fine-tuning is expected to be rather high. Fig.\ref{FT} reproduces our results:
for $(m_0/M_{1/2})=0.5$ and 1.6 we get, as expexted, low values for $\Delta^\Omega(K)\lesssim3$. On the other hand, the
sharp onset of coannihilations for $(m_0/M_{1/2})=2.5,4.0$ entails
dramatic $K$ fine-tuning peaks, corresponding to the values of $K$ such that $m_{\neut}\simeq m_{\sneu}$ and $m_{\neut}\simeq m_{\sbot}$ respectively. In the case of the sbottom, as suggested
by fig.\ref{spectraSneuSbot}, the steepness of $m_{\sbot}(K)$
renders the fine-tuning needed to get $m_{\sbot}\simeq m_{\neut}$
particularly large.\\ 
\noindent To conclude, we find that the extended
sfermion coannihilation channels of mNUSM models are
characterized by a large amount of fine-tuning in the
non-universality parameter $K$ in the narrow upper part of the coannihilation branch depicted in fig.\ref{corridorMAIN}. Low fine-tuning of $K$ is instead required in the lower part of the branch, at $K\lesssim0.2$ and $1\lesssim(m_0/M_{1/2})\lesssim2$ and in the standard stau coannihilation strip. Nonetheless, we stress that these
scenarios, though motivated in the context of $SU(5)$ SUSY GUTs,
are {\em ad hoc} sketchily simplified, reducing to an only
parameter $K$ the scalar non-universality variables, in order to
focus on peculiar coannihilation phenomena. In  ``realistic''
models one could expect to reproduce more naturally, hence with a smaller amount of
fine-tuning, the new outlined coannihilation regions.

\section{Conclusions}

In this paper we studied the cosmological and phenomenological consequences of top-down $b$-$\tau$ Yukawa coupling unification with minimal non-universal boundary conditions in the sfermion masses, inspired by $SU(5)$ GUT. We showed that the $\mu>0$ case is ruled out by the $b$-quark mass bound. We stress that this result holds also in the particular case of full universality (CMSSM). As for the $\mu<0$ case, $b$-$\tau$ YU is compatible with the set of all known cosmo-phenomenological constraints, among which the recent results from WMAP, for $31\lesssim\tan\beta\lesssim45$. A large parameter space region is also found to be consistent with the 2-$\sigma$ range of $\delta a_\mu$ determined from $\tau$ decay data. Further, if one resorts to a super-conservative approach \cite{superconservative} to  $\delta a_\mu$, the whole allowed regions discussed would fulfill the resulting bound.\\
We found that the SUSY spectrum allows for new types of coannihilations, namely neutralino-sbottom and neutralino-tau sneutrino-stau, that we analysed in detail. We fixed three benchmark scenarios for the three possible coannihilation patterns, including the CMSSM-like case of neutralino-stau, and we showed for these three cases the behavior of the cosmological and phenomenological constraints as functions of the LSP mass. We then discussed the cosmologically allowed regions for the two types of new coannihilations at various values of $\tan\beta$, and the main channels contributing to the neutralino relic density suppression. An analytical approximate treatment of these channels is given in the Appendix.\\
\vspace*{-0.7cm}
\subsection*{Acknowledgments}
\vspace*{-0.2cm} I would like to thank S.~Bertolini, S.~Petcov,
P.~Ullio and C.E.~Yaguna for many useful discussions, comments and
for support. I also acknowledge C.~Pallis for collaboration during
the early stages of this work, and A.~Djouadi and V.~Spanos for
helpful remarks. I am grateful to the Referee for useful
suggestions and comments.

\clearpage

\appendix

\section{Neutralino relic density calculation in the presence of coannihilations}

The starting point to compute the neutralino relic density is the generalization of the Boltzmann equation to a set of $N$ coannihilating species \cite{BGSalati,GriestSeckel}
\begin{equation}
\frac{dn}{dt}\ =\ -3Hn-\langle\sigma_{\rm eff}v\rangle\left(n^2-n^2_{\rm eq}\right),\label{boltzy}
\end{equation}
where $H$ is the Hubble constant, $n\equiv\sum_{i=1}^N\ n^i$ is the total number density summed over all coannihilating particles, $n_{\rm eq}$ is the equilibrium number density, which in the Maxwell-Boltzmann approximation, valid in the present cases, reads
\begin{equation}
n_{\rm eq}\ =\ \frac{T}{2\pi^2}\sum_{i=1}^N g_i\ m_i^2\ K_2\left(\frac{m_i}{T}\right)
\end{equation}
$g_i$ being the internal degrees of freedom of particle $i$ of
mass $m_i$, $T$ the photon temperature and $K_2(x)$ a modified
Bessel function. In eq.(\ref{boltzy}) $\sigma_{\rm eff}$ is the
effective cross section, defined as
\begin{equation}
\sigma_{\rm eff}\ =\ \sum_{i,j=1}^N\ \sigma_{ij}\ r_i\ r_j.
\end{equation}
In its turn, $\sigma_{ij}$ is the total cross section for the processes involving $ij$ (co-)annihilations, averaged over initial spin and particle-antiparticle states. The coefficients $r_i$, in the reasonable approximation \cite{GriestSeckel} where the ratio of the number density of species $i$ to the total number density maintains its equilibrium before, during and after freeze out, are defined as
\begin{equation}
r_i\ \equiv \ \frac{n^i_{\rm eq}}{n_{\rm eq}}=\ \frac{g_i\ \left(1+\Delta_i\right)^{3/2}{\rm e}^{-\Delta_i\ x}}{g_{\rm tot}}
\label{rcoeff}
\end{equation}
where
\begin{equation}
g_{\rm tot}=\sum_{i=1}^N\ g_i\ \left(1+\Delta_i\right)^{3/2}{\rm e}^{-\Delta_i\ x},\quad \Delta_i=\frac{m_i-m_{\neut}}{m_{\neut}},\quad x\ \equiv\ \frac{m_{\neut}}{T}.
\end{equation}
From (\ref{rcoeff}) it is apparent that only species which are quasi degenerate in mass with the LSP can effectively contribute to the coannihilation processes, since large mass differences are exponentially suppressed. Once numerically determined the freeze-out temperature $T_F$ \cite{GriestKamTurner}, the LSP relic density at the present cosmic time can be evaluated by \cite{GriestSeckel}
\begin{equation}
\Omega_{\neut}~h^2\approx\frac{1.07 \times 10^9
~{\rm GeV}^{-1}}{g_*^{1/2}M_{P}~x_F^{-1}~
\hat\sigma_{\rm eff}},
\label{omega}
\end{equation}
where $M_P=1.22 \times 10^{19}$ GeV is the Planck scale, $g_{*}\approx81$ is number of effective degrees of freedom at freeze out, $x_F=m_{\neut}/T_F$ and
\begin{equation}
\hat\sigma_{\rm eff}\equiv x_F\int_{x_F}^{\infty}
\langle\sigma_{\rm eff}v\rangle x^{-2}dx~.
\label{sigmaeff3}
\end{equation}
keeps track of the efficiency of post-freeze-out annihilations.\\
In many cases, one can approximate the thermally averaged product of the relative velocity and the cross section of the (co-)annihilating particles through a Taylor expansion
\begin{equation}
\sigma_{ij}v\ =\ a_{ij}+b_{ij}v^2
\end{equation}
This approximation is not accurate near $s$-channel poles and final-state thresholds, as pointed out in ref. \cite{GriestSeckel, GondoloGelmini}. In all other cases, and namely in the great part of the ones studied in the present paper (an exception is given in fig.\ref{omegaUBC}), one can proceed and calculate
\begin{equation}
\hat\sigma_{\rm eff}=\sum_{i,j}(\alpha_{ij}a_{ij}+
\beta_{ij}b_{ij})\equiv
\sum_{ij}\hat\sigma_{ij}~,
\label{sigmaeff}
\end{equation}
where the sum is extended to all the possible pairs of initial sparticle states, and the coefficients
\begin{equation}
\alpha_{ij}=x_F\int_{x_F}^\infty
\frac{dx}{x^2}r_i(x)r_j(x)~,
~\beta_{ij}=6x_F\int_{x_F}^\infty
\frac{dx}{x^3}r_i(x)r_j(x).
\label{alphabeta}
\end{equation}
We list in Tables \ref{CoanProc1} - \ref{CoanProc4} all the possible annihilation and coannihilation processes involving the neutralino, the stau, the sbottom and the tau sneutrino. Only a subset of these reactions effectively contribute to the reduction of the neutralino relic density, as described in sec.\ref{neutsbot} and \ref{neutstausneut}. In particular, contrary to the case of neutralino annihilation, the largest contributions to (\ref{sigmaeff}) arising from coannihilation processes come from the $a_{ij}$ coefficients. In sec.\ref{acoefficients} we give the analytical form of the $a_{ij}$ for the ``new'' coannihilations which arise in the present context of mNUSM. Namely, we study the most relevant processes in the cases of sbottom-neutralino, sbottom-sbottom and stau-tau sneutrino coannihilations.\\
A numerical check of the formul\ae \ given in sec.\ref{acoefficients}, consisting in a comparison between the computation outlined in this Appendix and the numerical result given by {\tt micrOMEGAs} confirmed the expected validity of this approximate treatment to a satisfactory extent, in the regimes not affected by the $A$-pole effects.

\newpage

\subsection{The coannihilation processes}

We report in Tables \ref{CoanProc1} - \ref{CoanProc4} the complete list of all (Co-)annihilation processes involving the lightest neutralino, sbottom and stau as well as the tau sneutrino. For a given couple of initial sparticles we list both all the possible final states and the tree level channels relative to any final state. c means \emph{4-particles contact interaction}, while $s(X),\ t(X)$ and $u(X)$ mean an $s$, $t$ or $u$ channel where $X$ is the exchanged (s)particle. $d$ and $u$ indicate respectively the down and up-type quarks, while $l$ and $\nu$ the charged leptons and the neutrinos of any family, where not differently specified. $f$ stands for a generic fermion (quark or lepton) \\
\vspace{0.5cm}

\begin{table}[!h]
\begin{center}
\begin{tabular}{| c |  c | c |}

\hline
& & \\[-0.3cm]

Initial state & Final states & Tree level channels \\[0.2cm]

\hline\hline
& & \\[-0.3cm]
$\neut\ \neut$ & $h\ h$, \  $h\ H$, \  $H\ H$ ,\ $A\ A$,\ $Z\ Z$,\ $A\ Z$,\  & s($h$), \  s($H$),\ t($\tilde\chi^0_i$), \ u($\tilde\chi^0_i$)\\[0.2cm]
 & $h[H]\ A$,\ $h[H]\ Z$ & s($A$),\ s($Z$),\ t($\tilde\chi^0_i$), \ u($\tilde\chi^0_i$)\\[0.2cm]
 & $W^+\ W^-$, $H^+\ H^-$ & s($h$), \  s($H$), \  s($Z$),\ t($\tilde\chi^-_j$), \ u($\tilde\chi^-_j$)\\[0.2cm]
 & $W^\pm\ H^\pm$ & s($h$), \  s($H$), \  s($A$),\ t($\tilde\chi^-_j$), \ u($\tilde\chi^-_j$)\\[0.2cm]
 & $f\ \overline{f}$ &  s($h$), \  s($H$),\ s($A$),\ s($Z$),\ t($\tilde f$), \ u($\tilde f$)\\[0.2cm]
\hline
& & \\[-0.3cm]
$\neut \ \sbot$ & $b\  h[H]$,\ $b\  Z$ & s($b$), \ t($\tilde b_{1,2}$), \ u($\tilde\chi^0_i$)\\[0.2cm]
 & $b\  A$ & s($b$),\ t($\tilde b_{2}$), \ u($\tilde\chi^0_i$)\\[0.2cm]
 & $b\  \gamma$,\ $b\ g$ & s($b$), \ t($\tilde b_{1}$)\\[0.2cm]
 & $t \ H^-$, $t \ W^-$ &
s($b$), \ t($\tilde t_{1,2}$), \ u($\tilde\chi^-_j)$\\[0.2cm]
\hline
& & \\[-0.3cm]
$\neut \ \stau$ & $\tau\  h[H]$,\ $\tau\  Z$ & s($\tau$), \ t($\tilde\tau_{1,2}$), \ u($\tilde\chi^0_i$)\\[0.2cm]
 & $\tau\  A$ & s($\tau$),\ t($\tilde\tau_{2}$), \ u($\tilde\chi^0_i$)\\[0.2cm]
 & $\tau\  \gamma$ & s($\tau$), \ t($\tilde\tau_{1}$)\\[0.2cm]
 & $\nu_\tau \ H^-$,\ $\nu_\tau \ W^-$ &
s($\tau$), \ t($\tilde\nu_\tau$), \ u($\tilde\chi^-_j)$\\[0.2cm]
\hline
& & \\[-0.3cm]
$ \neut \ \tilde{\nu}_\tau$ & $\nu_\tau\  h[H]$,\ $\nu_\tau \ Z$ & s($\nu_\tau$), \ t($\tilde\nu_\tau$), \ u($\tilde\chi^0_i$)\\[0.2cm]
 & $\nu_\tau\  A$ &  \ u($\tilde\chi^0_i$)\\[0.2cm]
 & $\tau\  H^+$,\ $\tau\  W^+$ &
s($\nu_\tau$), \ t($\tilde\tau_{1,2}$), \ u($\tilde\chi^-_j)$\\[0.2cm]
\hline
\end{tabular}

\vspace{0.5cm}

\caption{\emph{Neutralino annihilations and coannihilation with sbottom, stau and sneutrino.}}\label{CoanProc1}

\end{center}

\end{table}

\begin{table}[!h]
\begin{center}
\begin{tabular}{| c |  c | c |}

\hline
& & \\[-0.3cm]

Initial state & Final states & Tree level channels \\[0.2cm]

\hline\hline
& & \\[-0.3cm]
$\sbot \ \sbot$ & $b\  b$  & t($\tilde{\chi}^0_i$), \   u($\tilde{\chi}^0_i$),\ t($\tilde g$),\ u($\tilde g$)\\[0.2cm]
\hline
& & \\[-0.3cm]
$ \sbot \ \sbot^*$ & $h\ h$, \  $h\ H$, \  $H\ H$ , $Z\ Z$ & s($h$), \  s($H$), \  t($\tilde b_{1,2}$), \ u($\tilde b_{1,2}$),  c\\[0.2cm]

 & $A\ A$ & s($h$), \  s($H$), \  t($\tilde b_{2}$), \ u($\tilde b_{2}$), \  c\\[0.2cm]

 & $Z\ A$ & s($h$), \  s($H$), \  t($\tilde b_{2}$), \ u($\tilde b_{2}$)\\[0.2cm]

 & $A\ h[H]$ & s($Z$), \ t($\tilde b_{2}$), \ u($\tilde b_{2}$)\\[0.2cm]

 & $Z\ h[H]$& s($Z$), \ t($\tilde b_{1,2}$), \ u($\tilde b_{1,2}$)\\[0.2cm]

 & $W^+\ W^-$, $H^+\ H^-$ & s($h$), \  s($H$), \  s($Z$), \  t($\tilde t_{1,2}$), \  c,\ s($\gamma$)\\[0.2cm]

 & $W^\pm\  H^\mp$ & s($h$), \  s($H$), \  t($\tilde t_{1,2}$)\\[0.2cm]

 & $t\ \overline{t}$,\  $u\ \overline{u}$& s($h$), \  s($H$), \ s($Z$),\ s($\gamma$),\ s($g$),\ t($\tilde{\chi}^-_j$)\\[0.2cm]

 & $b\ \overline{b}$ & s($h$), \  s($H$), \ s($Z$),\ s($\gamma$),\ s($g$),\ t($\tilde{\chi}^-_j$),\ t($\tilde g$)\\[0.2cm]

 & $d\ \overline{d}$ & s($h$), \  s($H$), \ s($Z$),\ s($\gamma$),\ s($g$)\\[0.2cm]

 & $l\ \overline{l}$ & s($h$), \  s($H$), \ s($Z$),\ s($\gamma$)\\[0.2cm]

 & $\nu\ \overline{\nu}$ & s($Z$)\\[0.2cm]

 & $\gamma\   \gamma$,\ $\gamma \ Z$,\ $\gamma\  g$,\ $Z g$&  t($\tilde b_{1}$), \ u($\tilde b_{1}$),\ c\\[0.2cm]

 & $\gamma\  h[H]$,\ $g\ h[H]$ & t($\tilde b_{1}$), \ u($\tilde b_{1}$)\\[0.2cm]

 & $g\ g$ & s($g$),\  t($\tilde b_{1}$), \ u($\tilde b_{1}$)  ,\ c \\[0.2cm]
\hline
& & \\[-0.3cm]
$\sbot\ \stau$  & $b\ \tau$ & t($\tilde{\chi}^0_i$)\\[0.2cm]
\hline
 & & \\[-0.3cm]
$\sbot\ \stau^*$ & $b\  \overline{\tau}$ & t($\tilde{\chi}^0_i$)\\[0.2cm]
 &  $t\ \overline{\nu}_\tau$ & t($\tilde{\chi}^-_j$)\\[0.2cm]
\hline
& & \\[-0.3cm]
$\sbot\ \sneu$ & $b\ \nu_\tau$ & t($\tilde{\chi}^0_i$)\\[0.2cm]
 &  $t\ \tau$ & t($\tilde{\chi}^-_j$)\\[0.2cm]
\hline
& & \\[-0.3cm]
$\sbot\ \sneu^*$& $b\ \overline{\nu}_\tau$ & t($\tilde{\chi}^0_i$)\\[0.2cm]
\hline

\end{tabular}

\vspace{0.5cm}

\caption{\emph{Sbottom annihilations and coannihilations with stau and sneutrino.}}\label{CoanProc2}

\end{center}

\end{table}

\begin{table}[!h]
\begin{center}
\begin{tabular}{| c |  c | c |}

\hline
& & \\[-0.3cm]

Initial state & Final states & Tree level channels \\[0.2cm]

\hline\hline

& & \\[-0.3cm]
$\stau \ \stau$ & $\tau\  \tau$  & t($\tilde{\chi}^0_i$), \   u($\tilde{\chi}^0_i$)\\[0.2cm]
\hline
& & \\[-0.3cm]
$ \stau \ \stau^*$ & $h\ h$, \  $h\ H$, \  $H\ H$ , $Z\ Z$ & s($h$), \  s($H$), \  t($\tilde\tau_{1,2}$), \ u($\tilde\tau_{1,2}$),  c\\[0.2cm]

 & $A\ A$ & s($h$), \  s($H$), \  t($\tilde\tau_{2}$), \ u($\tilde\tau_{2}$), \  c\\[0.2cm]

 & $A\ Z$ & s($h$), \  s($H$), \  t($\tilde\tau_{2}$), \ u($\tilde\tau_{2}$)\\[0.2cm]

 & $A\ h[H]$ & s($Z$), \ t($\tilde\tau_{2}$), \ u($\tilde\tau_{2}$)\\[0.2cm]

 &$Z\ h[H]$ & s($Z$), \ t($\tilde\tau_{1,2}$), \ u($\tilde\tau_{1,2}$)\\[0.2cm]

 & $W^+\ W^-$, $H^+\ H^-$ & s($h$), \  s($H$), \  s($Z$), \  u($\tilde\nu_\tau$), \  c,\ s($\gamma$)\\[0.2cm]

 & $W^\pm\ H^\mp$ & s($h$), \  s($H$), \  u($\tilde\nu_\tau$)\\[0.2cm]

 & $u\ \overline{u}$,\ $d\ \overline{d}$,\  $l\ \overline{l}$& s($h$), \  s($H$), \ s($Z$),\ s($\gamma$)\\[0.2cm]

 & $\tau\ \overline{\tau}$ & s($h$), \  s($H$),\ s($Z$), \  t($\tilde{\chi}^0_i$),\ s($\gamma$)\\[0.2cm]

 & $\nu_\tau\ \overline{\nu}_\tau$ & s($Z$), \  t($\tilde{\chi}^-_j$)\\[0.2cm]

 & $\nu\ \overline{\nu}$ & s($Z$)\\[0.2cm]

 & $\gamma\   \gamma$,\ $\gamma \ Z$& c,\ t($\tilde\tau_{1}$), \ u($\tilde\tau_{1}$)\\[0.2cm]
 & $\gamma\  h[H]$ & t($\tilde\tau_{1}$), \ u($\tilde\tau_{1}$)\\[0.2cm]
\hline
& & \\[-0.3cm]
$ \stau \ \sneu $ & $\nu_\tau \ \tau$ & t($\tilde{\chi}^0_i$), \  u($\tilde{\chi}^-_j$)\\[0.2cm]
\hline
& & \\[-0.3cm]

$ \stau\ \sneu^* $ & $\overline{u}\ d$,\ $\overline{\nu}\ l$& s($H^-$), \  s($W^-$)\\[0.2cm]

 & $\overline{\nu_\tau}\ \tau$ & s($W^-$), \ s($H^-$),\  t($\tilde{\chi}^0_i$)\\[0.2cm]

 & $Z\ W^-$ & s($W^-$), \  u($\tilde{\tau}_{1,2}$), \  t($\sneu$), \  c\\[0.2cm]

 & $\gamma\ W^-$ & s($W^-$), \  u($\tilde{\tau}_{1}$), \  c\\[0.2cm]

 & $Z\ H^-$ & s($H^-$), \ u($\tilde{\tau}_{1,2}$) , \  t($\sneu$)\\[0.2cm]

 & $\gamma\ H^-$ & s($H^-$), \  u($\tilde{\tau}_{1}$)\\[0.2cm]

 & $W^-\ h$, \  $W^-\ H$ & s($H^-$), \  s($W^-$), \  u($\tilde{\tau}_{1,2}$) , \  t($\sneu$)\\[0.2cm]

 & $W^-\ A$ & s($H^-$),\  u($\tilde{\tau}_{2}$)\\[0.2cm]

 & $H^-\ h$, \  $H^-\ H$  & s($H^-$), \  s($W^-$), \  u($\tilde{\tau}_{1,2}$) , \  t($\sneu$), \  c\\[0.2cm]

 & $H^-\ A$ & s($W^-$),\  u($\tilde{\tau}_{2}$),\  c\\[0.2cm]
\hline
\end{tabular}

\vspace{0.5cm}

\caption{\emph{Stau annihilations and coannihilations with sneutrino.}}\label{CoanProc3}
\end{center}
\end{table}

\clearpage

\begin{table}[!th]
\begin{center}
\begin{tabular}{| c |  c | c |}
\hline
& & \\[-0.3cm]

Initial state & Final states & Tree level channels \\[0.2cm]

\hline\hline

& & \\[-0.3cm]

$\sneu \ \sneu$ & $\nu_\tau\  \nu_\tau$ & t($\tilde{\chi}^0_i$), \   u($\tilde{\chi}^0_i$)\\[0.2cm]

\hline
& & \\[-0.3cm]
$ \sneu \ \sneu^* $ & $h\ h$, \  $h\ H$, \  $H\ H$ , $Z\ Z$ & s($h$), \  s($H$), \  t($\tilde\nu_\tau$), \ u($\tilde\nu_\tau$),  c\\[0.2cm]
 & $A\ A$ & s($h$), \  s($H$), \  c\\[0.2cm]

 & $Z\ A$ & s($h$), \  s($H$)\\[0.2cm]

 & $Z\ h[H]$& s($Z$), \ t($\tilde\nu_\tau$), \ u($\tilde\nu_\tau$)\\[0.2cm]

 & $W^+\ W^-$, $H^+\ H^-$ & s($h$), \  s($H$), \  s($Z$), \  t($\tilde\tau_{1,2}$), \  c\\[0.2cm]

 & $W^\pm \ H^\mp$ & s($h$), \  s($H$), \  t($\tilde\tau_{1,2}$)\\[0.2cm]
 & $u\ \overline{u}$,\ $d\ \overline{d}$,\ $l\ \overline{l}$,  & s($h$), \  s($H$), \
s($Z$)\\[0.2cm]

 & $\tau\ \overline{\tau}$ &s($h$), \  s($H$), \ s($Z$), \  t($\tilde{\chi}^0_i$)\\[0.2cm]

 &$\nu_\tau\ \overline{\nu}_\tau$ & s($Z$), \  t($\tilde{\chi}^0_i$)\\[0.2cm]

 &  $A\ h[H]$,\ $\nu\ \overline{\nu}$ & s($Z$)\\

\hline

\end{tabular}

\vspace{0.2cm}
\caption{\emph{Sneutrino annihilations.}}\label{CoanProc4}
\vspace{-1.cm}
\end{center}
\end{table}


\subsection{Relevant approximate formul\ae \ for the relic density calculation in the coannihilation regions}
\label{acoefficients}

The $a$ coefficients for the pair annihilation of neutralinos can be readily and completely derived from ref \cite{RoskNeut}, while those concerning the stau annihilations and coannihilations can be calculated from the formul\ae \ of ref. \cite{RoskCoan} and \cite{EllisStau1, EllisNUHM}. Some instances of $a$ coefficients computations can be found in ref. \cite{pallisstau}, with the corrections given in \cite{YQU}. We follow here the notation set in \cite{pallisstau}. As regards the tau sneutrino (Co-)annihilations, the relative $a$ coefficients can be readily derived from those of the stau via suitable replacements in the couplings (e.g. $g_{\stau\stau h}\rightarrow g_{\sneu\sneu h}$ etc.), in the masses of the (s-)particles (e.g. $m_{\stau}\rightarrow m_{\sneu}$) and setting the mixing angle between the stau mass eigenstates to zero. We do not give any explicit formula for the sbottom-stau and the sbottom-sneutrino coannihilations since even in the transition regions where $m_{\sbot}\simeq m_{\sneu}\simeq m_{\stau}$ these processes do not give any sizeable contribution to the neutralino relic density suppression (namely, their total contribution is always less than 0.5\%). Therefore we are left with the ``new'' (co-)annihilation processes involving respectively \emph{neutralino-sbottom}, \emph{sbottom-sbottom} and \emph{stau-tau sneutrino}.\\
We give here the explicit formul\ae \ for the dominant processes, as discussed  in sec.\ref{neutsbot} and \ref{neutstausneut}. In the kinematic part of the $a$-coefficients we neglect the mass of the final Standard Model Particles up to the $b$-quark included. Since the mass of the final SM particles $m_{SM}$ appears there as $m_{SM}^2/m_{SUSY}^2$, the corrections are in fact always negligible. On the other hand, we keep trace of both $m_b$ and $m_\tau$ if the couplings or the whole amplitudes are proportional to them, as it is the case in some of the considered processes. Moreover, in the processes involving the sbottom we followed the approximations of ref. \cite{EllisStopCoann} neglecting the terms in $\alpha_{em}$ and $\alpha_{W}$. Bar over a mass means that the mass is divided by the sum of the masses of the incoming sparticles. $S$ stands for the three neutral physical Higgs $h$,\ $H$ and $A$, and $m_{S}$ for the respective masses, $s_{\sss X}\equiv\sin\theta_{\sss X}$, $c_{\sss X}\equiv\cos\theta_{\sss X}$ and for the other symbols we follow the notation of ref. \cite{GunionHaber}.

\subsubsection{Couplings}

\begin{table}[!hb]
\begin{center}
\begin{tabular}{| c | c |}

\hline

{\bf $g$ Symbol} & {\bf Expression} \\
\hline & \\[-0.22cm]
$g^L_{\sbot b \neut_i}$ & ${\displaystyle\frac{-g_2}{\sqrt{2}}\left(c_{\tilde b}\left(N_{i2}-\frac{\tan\theta_W}{6}N_{i1}\right)+s_{\tilde b}\frac{m_b N_{i3}}{m_W \cos\beta}\right)}$
\\[0.2cm]
\hline & \\[-0.22cm]
$g^R_{\sbot b \neut_i}$ & ${\displaystyle\frac{-g_2}{\sqrt{2}}\left(c_{\tilde b}\frac{m_b N_{i3}}{m_W \cos\beta}+\frac{2}{3}s_{\tilde b}\tan\theta_W N_{i1}\right){\rm sign}(m_{\tilde\chi^0_i})}$
\\[0.2cm]
\hline & \\[-0.3cm]
$g_{\sbot\sbot g}$=$g_{b b g}$=$g_{ggg}$ & $-g_s$
\\[0.15cm]
\hline & \\[-0.22cm]
$g^L_{\sbot b g}$ & $\sqrt{2} g_s \ s_{\tilde b}$
\\[0.2cm]
\hline & \\[-0.22cm]
$g^R_{\sbot b g}$ & $-\sqrt{2} g_s \ c_{\tilde b}$
\\[0.2cm]
\hline & \\[-0.22cm]
$g^{\oplus}_{\sbot b \tilde\chi_i}$ & $\frac{1}{2}\left( g^L_{\sbot b \tilde\chi_i}+g^R_{\sbot b \tilde\chi_i}  \right)$
\\[0.2cm]
\hline & \\[-0.22cm]
$g^{\ominus}_{\sbot b \tilde\chi_i}$ & $\frac{1}{2}\left( g^L_{\sbot b \tilde\chi_i}-g^R_{\sbot b \tilde\chi_i}  \right)$
\\[0.2cm]
\hline & \\[-0.3cm]
$g_{\sbot\sbot gg}$ & $g_s^2$
\\[0.15cm]
\hline & \\[-0.22cm]
$g_{ h[H] \tilde\tau_1 \tilde\tau_1}$ &
${\displaystyle-\frac{g_2}{2c_W} m_Zs_{\alpha+\beta}[-c_{\alpha+\beta}]
\Big((1-2s^2_W) s_{\tilde\tau}^2+2s^2_W c_{\tilde\tau}^2\Big)+
}$\\[0.2cm]
& ${\displaystyle +\frac{g_2m_\tau}{M_Wc_\beta}\Big[m_\tau s_\alpha[-c_\alpha]
+s_{\tilde\tau} c_{\tilde\tau}\Big(A_\tau s_\alpha[-c_\alpha]+
\mu c_\alpha[s_\alpha]\Big)\Big]}$
\\[0.2cm]
\hline & \\[-0.32cm]
$g_{A \tilde\tau_1 \tilde\tau_1}$ & 0
\\[0.1cm]
\hline & \\[-0.22cm]
$g_{W\tilde\nu_\tau\tilde\tau_1}$ &
${\displaystyle \frac{g_2 s_{\tilde\tau}}{\sqrt{2}}}$
\\[0.2cm]
\hline & \\[-0.22cm]
$g_{W\tilde\nu_\tau\tilde\tau_2}$ &
${\displaystyle -\frac{g_2 c_{\tilde\tau}}{\sqrt{2}}}$
\\[0.2cm]
\hline & \\[-0.22cm]
$g_{H^+ h[H]W^-}$ &
${\displaystyle -\frac{g_2}{2}c_{\alpha-\beta}[s_{\alpha-\beta}]}$
\\[0.2cm]
\hline & \\[-0.22cm]
$g_{H^+ AW^-}$ &
${\displaystyle \frac{g_2}{2}}$
\\[0.2cm]
\hline & \\[-0.22cm]
$g_{H^+\tilde\nu_\tau \tilde\tau_1}$ &
${\displaystyle \frac{g_2 M_W s_{\tilde\tau}}{\sqrt{2}}s_{2\beta}}$
\\[0.2cm]
\hline

\end{tabular}

\vspace{0.3cm}

\caption{\emph{Relevant couplings used in the formul\ae \ of sec.\ref{FormulaeFirst} (I)}}\label{Couplings}
\end{center}
\end{table}


\begin{table}
\begin{center}
\begin{tabular}{| c | c |}

\hline

{\bf $g$ Symbol} & {\bf Expression} \\

\hline & \\[-0.22cm]
$g_{ h[H]W^+W^-}$ &
${\displaystyle g_2M_Ws_{\beta-\alpha}[c_{\beta-\alpha}]}$
\\[0.2cm]
\hline & \\[-0.32cm]
$g_{AW^+W^-}$ &
0\\[0.1cm]
\hline & \\[-0.22cm]
$g_{ h[H]\tilde\tau_1\tilde\tau_2}$ &
${\displaystyle \frac{g_2}{2c_W}m_Z s_{\alpha+\beta}[-c_{\alpha+\beta}]
(1-4s^2_W)s_{\tilde\tau} c_{\tilde\tau}}$
\\[0.2cm]
& ${\displaystyle +g_2m_\tau(c^2_{\tilde\tau}-s^2_{\tilde\tau})\frac{A_\tau s_{\alpha}[-c_{\alpha}]
+\mu c_{\alpha}[s_{\alpha}]}{2M_W c_\beta}}$
\\[0.2cm]
\hline & \\[-0.22cm]
$g_{A\tilde\tau_1\tilde\tau_2}$ &
${\displaystyle g_2\ m_\tau\ \frac{A_\tau\tan_\beta+\mu}{2M_W}}$
\\[0.2cm]
\hline & \\[-0.22cm]
$g_{WZ\tilde\nu_\tau \tilde\tau_1}$ &
${\displaystyle \frac{g_2^2 s_W^2}{\sqrt{2} c_W}\ s_{\tilde\tau}}$
\\[0.2cm]
\hline & \\[-0.22cm]
$g_{WWZ}$ &
${\displaystyle g_2\ c_W}$
\\[0.2cm]
\hline & \\[-0.22cm]
$g_{Wtb}$ &
${\displaystyle -\frac{g_2}{\sqrt{2}}}$
\\[0.2cm]
\hline & \\[-0.22cm]
$g_{H^+ tb}^L$ &
${\displaystyle \frac{g_2}{\sqrt{2}M_W}\left(m_t \ \cot \beta\right)}$
\\[0.2cm]
\hline & \\[-0.22cm]
$g_{H^+ tb}^R$ &
${\displaystyle \frac{g_2}{\sqrt{2}M_W}\left(m_b \ \tan \beta\right)}$
\\[0.2cm]
\hline & \\[-0.22cm]
$g^L_{\tilde\tau_1 \tau \tilde\chi^0_i} $ &
${\displaystyle s_{\tilde\tau}\frac{g_2}{\sqrt{2}M_W}\frac{N^*_{i3}m_\tau}{c_\beta}+c_{\tilde\tau}\left(\frac{\sqrt{2}g_2}{c_W}s_W^2N^{\prime *}_{i2}-\sqrt{2}g_1c_WN^{\prime *}_{i1}\right)}$
\\[0.2cm]
\hline & \\[-0.22cm]
$g^R_{\tilde\tau_1 \tau \tilde\chi^0_i}$ &
${\displaystyle -s_{\tilde\tau}\sqrt{2}\left(g_1c_WN^\prime_{i1}+\frac{g_2}{c_W}\left(\frac{1}{2}-s_W^2\right)N^\prime_{i2}\right)-c_{\tilde\tau}\frac{g_2}{\sqrt{2}M_W}\frac{N_{i3}m_\tau}{c_\beta}}$
\\[0.2cm]
\hline & \\[-0.32cm]
$g^L_{\tilde\nu_\tau \nu_\tau \tilde\chi^0_i}$ &
${\displaystyle 0}$
\\[0.1cm]
\hline & \\[-0.22cm]
$g^R_{\tilde\nu_\tau \nu_\tau \tilde\chi^0_i}$ &
${\displaystyle -\frac{g_2}{\sqrt{2}M_W}N^\prime_{i2}}$
\\[0.2cm]
\hline & \\[-0.22cm]
$g^L_{\tilde\tau_1 \nu_\tau \tilde\chi^-_j}$ &
${\displaystyle \frac{g_2}{\sqrt{2}M_W}\frac{U_{j2}m_\tau}{c_\beta}c_{\tilde\tau}}$
\\[0.2cm]
\hline & \\[-0.22cm]
$g^R_{\tilde\tau_1 \nu_\tau \tilde\chi^-_j}$ &
${\displaystyle g_2U_{j1}s_{\tilde\tau}}$
\\[0.2cm]
\hline & \\[-0.22cm]
$g^L_{\tilde\nu_\tau \tau \tilde\chi^-_j}$ &
${\displaystyle \frac{g_2}{\sqrt{2}M_W}\frac{U^*_{j2}m_\tau}{c_\beta}}$
\\[0.2cm]
\hline & \\[-0.22cm]
$g^R_{\tilde\nu_\tau \tau \tilde\chi^-_j}$ &
${\displaystyle -g_2V_{j1}}$
\\[0.2cm]
\hline

\end{tabular}

\vspace{0.5cm}

\caption{\emph{Relevant couplings used in the formul\ae \ of sec.\ref{FormulaeFirst} (II)}}
\end{center}
\end{table}

\clearpage


\subsubsection{$\neut-\sbot$ coannihilations}
\label{FormulaeFirst}


\begin{table*}[!h]
\begin{tabular}{c|l}

\hline

\begin{rotate}{90}{\hspace{-1.5cm} $\neut \ \ \sbot \rightarrow\ \ b \ \ g $ }
\end{rotate}

&

\begin{tabular}{l}

\begin{minipage}[t]{13cm}

\begin{eqnarray*}
&&\frac{1}{24\pi m_{\sbot}^2}\Bigg\{\Big(g^L_{\sbot b \neut_1}+g^R_{\sbot b \neut_1}\Big)\Big[g^2_{\sbot\sbot g}(\bar\mneut-\bar m_{\sbot})+g^2_{bbg}\bar m_{\sbot}\Big] \\
&&\\
&&\\
&&- 4 g_{bbg}g_{\sbot\sbot g}\Big(\left(g^L_{\sbot b \neut_1}\right)^2+\left(g^R_{\sbot b \neut_1}\right)^2\Big)(\bar\mneut-\bar m_{\sbot})\Bigg\}\\
&&\\
\end{eqnarray*}

\end{minipage}\\
\end{tabular}\\

\hline

\end{tabular}

\end{table*}



\subsubsection{$\sbot-\sbot$ annihilations}


\begin{tabular}{c|l}

\hline

\begin{rotate}{90}{\hspace{-1cm} $\sbot \ \ \sbot \rightarrow\ \ b \ \ b $ }
\end{rotate}

&

\begin{tabular}{l}

\begin{minipage}[t]{13cm}

\begin{eqnarray*}
&&\frac{\left(\left(g_{\sbot b g}^L\right)^2+\left(g_{\sbot b g}^R\right)^2\right)\ m_{\tilde g}^2\ +\ g_{\sbot b g}^L\ g_{\sbot b g}^R m_{\sbot}^2}{54\ \pi\ \left(m_{\tilde g}^2+m_{\sbot}^2\right)^2}\\
&&\\
&&\\
&&+\ \sum_{i,j=1}^4\ \Bigg\{ \frac{m_{\tilde\chi_i}\ m_{\tilde\chi_j}\ \left(\left(g^{\oplus}_{\sbot b \tilde\chi_i}\right)^2+\left(g^{\ominus}_{\sbot b \tilde\chi_i}\right)^2\right)\left(\left(g^{\oplus}_{\sbot b \tilde\chi_j}\right)^2+\left(g^{\ominus}_{\sbot b \tilde\chi_j}\right)^2\right)}{2\ \pi\ \left(m_{\tilde\chi_i}^2+m_{\sbot}^2 \right)\ \left(m_{\tilde\chi_j}^2+m_{\sbot}^2 \right)}\\
&&\\
&&\\
&&+\ \frac{4\ m_{\tilde\chi_i}\ m_{\tilde\chi_j}\ g^{\oplus}_{\sbot b \tilde\chi_i}\ g^{\ominus}_{\sbot b \tilde\chi_i}\ g^{\oplus}_{\sbot b \tilde\chi_j}\ g^{\ominus}_{\sbot b \tilde\chi_j}\ +\ m_{\sbot}^2\left(g^{\oplus}_{\sbot b \tilde\chi_i}\right)^2\left(g^{\ominus}_{\sbot b \tilde\chi_j}\right)^2}{4\ \pi\ \left(m_{\tilde\chi_i}^2+m_{\sbot}^2 \right)\ \left(m_{\tilde\chi_j}^2+m_{\sbot}^2 \right)}\Bigg\}\\
&&\\
\end{eqnarray*}

\end{minipage}\\
\end{tabular}\\

\hline
\hline\\

\begin{rotate}{90}{\hspace{-1cm} $\sbot \ \ \sbot^* \rightarrow\ \ g \ \ g $ }
\end{rotate}

&

\begin{tabular}{l}

\begin{minipage}[t]{13cm}

\begin{eqnarray*}
&&\frac{81\ g_{ggg}^2\ g_{\sbot\sbot g}^2\ +\ 56\ g_{\sbot\sbot g g}\ \left(2g_{\sbot\sbot g g}-g_{\sbot\sbot g}\right)}{1728\ \pi\ m_{\sbot}^2}\\
&&\\
\end{eqnarray*}

\end{minipage}\\
\end{tabular}\\
\hline
\end{tabular}


\clearpage

\subsubsection{$\stau-\sneu$ coannihilations}


\begin{table*}[!h]

\begin{tabular}{c|l}

\hline

\begin{rotate}{90}{\hspace{-1.3cm}$\tilde\nu_\tau\ \ \tilde\tau_2^*\ \rightarrow\ S \ W^+ $ }
\end{rotate}

&

\begin{tabular}{l}

\begin{minipage}[t]{13cm}

\begin{eqnarray*}
&& \frac{\left( \big[ 1-\big(\bar M_W + \bar m_{S}\big)^2 \big]\big[ 1-\big(\bar
M_W - \bar m_{S}\big)^2 \big] \right)^{3/2}}{128\ \pi\  M_W^2\  m_{\tilde\nu_\tau}
m_{\tilde\tau_1} } \Bigg[\ \frac{2\ g_{S\tilde\tau_1 \tilde\tau_1}\ g_{W\tilde\nu_\tau\tilde\tau_1}\
m_{\tilde\nu_\tau}}{m_{\tilde\nu_\tau}\bar
m_{S}^2-m_{\tilde\tau_1}(1-\bar M_W^2)}\\
&&\\
&& + \ \frac{2\ g_{H^+SW^-}\ g_{H^+\tilde\nu_\tau
\tilde\tau_1}\bar m_{\tilde\nu_\tau}}{1-\bar m_{H^\pm}^2}
+ \frac{g_{W\tilde\nu_\tau\tilde\tau_1}\ g_{SW^+W^-}\ \left((\bar
m_{\tilde\nu_\tau}-\bar m_{\tilde\tau_1})^2-\bar M_W^2\right)}{\bar
M_W^2\left(1-\bar M_W^2\right)}\\
&&\\
&&+\frac{2\ g_{S\tilde\tau_1\tilde\tau_2}\
g_{W\tilde\nu_\tau\tilde\tau_2}\ m_{\tilde\nu_\tau}}{m_{\tilde\nu_\tau}\bar
m_{S}^2-\bar m^2_{\tilde\nu_\tau}m_{\tilde\tau_1}-m_{\tilde\nu_\tau}(\bar
m_{\tilde\tau_2}^2+\bar m_{\tilde\tau_1}^2)-m_{\tilde\tau_1}(\bar
m^2_{\tilde\tau_2}-\bar M_W^2)}\Bigg]^2  \\
\end{eqnarray*}

\end{minipage}\\
\\
\end{tabular}\\
\hline
\hline

\begin{rotate}{90}{\hspace{-1.8cm} $\tilde\nu_\tau\ \ \tilde\tau_2^*\ \rightarrow\ Z \ W^+ $}
\end{rotate}

&

\begin{tabular}{l}

\begin{minipage}[t]{13cm}

\begin{eqnarray*}
&& \frac{\left( \big[ 1-\big(\bar M_W + \bar M_Z\big)^2 \big]\big[ 1-\big(\bar
M_W - \bar M_Z\big)^2 \big] \right)^{1/2}}{32\ \pi\  m_{\tilde\nu_\tau}\
m_{\tilde\tau_2} }\Bigg[\left(\frac{\bar M_W^2+\bar M_Z^2-1}{2\bar M_Z \bar M_W}\right)^2+2\Bigg]\\
&&\\
&&\times \Bigg[\ 7\ g_{WZ\tilde\nu_\tau \tilde\tau_1}^2 \ +\ 7\ g_{W\tilde\nu_\tau \tilde\tau_1}^2\ g_{WWZ}^2\left(1-\frac{M_Z^2}{M_W^2}\right)\\
&&\\
&& +\ 4\ g_{WZ\tilde\nu_\tau \tilde\tau_1}\ g_{W\tilde\nu_\tau \tilde\tau_1}\ g_{WWZ} \left(1-\frac{M_Z^2}{M_W^2}\right)^2\ \Bigg]\\
 \end{eqnarray*}

\end{minipage}\\
\\
\end{tabular}\\

\hline
\hline

\begin{rotate}{90}{\hspace{-1cm}$\tilde\nu_\tau\ \ \tilde\tau_2^*\ \rightarrow\ b \bar t  $}
\end{rotate}

&

\begin{tabular}{l}

\begin{minipage}[t]{13cm}

\begin{eqnarray*}
&& \frac{\Big( 1- \bar m_t^2 \Big)^{2}}{16\ \pi\  m_{\tilde\nu_\tau} m_{\tilde\tau_1} }\Bigg[\ g_{W\tilde\nu_\tau\tilde\tau_1}^2 \ g_{Wtb}^2 \ \frac{\bar m_b \bar m_t}{\bar M_W^4}+\\
&&\\
&& +\ \frac{\bar g_{H^+\tilde\nu_\tau\tilde\tau_1}^2}{\left(1-\bar m_{H^\pm}^2\right)^2}\Big(\ \Big[\big|g_{H^+tb}^L\big|^2+\big|g_{H^+tb}^R\big|^2\Big] \ \bar m_b\ \bar m_t +\  \left( g_{H^+tb}^L \ g_{H^+tb}^R\ \right)\left(\bar m_t^2 -1\right)\Big)\\
&&\\
&& +\frac{\bar g_{H^+\tilde\nu_\tau\tilde\tau_1}\ g_{W\tilde\nu_\tau\tilde\tau_1} \ g_{Wtb}}{\bar M_W^2\ \left(1-\bar m_{H^\pm}^2\right) }\ \Big(  g_{H^+tb}^L \bar m_t  \left(1-\bar m_t^2\right) -\ g_{H^+tb}^R \bar m_b  \left(1 +\bar m_t^2 \right)  \Big)
\Bigg]\\
\end{eqnarray*}

\end{minipage}\\
\\
\end{tabular}\\
\hline

\end{tabular}

\end{table*}


\newpage

\begin{table*}[!h]

\begin{tabular}{c|l}

\hline

\begin{rotate}{90}{\hspace{-1.5cm} $\tilde\nu_\tau\ \ \tilde\tau_2\ \rightarrow\ \nu_\tau  \ \tau  $ }
\end{rotate}

&

\begin{tabular}{l}

\begin{minipage}[t]{13cm}

\begin{eqnarray*}
&&\left(\frac{m_\tau^2}{32\ \pi\  m_{\tilde\nu_\tau} m_{\tilde\tau_1}}\right)\
\Bigg\{\ \sum_{i,k=1}^4\ \frac{m_{\tilde\nu_\tau}^2\ \mathrm{Re}\Bigg[g^L_{\tilde\tau_1 \tau \tilde\chi^0_i}\ \left(g^L_{\tilde\tau_1 \tau \tilde\chi^0_k}\right)^*\ g^R_{\tilde\nu_\tau \nu_\tau \tilde\chi^0_i}\ \left(g^R_{\tilde\nu_\tau \nu_\tau \tilde\chi^0_k}\right)^*\Bigg]}{\left(m_{\tilde\chi^0_i}^2+m_{\tilde\nu_\tau} m_{\tilde\tau_1}\right)\left(m_{\tilde\chi^0_k}^2+m_{\tilde\nu_\tau} m_{\tilde\tau_1}\right)}\\
&&\\
&&\\
&& +\ \sum_{i,k=1}^4\ \frac{m_{\tilde\nu_\tau}^2\ \mathrm{Re}\Bigg[\ g^L_{\tilde\nu_\tau \nu_\tau \tilde\chi^0_i}\ \left(g^L_{\tilde\nu_\tau \nu_\tau \tilde\chi^0_k}\right)^*\ g^R_{\tilde\tau_1 \tau \tilde\chi^0_i}\ \left(g^R_{\tilde\tau_1 \tau \tilde\chi^0_k}\right)^*\Bigg]}{\left(m_{\tilde\chi^0_i}^2+m_{\tilde\nu_\tau} m_{\tilde\tau_1}\right)\left(m_{\tilde\chi^0_k}^2+m_{\tilde\nu_\tau} m_{\tilde\tau_1}\right)}\\
&&\\
&&\\
&&+\sum_{j,l=1}^2\ \frac{ m_{\tilde\tau_1}^2\ \mathrm{Re}\Bigg[g^L_{\tilde\tau_1 \nu_\tau \tilde\chi^-_j}\ \left(g^L_{\tilde\tau_1 \nu_\tau \tilde\chi^-_l}\right)^*\ g^R_{\tilde\nu_\tau \tau \tilde\chi^-_j}\ \left(g^R_{\tilde\nu_\tau \tau \tilde\chi^-_l}\right)^*\ \Bigg]}{\Big(m_{\tilde\chi^-_j}^2+m_{\tilde\nu_\tau} m_{\tilde\tau_1}\Big)\Big(m_{\tilde\chi^-_l}^2+m_{\tilde\nu_\tau} m_{\tilde\tau_1}\Big)}\\
&&\\
&&\\
&&+\sum_{j,l=1}^2\ \frac{ m_{\tilde\tau_1}^2\ \mathrm{Re}\Bigg[\ g^L_{\tilde\nu_\tau \tau \tilde\chi^-_j}\ \left(g^L_{\tilde\nu_\tau \tau \tilde\chi^-_l}\right)^*\ g^R_{\tilde\tau_1 \nu_\tau \tilde\chi^-_j}\ \left(g^R_{\tilde\tau_1 \nu_\tau \tilde\chi^-_l}\right)^*\Bigg]}{\Big(m_{\tilde\chi^-_j}^2+m_{\tilde\nu_\tau} m_{\tilde\tau_1}\Big)\Big(m_{\tilde\chi^-_l}^2+m_{\tilde\nu_\tau} m_{\tilde\tau_1}\Big)}\\
&&\\
&&\\
&&-\sum_{i=1}^4\sum_{j=1}^2\frac{m_{\tilde\tau_1}m_{\tilde\nu_\tau}\mathrm{Re}\left(\left(
g^L_{\tilde\nu_\tau \tau \tilde\chi^-_j}\right)^*\ g^L_{\tilde\tau_1 \tau \tilde\chi^0_i}\ g^R_{\tilde\nu_\tau \nu_\tau \tilde\chi^0_i}\ \left(g^R_{\tilde\tau_1 \nu_\tau \tilde\chi^-_j}\right)^*\ \right)}{\left(m_{\tilde\chi^0_i}^2+m_{\tilde\nu_\tau}m_{\tilde\tau_1}\right)\left(m_{\tilde\chi^-_j}^2+m_{\tilde\nu_\tau}m_{\tilde\tau_1}\right)}\\
&&\\
&&\\
&&-\sum_{i=1}^4\sum_{j=1}^2\frac{m_{\tilde\tau_1}m_{\tilde\nu_\tau}\mathrm{Re}\left(\ g^L_{\tilde\nu_\tau \nu_\tau \tilde\chi^0_i}\ \left(g^L_{\tilde\tau_1 \nu_\tau \tilde\chi^-_j}\right)^*\ \left(g^R_{\tilde\nu_\tau \tau \tilde\chi^-_j}\right)^*\ g^R_{\tilde\tau_1 \tau \tilde\chi^0_i}
\right)}{\left(m_{\tilde\chi^0_i}^2+m_{\tilde\nu_\tau}m_{\tilde\tau_1}\right)\left(m_{\tilde\chi^-_j}^2+m_{\tilde\nu_\tau}m_{\tilde\tau_1}\right)}\Bigg\}\\
\end{eqnarray*}

\end{minipage}\\[0.5cm]
\end{tabular}\\
\hline

\end{tabular}

\end{table*}

\end{document}